
\input harvmac
%
%
%
%
\ifx\answ\bigans
\else
\output={
  \almostshipout{\leftline{\vbox{\pagebody\makefootline}}}
  \advancepageno
}
\fi
%
%
%
\def\mayer{\vbox{\baselineskip=14truept
\sl\centerline{Department of Physics 0319}%
\centerline{University of California, San Diego}
\centerline{9500 Gilman Drive}
\centerline{La Jolla, CA 92093-0319}}}
%
%

%
%
%
%
\def\abstract#1{\centerline{\bf Abstract}\nobreak\medskip\nobreak\par
#1}
%
%
%
%
%
%
%
%

%
\def\inv{^{\raise.15ex\hbox{${\scriptscriptstyle -}$}\kern-.05em 1}}
\def\lbar{{\lower.35ex\hbox{$\mathchar'26$}\mkern-10mu\lambda}}


%
%
%
%
\def\dsl{\,\raise.15ex\hbox{/}\mkern-13.5mu D}

\def\delsl{\raise.15ex\hbox{/}\kern-.57em\partial}
\def\Ksl{\hbox{/\kern-.6000em\rm K}}
\def\Asl{\hbox{/\kern-.6500em \rm A}}
\def\Dsl{\hbox{/\kern-.6000em\rm D}} 
\def\Qsl{\hbox{/\kern-.6000em\rm Q}}
\def\gradsl{\hbox{/\kern-.6500em$\nabla$}}
%
%
\def\lspace{\ifx\answ\bigans{}\else\qquad\fi}
\def\lbspace{\ifx\answ\bigans{}\else\hskip-.2in\fi} 
%
%
\def\boxeqn#1{\vcenter{\vbox{\hrule\hbox{\vrule\kern3pt\vbox{\kern3pt
        \hbox{${\displaystyle #1}$}\kern3pt}\kern3pt\vrule}\hrule}}}
%
%
\def\mbox#1#2{\vcenter{\hrule \hbox{\vrule height#2in
\kern#1in \vrule} \hrule}}
%
%
%
%
\def\CA{{\cal A}}  \def\CC{{\cal C}} \def\CD{{\cal
D}}
   \def\CH{{\cal
H}}

%
%
%
%
%

%

\def\bar#1{\overline{#1}}

\def\darr#1{\raise1.5ex\hbox{$\leftrightarrow$}\mkern-16.5mu #1}

%
%
\def\frac#1#2{{\textstyle{#1\over #2}}} 

%
%
%
%

\def\Tr{\mathop{\rm Tr}}

%
%
%
%

%
%
\def\ltap{\ \raise.3ex\hbox{$<$\kern-.75em\lower1ex\hbox{$\sim$}}\ }
\def\gtap{\ \raise.3ex\hbox{$>$\kern-.75em\lower1ex\hbox{$\sim$}}\ }
\def\gl{\ \raise.5ex\hbox{$>$}\kern-.8em\lower.5ex\hbox{$<$}\ }
\def\roughly#1{\raise.3ex\hbox{$#1$\kern-.75em\lower1ex\hbox{$\sim$}}
}
%
%

%

%

\def\npb#1#2#3{{Nucl. Phys. } B{#1} (#2) #3}

\def\prl#1#2#3{{Phys. Rev. Lett. } {#1} (#2) #3}
\def\physrev#1#2#3{{Phys. Rev. } {#1} (#2) #3}

\relax

\def\dpart{\partial\kern .5ex\llap{\raise
1.7ex\hbox{$\leftrightarrow$}}\kern -.7ex {_\mu}}

\def\frac#1#2{{\textstyle{#1 \over #2}}}

\def\s2weak{\sin^2\theta_{\rm w}}

\def\({\left(}\def\){\right)}

\def\mayer{\vbox{\baselineskip=14truept\sl\centerline{Department of Physics,
9500 Gilman Drive 0319}
\centerline{University of California, San Diego}
\centerline{La Jolla, CA 92093-0319}}}

\def\Queens{\vbox{\baselineskip=14truept
\sl\centerline{Department of Physics, Stirling Hall}
\centerline{Queen's University}
\centerline{Kingston, Canada, K7L 3N6}}}
\def\Duke{\vbox{\baselineskip=14truept
\sl\centerline{Department of Physics}
\centerline{Duke University, Durham, NC 27706}}}

\def\[{\left[}
\def\]{\right]}
\def\({\left(}
\def\){\right)}

\noblackbox
 at 12truept
\vskip 1.in
\centerline{{\titlefont{Strong and Electromagnetic Decays}}}
\medskip
\centerline{{\titlefont{of the Baryon Decuplet}}}
\bigskip
\centerline{Malcolm N.\ Butler}
\bigskip
\Queens
\bigskip
\centerline{Martin J.\ Savage\footnote{$^{\dagger}$}
{SSC Fellow}}
\bigskip
\mayer
\bigskip
\centerline{Roxanne P.\ Springer}
\bigskip
\Duke
\bigskip
\centerline{{\bf Abstract}}
\medskip

To explore further the role that QCD symmetries play on hadron dynamics, we
discuss the radiative decay of the baryon decuplet to the baryon octet and
compute the leading nonanalytic SU(3) violating corrections induced at the
one loop level.  In the limit of exact flavour SU(3) symmetry two of the
possible decay modes are forbidden and the rates for these two decays are
dominated by loop corrections.  There is only one SU(3) conserving contact
term at leading order in chiral perturbation theory for the radiative
decays.  We determine its coefficient from the observed branching ratio for
$\Delta\rightarrow N\gamma$ and from the present upper limit on
$\Xi^{*0}\rightarrow\Xi^0\gamma$.  We then predict the branching fractions
for $\Sigma^{*-}\rightarrow\Sigma^-\gamma$,
$\Sigma^{*0}\rightarrow\Lambda\gamma$,
$\Sigma^{*0}\rightarrow\Sigma^0\gamma$,
$\Sigma^{*+}\rightarrow\Sigma^+\gamma$, $\Xi^{*0}\rightarrow\Xi^0\gamma$
and
$\Xi^{*-}\rightarrow\Xi^-\gamma$.  Some of the decay modes are predicted to
have branching fractions near the current experimental upper limit and could
possibly be observed in the near future.  The leading corrections to the
octet-decuplet-meson strong coupling constant, $\CC$, are computed at
one-loop and the strong decay rates of the $\Delta, \Sigma^*$ and $\Xi^*$
allow us to determine $|\CC| = 1.2\pm 0.1$ and $\CH = -2.2\pm 0.6$, in
remarkable agreement with predictions based on  spin-flavour symmetry.
We calculate the $\Delta$ pole graph contribution to the polarisability of
the nucleon to order 1/$\Lambda_\chi^2$ in chiral perturbation theory.

\vfill
\hbox{\hbox{ UCSD/PTH 92-37, QUSTH-92-05,
Duke-TH-92-44\hskip 1in November 1992} }
\eject

\newsec{Introduction}
\medskip

The approximate $SU(3)_L\otimes SU(3)_R$ chiral symmetry of the strong
interactions has recently been incorporated into a lagrangian that for the
first time can consistently describe the dynamics of baryons near their
mass-shell \ref\heavybaryon{E.\ Jenkins and A.\ Manohar, Phys.\ Lett.\ {\bf
255B} (1991) 558 ; Phys.\ Lett.\ {\bf 259B} (1991) 353.}.  By performing a
field redefinition, analogous to that used in heavy quark field theory (for a
review see \ref\mbwise{M.B. Wise,``New Symmetries of the Strong
Interaction," lectures presented at the Lake Louise Winter Institute, Feb.
1991}), the large momentum associated with the classical trajectory of the
baryon is removed, and one can write down a lagrangian that has consistent
power counting in a loop expansion, and expansions in the baryon mass and
chiral symmetry breaking scale.  Calculations done using this lagrangian,
therefore, probe the properties of QCD and help us understand how these
manifest themselves in physical measurements.  The energy range of validity
for the method of heavy baryon chiral perturbation theory that will be used
here encompasses many interactions that will be probed, for instance, at
CEBAF.

In order to more deeply understand the significance of QCD symmetries, and
their effects on baryon and meson processes, it is important to perform
calculations on interactions that will be sensitive to these symmetries.
One such involves the radiative decays from the baryon decuplet to the
baryon octet.  These are presented here.  To understand the influence of
meson loops on these decay rates, we study the SU(3) violating loop
processes that probe them directly.  Because the measurements of these
radiative decay rates will be done soon, we believe it is pertinent to study
them now, in a way that is model independent, and therefore more
systematically accurate than attempts that have been made to date.

The strong decays of the baryon decuplet are calculated here to one loop.
They also probe the underlying symmetries obeyed by baryons and mesons,
and
may lead to some (model independent) explanation of why the spin-flavor
SU(6) symmetry, as implemented by the quark model, is so predictive.

\newsec{Heavy Baryon Chiral Perturbation Theory}

In this section we review the formalism of heavy baryon chiral perturbation
theory (HBChPT) developed by Manohar and Jenkins\heavybaryon.  The
dynamics
of the pseudogoldstone bosons arising from the breaking of
SU(3)$_L\otimes$SU(3)$_R$ chiral symmetry to $SU(3)_V$ symmetry are
described by the lagrangian
\eqn\gold{{\cal L}_M = {f_\pi^2\over 8}Tr[ D^\mu\Sigma^\dagger D_\mu\Sigma
]
+ .....\ \ .} With the normalizations used here, $f_\pi \approx135$ MeV is
the pion decay constant and $\Sigma$ is the exponential of the
pseudogoldstone boson field,
\eqn\sig{\Sigma = \exp(2iM/f_\pi)\ \ \ \ ,}
where $M$ is the pseudogoldstone boson octet
\eqn\octet{M =
\left(\matrix{{1\over\sqrt{6}}\eta+{1\over\sqrt{2}}\pi^0&\pi^+&K^+\cr
\pi^-&{1\over\sqrt{6}}\eta-{1\over\sqrt{2}}\pi^0&K^0\cr
K^-&\overline{K}^0&-{2\over\sqrt{6}}\eta\cr}\right)\ \ \ \ .}
The covariant derivative includes the $U(1)_Q$ of
electromagnetism
\eqn\covar{D_\mu = \partial_\mu + i\CA_\mu [Q, \,\,\, ] \ \ \ \ \ ,}
where $\CA$ is the electromagnetic field and $Q$ is the SU(3) charge
operator.

The pseudogoldstone bosons have a well defined transformation under chiral
symmetry, $\Sigma\rightarrow L\Sigma R^\dagger$, but a unique
transformation
does not exist for baryon fields.  To introduce baryon fields consistently
into the lagrangian, the field
 $\xi$ is defined,
$\xi^2=\Sigma$, so that
\eqn\squig{\xi = \exp(iM/f_\pi)\ \ \ \ .} This transforms under chiral
symmetry as
\eqn\xitransform{\xi \rightarrow L \xi U^\dagger = U \xi
R^\dagger ,}
where $U$ is defined implicitly by  \xitransform. From this,
vector fields with definite parity can be defined:
\eqn\av{ V^\mu = {1\over2}\left(\xi
D^\mu\xi^\dagger+\xi^\dagger D^\mu\xi\right),\quad A^\mu =
{i\over2}\left(\xi D^\mu\xi^\dagger-\xi^\dagger D^\mu\xi\right)\ \ \ ,}
where $A_\mu$ transforms homogeneously and $V_\mu$ transforms
inhomogeneously under chiral symmetry.

The work of ref.\heavybaryon\ shows that processes with baryons that are
nearly on their mass shell can be consistently described by a chiral
lagrangian involving fields of definite velocity.  In this representation
the chiral lagrangian becomes an expansion involving both $k/\Lambda_\chi$
and $k/M_B$, where $\Lambda_\chi$ is the chiral symmetry breaking scale (of
order 1 GeV), $k$ is the residual off-shellness of the nucleon, and $M_B$ is
the mass of the baryon.  These expansion parameters are both much less
than
unity, as required for such an expansion to be useful in calculations.  In
contrast, an expansion in $P/\Lambda_\chi$ (of order unity) where $P$ is the
full momentum of the baryon, and each order in the derivative expansion is
equally important, does not satisfy this requirement.  The HBChPT allows for
a systematic expansion in powers of $1/M_B$ and $1/\Lambda_\chi$ (for a
review see \ref\jmhungary{E.  Jenkins and A. Manohar, Proceedings of the
workshop on``Effective Field Theories of the Standard Model '', ed.  U.
Meissner, World Scientific (1992)}).  The lagrangian describing the strong
interactions of the lowest lying octet of baryons with the pseudogoldstone
bosons to lowest order in $1/M_B$, lowest order in derivatives, and lowest
order in the quark mass matrix is
\eqn\chirallag{\eqalign{{\cal L}_v^8 &= i \Tr \bar B_v \left(v\cdot \CD
\right)B_v +2D\, \Tr\, \bar B_v\, S_v^\mu\, \{ A_\mu, B_v \} +2F\,
\Tr\,  B_v\, S_v^\mu\, [A_\mu, B_v]\cr}\ \ \ \ ,} where \eqn\bv{ B_v =
\pmatrix { {1\over\sqrt2}\Sigma_v^0 + {1\over\sqrt6}\Lambda_v &
\Sigma_v^+ & p_v\cr \Sigma_v^-& -{1\over\sqrt2}\Sigma_v^0 +
{1\over\sqrt6}\Lambda_v&n_v\cr \Xi_v^- &\Xi_v^0 &-
{2\over\sqrt6}\Lambda_v
\cr }\ \ \ ,} and the covariant chiral derivative is $\CD_\mu = \partial_\mu
+ [ V_\mu ,\ \ ]$.  The subscript $v$ indicates that the fields are
redefined to remove baryon mass factors, and are of definite velocity.
Therefore, only factors of the residual off--shell momentum are generated by
derivatives acting on the baryon field.  The spin operator, $S_v$, acts on
the baryon fields.  The strong interaction coupling constants $F$ and $D$
have been determined at the one--loop level in chiral perturbation theory
\heavybaryon\ to be $F = 0.40\pm 0.03$ and $D = 0.61\pm 0.04$.  The strong
interactions of the lowest lying decuplet of baryon resonances are
incorporated to the same order in $1/M_B$, derivatives, and the quark mass
matrix by
\eqn\ltv{\eqalign{{\cal L}^{10}_v &= - i \,\bar
T_v^{\mu} (v \cdot \CD) \  T_{v\,\mu} + \Delta m \,\bar T_v^{\mu}\,
T_{v\,\mu} + \CC\,\left(\bar T_v^{\mu} A_{\mu} B_v + \,\bar B_v A_{\mu}
T_v^{\mu}\right)\cr &+2\CH \,\bar T_v^{\mu} S_{v \,\nu} A^{\nu}  T_{v
\, \mu}}\ \ \ \ ,}
where the elements of $T_v$ are (suppressing the
Lorentz index and velocity subscript)
\eqn\decmem{\eqalign{T^{111} & =
\Delta^{++}, \ \ T^{112} = {1\over\sqrt{3}}\Delta^{+}, \ \ T^{122} =
{1\over\sqrt{3}}\Delta^{0}, \ \ T^{222} = \Delta^{-}, \ \ T^{113} =
{1\over \sqrt{3}}\Sigma^{*+}\cr T^{123} & =
{1\over\sqrt{6}}\Sigma^{*0}, \ \ T^{223} = {1\over\sqrt{3}}\Sigma^{*-},
\ \ T^{133} = {1\over\sqrt{3}}\Xi^{*0}, \ \ T^{233} =
{1\over\sqrt{3}}\Xi^{*-}, \ \ T^{333} = \Omega^-}.}
 The mass splitting between the decuplet of resonances and the octet of
baryons is $\Delta m$.  The coupling constant $\CH$ cannot be extracted
directly from experimental data but because the decuplet baryons appear in
loop processes it can be extracted indirectly.  The coupling constant,
$\CC$, can be found from the strong decay of decuplet baryons to octet
baryons.

\newsec{Strong Decays of the Decuplet}

The strong decay of the decuplet of baryon resonances to the baryon octet is
determined by the coupling constant $\CC$ of \ltv.  Experimentally, the
values of $\CC$ extracted from the decay rates of the processes
$\Delta\rightarrow N\pi , \Sigma^*\rightarrow\Sigma\pi ,
\Sigma^*\rightarrow\Lambda\pi$, and $\Xi^*\rightarrow\Xi\pi$ are
centered about $1.5,
1.4, 1.4$, and $1.3$, respectively.  These, as expected, have SU(3)
violation at the $30\%$ level.  The tree level amplitude for these processes
is SU(3) conserving, which prompts us to look at loop corrections.  The
leading SU(3) violation to these decays arises from the graphs shown in
\fig\tbpi{1PI graphs for the strong decay of a decuplet baryon to an octet
 baryon.}.  We retain the leading nonanalytic terms of the form $m_K^2\log
(m_K^2/\mu^2)$, choosing the renormalisation point $\mu$ to be the chiral
symmetry breaking scale $\Lambda_\chi$ .  We include the contribution of
graphs involving $\eta$ mesons, by relating them to those
 involving $K$ mesons, through the Gell-Mann-Okubo mass formula
 $m_\eta^2={4\over 3}m_K^2$.  The corrections to this are of order
 $m_q/\Lambda_\chi$,
where $m_q$ is the current quark mass.

We find that with the nonanalytic contributions, the relevant vertex at
one--loop is
\eqn\tbpvert{\CC\bigg(1-(\alpha_{TB}+\bar\lambda_{TB}){m_K^2\over 16\pi^2
f_K^2}
\log(m_K^2/\Lambda_\chi^2)\bigg)(\overline T^\nu A_\nu B+\overline B A_\nu
T^\nu)\>,} where $T$ is a baryon resonance in the decuplet, $B$ is an octet
baryon, $A_v$ is the axial field defined earlier, and $f _K$ is the kaon
decay constant.  The coefficient $\alpha_{TB}$ comes from the one-particle
irreducible graphs shown in \tbpi\ , and $\bar\lambda_{TB}={1\over
2}(\lambda_T+\lambda_B)$ is the contribution from the wavefunction
renormalisation graphs shown in
\fig\wfrn{Graphs contributing to wavefunction renormalisation of the
decuplet and octet baryons.}.  The expressions for $\alpha_{TB}$ and
$\lambda_{TB}$ are given in Appendix A.  We use the values of $F=0.40$
and
$D=0.61$ extracted at one--loop from axial current matrix
elements\heavybaryon, and fit $\CC$ and $\CH$ to the four decays:
$\Delta^{++}\rightarrow p\pi^+ ,
\Sigma^{*+}\rightarrow\Sigma^0\pi^+ , \Sigma^{*+}\rightarrow\Lambda\pi^+$,
and  $\Xi^{*0}\rightarrow\Xi^-\pi^+$ .
Plotted in
\fig\tbpiplot{The decuplet-octet-meson coupling constant $\CC$ as a function
of the decuplet-decuplet-meson coupling constant $\CH$ for the decays
$\Delta^{++}\rightarrow p\pi^+ ,
\Sigma^{*+}\rightarrow\Sigma^0\pi^+ , \Sigma^{*+}\rightarrow\Lambda\pi^+$
and $\Xi^{*0}\rightarrow\Xi^-\pi^+$.  The width of each line represents the
$1\sigma$ error arising from experimental determinations of the baryon
resonant width.  There is no theoretical uncertainty included in the
curves.  Note that the sign of $\CC$ is chosen for ease of comparison to
SU(6) predictions, but is in fact undetermined in HBChPT.} is $\CC$ as a
function of $\CH$ for each decay mode.  The four rates can be consistently
described in terms of one value for the coupling constant $\CC$ if $1.1 <
|\CC| < 1.3$.  This requires that the coupling constant $\CH$ is constrained
to lie in the interval
 $-2.8 < \CH < -1.6$.  This is consistent with the constraint found in
ref.\heavybaryon\ of $\CH = -1.9\pm 1.0$.  These two determinations are
independent in the sense that they arise from experimental data on
completely different processes.  So we have some indication that our
computation of the allowed region for $\CH$ is valid.

The above results impact discussions on the spin-flavor SU(6) symmetry.
The
heavy baryon octet and decuplet fields may be combined into a single {\bf
56} representation of an SU(6) spin-flavour group \heavybaryon\jmhungary
\ref\carone{C. Carone and H. Georgi, \npb{375}{1992}{243}}.  The predictions
arising from the application of this symmetry work well for many
interactions, despite the fact that from a QCD standpoint this symmetry
should be completely destroyed by the strong dynamics.  If it is assumed
that the goldstone boson field, $A_\mu$, transforms as a {\bf 35} of SU(6)
(instead of a {\bf 405} or {\bf 2695}), then the following relations emerge
between coupling constants (the same relations can be found from
non-relativistic quark models):
\eqn\susix{\eqalign{&F={2\over 3}D\ \ \ \ \ \ \CC=-2D\cr &\CH=-3D}\ \ \ \ .}
The computations performed in ref.\heavybaryon\ show that the first
relation, $F={2\over 3}D$, is very well satisfied in HBChPT.  We cannot
extract the sign of $\CC$ (it has no physical consequence in HBChPT),
 but find the relation, $|\CC|=2D$, is also very well satisfied (with the
central values of $|\CC| = 1.2$ and $2D=1.2$.)  The third relation,
$\CH=-3D$, would indicate that $\CH\sim -1.8$.  This is within the $1\sigma$
region extracted for $\CH$ from \tbpiplot.  It seems that the relations
between coupling constants that come from invoking an SU(6) spin-flavour
symmetry are all well satisfied in HBChPT calculations.  This lends credence
to the idea, discussed in ref.\jmhungary\ , that there may well be an
underlying SU(6) symmetry manifest in the hadronic sector.

\newsec{Radiative Decays of the Decuplet}

Radiative decays of the decuplet of baryon resonances provide insight into
the
electromagnetic interaction of baryons complementary to that obtained from
static properties of the lowest lying baryon octet.
The Coleman-Glashow relations for the magnetic moments of octet baryons
\ref\colglas{S. Coleman and S.L. Glashow, \prl{6}{1961}{423}.}
in the limit of exact SU(3) symmetry
work remarkably well and require that loop-induced SU(3) violation be small
\ref\pagels{D.G. Caldi and H. Pagels, \physrev{D10}{1974}{3739}}
\ref\sdgang{E. Jenkins, M. Luke, A.V. Manohar and M.J. Savage,
UCSD/PTH-92-34}.  In contrast, little is known about the radiative decays of
the decuplet of baryon resonances.  The branching ratio for the radiative
decay of the $\Delta$ is inferred to be $0.0061\pm 0.0005$
\ref\pdg{Particle Data Group, \physrev{D45}{1992}{1}}, while only upper
limits exist for the branching ratios of the $\Sigma^*$ and $\Xi^*$
radiative decays.  Meanwhile, the theoretical investigations have been
limited to relativistic quark models and nonrelativistic potential models
\ref\quark{E. Kaxiras, E.J. Moniz and M. Soyeur, \physrev{D32}{1985}{695}},
where it was found that the decay rates were very sensitive to the details
of the model (Primakoff production of the resonances via hyperon beams has
also been examined in the context of the quark model
\ref\lipkin{H. Lipkin, \physrev{D7}{1973}{846}}).  We will use chiral
perturbation theory to make predictions for the radiative decay modes of the
decuplet of baryon resonances.

There is one SU(3) conserving, dimension five, counterterm that contributes
to the radiative decay of the baryon decuplet of the form
\eqn\trans{{\cal L}^{TB\gamma}_v = i \Theta{e\over \Lambda_\chi} \bar B_v
S^\mu_v Q T_v^\nu F_{\mu\nu}\ \ \ \ \ ,} where $Q$ is the electromagnetic
charge matrix,
\eqn\charge{Q = \left(\matrix{ 2/3&0&0\cr 0&-1/3&0\cr 0&0&-1/3}\right) \ \ \
\ ,} and $\Theta$ is an unknown coefficient that we will subsequently
determine. Note that, formally, there is an analogous counterterm
proportional 1/ $M_B$.  Here, this type of term is absorbed into the above
operator.
 In the limit of exact SU(3), this term forbids the decays
$\Sigma^{*-}\to\Sigma^-\gamma$ and $\Xi^{*-}\to\Xi^-\gamma$.  This is a
consequence of
 the $V$--spin invariance of both the strong and electromagnetic
interactions.  However, these decays are induced at loop level through the
SU(3) violation that occurs from keeping explicit the $K$ and $\pi$ masses
in the loop diagrams, and by insertions of the quark mass matrix in local
operators.

At one--loop, two graphs contribute to the radiative decay of decuplet
baryons.  These are shown in
\fig\graphs{Graphs contributing to the radiative decay of decuplet baryons.
The heavy line represents a decuplet baryon, the solid line an octet baryon,
the dashed line an octet meson, and the wavy line a photon. We work in
$\epsilon_\gamma\cdot v=0$ gauge where there is no direct coupling of the
photon to either the octet or decuplet baryon.}.  It is these graphs that
give the leading SU(3) violating terms in the decay rates, and are therefore
the dominant contribution to the decays $\Sigma^{*-}\to\Sigma^-\gamma$ and
$\Xi^{*-}\to\Xi^-\gamma$.  In calculating the two graphs in $4-2\epsilon$
dimensions, it is convenient to define
\eqn\integs{\eqalign{I_1&=\int_0^1dx x
\Gamma(\epsilon) I(-\epsilon,\omega_\gamma x-\Delta m,M^2)\cr
I_2&=\int_0^1dx(x-1)\Gamma(\epsilon)I(-\epsilon,\omega_\gamma x-\Delta m,
M^2)\cr}\ \ \ \ \ ,} where $\omega_\gamma$ is the energy of the photon
emitted during the decay, $\Delta m$ is the mass difference between the
initial decuplet baryon and the intermediate state baryon, $M$ is the mass
of the goldstone boson in the loop, and
\eqn\ieps{I(a,b,c)=\int_0^\infty d\lambda
(\lambda^2+2b\lambda+c)^a\>.}
Recursion relations and rules for the latter integral are discussed in
ref.~\jmhungary.   An SU(3) symmetric
$\overline{\rm MS}$ subtraction scheme is used to define the finite terms in
$I_1$ and $I_2$, and
we use
\eqn\subtract{\Gamma(\epsilon)I(-\epsilon,b,c) = b\left[
\log(c/\Lambda_\chi^2) -2\right]
-\sqrt{b^2-c}\log\left({b-\sqrt{b^2-c+i\epsilon}\over
b+\sqrt{b^2-c+i\epsilon}}\right)\ \ \ \ ,} where only finite pieces are
shown.  The matrix element for radiative decays can be written as
\eqn\matel{{\cal M} = X\ \ \overline{B}S\cdot k\ T\cdot\epsilon_\gamma\ \ \
+\ \ \ Y\ \ \overline{B} S\cdot\epsilon_\gamma\ k\cdot T\ \ \ \ ,} where $X$
and $Y$ include contributions from both one-loop graphs and the counterterm;
\eqn\xx{\eqalign{ X =\ & \ -ie\left[ Q_{TB}{\Theta\over \Lambda_\chi} +
{1\over 4\pi^2f^2} \bigg(\beta_{TBB}I_2-\beta_{TTB}[I_1-{2\over
3}I_2]\bigg)\right]\cr Y =\ & \ -ie\left[ -Q_{TB}{\Theta \over\Lambda_\chi}+
{1\over 4\pi^2f^2} \bigg(\beta_{TBB}I_1-\beta_{TTB}[I_2-{2\over
3}I_1]\bigg)\right]\cr}\ \ \ \ ,} where $\epsilon_\gamma$ is the photon
polarization vector, $\beta_{TBB}$ and $\beta_{TTB}$ are the
Clebsch--Gordan
coefficients relevant for each decay (from one-loop graphs with an octet and
decuplet baryon, respectively, in the intermediate state), and $Q_{TB}$ is
the Clebsch-Gordan coefficient in the counterterm (see Appendix B).  Note
that the integrals $I_1$ and $I_2$ depend on $\omega_\gamma$, $\Delta m$,
and $M$, and are therefore different for each intermediate state.  The
contribution from the one-loop wavefunction renormalisation graphs is not
included, being higher order in the chiral expansion.  Keeping explicit the
difference between $f_K$ and $f_\pi$ in the computations, we use $1/f_K^2$
for kaon loops and $1/f_\pi^2$ for pion loops, where $f_K=1.22 f_\pi$.  We
find the width for a given radiative decay mode $\Gamma (T\rightarrow
B\gamma)$ is
\eqn\rate{\Gamma  (T\rightarrow B\gamma) = {\omega_\gamma^3\over 8\pi}
\left[  |X|^2 + |Y|^2 + {2\over 3}(X^*Y + Y^*X)   \right]  \ \ \ .}

The $\Delta\to N\gamma$ branching ratio is believed to be $~0.56-0.66$\%
from studies of pion photoproduction.  We use this to constrain the value of
the counterterm $\Theta$ (remember that this counterterm is subtraction
scheme dependent when determined to order 1/$\Lambda_\chi^2$).  The
width
depends quadratically on $\Theta$, leading to two possible solutions.  One
of the solutions gives a branching ratio for $\Xi^{*0}\to\Xi^0\gamma$ that
is greater than the experimental upper limit of $4\%$, which eliminates this
as a physical solution.  Therefore, all of the SU(3) allowed and SU(3)
forbidden decay rates can be predicted as a function of $\CH$.  We found the
range of allowed values for $\CH$ in the previous section. We will here give
a range for each decay rate which includes both experimental uncertainties
on the widths of the resonances, and an estimate of the theoretical
uncertainties
 (about $30\%$) inherent in chiral perturbation theory.  In computing the
branching ratio for each decay we have used the following values for the
widths (in MeV) of the decuplet states\pdg :
\eqn\decwid{\eqalign{
\Gamma(\Sigma^{*0}) = 36\pm 5 \ \ \ &\ \ \ \Gamma(\Sigma^{*-}) = 39.4\pm
2.1\cr
\Gamma(\Sigma^{*+}) = 35.8\pm 0.8 \ \ \ &\ \ \ \Gamma(\Xi^{*0}) = 9.1\pm
0.5\cr
\Gamma(\Xi^{*-}) = 9.9\pm 1.8 \ \ \ & \cr}\ \ \ \ .} The estimates for the
branching fraction for each decay mode are given in Table 1.  The ones in
the lower box are SU(3) violating.

\input tables
\bigskip
\centerline{{\bf Table 1: Decuplet Branching Ratios (\%)}}
\medskip
\begintable
Decay Mode |  ~~Branching Ratio $(\%)$~~ | ~~Experimental Limit
$(\%)$~~\cr
$\Sigma^{*+}\rightarrow \Sigma^+\gamma$| $0.2- 0.6$ | $< 5$\crnorule
$\Sigma^{*0}\rightarrow \Sigma^0\gamma$| $0.04- 0.1$ |$ < 5$\crnorule
$\Sigma^{*0}\rightarrow \Lambda\gamma$| $0.8- 1.3$ | $< 5$\crnorule
$\Xi^{*0}\rightarrow \Xi^0\gamma$| $1.0- 3.0$ | $< 4$\cr
$\Sigma^{*-}\rightarrow \Sigma^-\gamma$| $0.004- 0.006$ | $<5$\crnorule
$\Xi^{*-}\rightarrow \Xi^-\gamma$| $0.01- 0.03$ | $< 4$
\endtable
\bigskip

The largest predicted branching ratio is for the $\Xi^{*0}$, which is near
the present experimental upper limit of $4\%$.  Similarly the branching
ratio for $\Sigma^{*0}\rightarrow \Lambda\gamma$ is close to the present
bound.  Although the SU(3) violating decays have much smaller branching
ratios, a
 measurement of one or both of these branching fractions would give
invaluable information on the nature of SU(3) violation in HBChPT.  Further,
it would also give an indication of whether our estimate derived from kaon
loops is reliable.  We expect that CEBAF will see these decay modes in the
coming years; the measurement will be difficult, but the return will be
enormous for its impact on our understanding of the role played by meson
exchanges, and the consequent manifestation of QCD symmetries in these
types
of processes.

\newsec{Pole Graph Contribution to Nucleon Polarisability}

We are now in a position to comment on the polarisability of the nucleon
which has been computed in chiral perturbation theory
at order $1/\Lambda_\chi^2$ with octet baryons as intermediate states
\ref\bern{V. Bernard {\it et al}, Preprint BUTP-92/15, CRN 92-24 (1992)}
and with the inclusion of the decuplet of baryon resonances
\ref\butl{M.N. Butler and M.J. Savage, UCSD/PTH 92-30, QUSTH 92-04
(1992)}.
These calculations suffer from an uncertainty in how the decuplet pole graphs
(\fig\pole{Pole graph contribution to the polarisability of the nucleon from
decuplet intermediate states.}) contribute to this process.
\foot{We thank W. Broniowski for pointing out the potential importance of
this graph to the magnetic susceptibility of the nucleon}
The importance of including the decuplet of resonances in nucleon
polarisability has been discussed in both the context of the Skyrme model and
in chiral perturbation theory by Broniowski and Cohen \ref\bc{W. Broniowski
and T.D. Cohen, University of Maryland Preprint U.of.MD PP 92-193;
T.D. Cohen and W. Broniowski, University of Maryland Preprint U.of.MD
PP 92-191.}.  Now, the same
graphs and vertices that appear in the radiative decay of the decuplet
baryons are also in the pole graph contribution to the polarisability.
Formally, we should only keep contributions at order $1/\Lambda_\chi^2$ in
the chiral expansion since contact terms at order
$1/\Lambda_\chi^3$ and higher are incalculable.  To include the pole graph
consistently, we must retain the loop corrected vertices in \pole\ (each
contributes a 1/$\Lambda_\chi$ {\sl and} a 1/$\Lambda_\chi^2$ term)
because
the pole itself enhances the naive contribution of the graph by
${\Lambda_\chi / \Delta m}$;
 terms that appear, from vertex counting, to go as 1/$\Lambda_\chi^3$ are
 actually of order 1/$\Lambda_\chi^2$ and must be kept.  In doing so, terms
 in the graph of order 1/$\Lambda_\chi^3$ (from naive vertex counting they
 are of order 1/$\Lambda_\chi^4$) are
present, but since they are of lower order they are expected to be small.
The pole graph with a $\Delta$ in the intermediate state does not contribute
to the electric
 susceptibility, $\alpha$, but does give a finite contribution to the
magnetic susceptibility, $\beta$.  We find that
\eqn\tbgbeta{\beta^{\rm pole} = -{1\over 24\pi\Delta m} Re\left( X^*Y
\right)\ \ \ \ \ ,} where the integrals in $X$ and $Y$ are now evaluated at
the off-shell point $\omega_\gamma = 0$ (where $Y=-X$).  The allowed
solution for the counterterm $\Theta$ indicates that the contribution from
the pole graphs to $\beta$ is
\eqn\nbeta{\beta^{\rm pole} = (1-5)\times 10^{-4} \ \ \ \ {\rm fm}^3\ \ \ \
.} This is consistent with the experimental determination of $\beta^{\rm
expt} = (3\pm 2\pm 2)\times 10^{-4}\ \ {\rm fm}^3$.
It should be noted here that $\beta$ may be less well determined
than the quoted experimental uncertainties suggest.
The extraction  of proton  polarisabilities  in these measurements use a
 result from dispersion
relations,\foot{We thank J. Schmiedmayer for pointing out the use of
dispersion relations in the experimental analyses. }
 $\alpha_p+\beta_p=14.2\times 10^{-4}$ fm$^3$
(ignoring a conservatively estimated uncertainty of $0.5\times10^{-4}$
fm$^3$),
as input to a theoretical cross--section, and then fit $\alpha_p-\beta_p$ to
their data
\ref\ppol{F.J.\ Fiederspiel {\it et al.}, Phys.\ Rev.\ Lett.\ {\bf 67} (1991)
1511}
(for a review see ref. \ref\lvov{A.I. L'vov, talk presented at
Brookhaven Workshop on Hadron Structure from Photoreactions at
Intermediate Energies, May 1992.}).
In principle though,
both expressions can be extracted from their data, with the result that
$\alpha_p$
does not change dramatically, but the uncertainty in $\beta_p$ completely
overwhelms the quoted central value.
For measurements of the neutron
polarisabilities, it is only possible to extract $\alpha_n$ from the
data, and the dispersion relation result of
$\alpha_n+\beta_n=15.8\pm0.5\times 10^{-4}$ fm$^3$ is then used to
estimate $\beta_n$\ref\npol{J.\ Schmiedmayer {\it et al.},
Phys.\ Rev.\ Lett.\ {\bf 66} (1991) 1015}.
What is now needed is an entirely experimental determination of
both $\alpha$ and $\beta$, independent of
 dispersion relations;  these are the numbers
that will allow us to better understand hadronic structure.

 The large theoretical
uncertainty comes from the approximations made in the $1/M_B$,
$1/\Lambda_\chi$, and loop expansions.  If we had naively used the vertices
found from the radiative decay {\sl without} evaluating at the off-shell
point we would have found $\beta^{\rm pole} = (4-8)\times 10^{-4} \ \ {\rm
fm^3} $.  This highlights the importance of retaining the momentum
dependence of the photon vertex.  Although the difference in the ranges
given between the off-shell and the on-shell vertex calculations does
not appear dramatic, in fact there is no overlap between the two.
For instance, when the off-shell calculation gives $5 \times 10^{-4}
 \ \ {\rm fm^3}$, the corresponding value for the on-shell case is
 $8 \times 10^{-4} \ \ {\rm fm^3}$. The electric susceptibility appears to be
dominated by long-distance pion loops, and the magnetic susceptibility by
the $\Delta$ pole graph.

\newsec{Conclusions}

We computed the radiative decay rates of decuplet baryons.  Two of the
decay
modes are forbidden in the limit of SU(3) but are induced at loop--level
with the dominant contribution to the decay amplitude arising from kaon
loops.  We have computed the leading nonanalytic SU(3) violating
contribution at order $m_s^{1/2}$ to all possible radiative decays of
decuplet baryons arising at one--loop.  The single (SU(3) symmetric)
counterterm is determined by using the observed branching ratio for
$\Delta\rightarrow N\gamma $ and the present upper limit on the branching
ratio of $\Xi^{*0}\rightarrow\Xi^0\gamma$.  From these constraints we have
made predictions of the branching ratios for
$\Sigma^{*+}\to\Sigma^+\gamma$,\ $\Sigma^{*0}\to\Sigma^0\gamma$, \
$\Sigma^{*0}\to\Lambda\gamma$, $\Xi^{*0}\to\Xi^0\gamma$,\ using the value
of
$\CH$ found at one-loop.  Branching ratios for the SU(3) violating decays
$\Sigma^{*-}\to\Sigma^-\gamma$ and $\Xi^{*-}\to\Xi^-\gamma$ are
independent
of the counterterm.  Some of these branching ratios should be observable in
the near future, particularly $\Sigma^{*+}\to\Sigma^+\gamma$,
$\Sigma^{*0}\to\Lambda\gamma$ and $\Xi^{*0}\rightarrow\Xi^0\gamma$, with
the
branching ratios predicted to be $\sim 1\%$.  It will be interesting to
learn if these predictions are verified by experimental observation.  In
particular, if the branching ratios of the two decay modes which proceed
only through SU(3) violation are found to be smaller than our predictions
then this will be further evidence that the leading nonanalytic contribution
from kaon loops overestimate the amount of SU(3) violation \sdgang.  It is
important that this is checked experimentally so we can determine if kaon
loops are suppressed as suggested by Gasser and Leutwyler
\ref\gass{J. Gasser and H. Leutwyler, Phys. Rep. {\bf 87} (1982) 77.}
in an approach they call improved chiral perturbation theory.

We have also discussed the strong decays of the decuplet baryons.  From
comparison with experimental measurements, we have found that the bare
value
of the decuplet-octet-meson coupling constant, $\CC$, lies in the interval
$1.1 < |\CC| < 1.3$, and that the decuplet-decuplet-meson coupling constant,
$\CH$, lies in the interval $-2.8 < \CH < -1.6$.  These couplings are in
remarkable agreement with predictions based on an SU(6) spin-flavour
symmetry between the heavy octet and decuplet baryon fields, if the
goldstone bosons are assigned to a {\bf 35} representation.  One might
conclude from this that there is, in fact, an underlying SU(6) spin-flavour
symmetry present in the hadronic sector.

 Finally, we have estimated the $\Delta$ pole contribution to the
polarisability of the nucleon including the momentum dependence of the
photon vertex.  This is consistent to order 1$/\Lambda_\chi^2$, dominates
the magnetic susceptibility of the nucleon, and is compatible with current
experimental data.  We believe the calculations presented here provide
another important probe in the study of the long-distance behavior of QCD.

\bigskip

MJS and MNB wish to thank the physics department at Duke University
for their kind hospitality while some of this work was performed.
We wish to thank E. Jenkins, A. Manohar and M. Luke for numerous
discussions.  MNB acknowledges the support of the Natural Science and
Engineering Research Council  (NSERC) of Canada.  MJS acknowledges the
support of a Superconducting Supercollider National Fellowship from the
Texas National Research Laboratory Commission under grant  FCFY9219.
RPS acknowledges the support of DOE grant DE-FG05-90ER40592.

\bigskip
\bigskip
\vfil\eject

\noindent{\bf Appendix A}
\medskip
\noindent A.1 \ \ \ Clebsch-Gordan coefficients for the 1PI contributions to
the renormalisation of $\CC$.

$$
\eqalign{\alpha_{\Delta N}=&\ {1\over 3} - {10\over 81}\CH^2 +
{20\over 9}F\CH-{10\over 27}D\CH -{1\over 18}\CC^2-{2\over 3}D^2-4DF
+2F^2\cr
\noalign{\bigskip\bigskip}
\alpha_{\Sigma^*\Lambda}=&\ {1\over 3} - {20\over 81}\CH^2+
{10\over 27}D\CH+{10\over 3}F\CH -{1\over 9}\CC^2
-{16\over 9}D^2-{16\over 3}DF\cr
\noalign{\bigskip\bigskip}
\alpha_{\Sigma^*\Sigma}=&\ {1\over 3}-{80\over 81}\CH^2+ {20\over
9}F\CH-{1\over 3}\CC^2 -4D^2 +4F^2 -{16\over 3}FD\cr
\noalign{\bigskip\bigskip}
\alpha_{\Xi^*\Xi}=&\ {1\over 3}-{20\over 27} \CH^2+{25\over 27}
D\CH+{35\over 9}F\CH -{5\over 18}\CC^2-4D^2-{20\over 3}DF\cr } $$

\bigskip
\bigskip
\noindent A.2 \ \ \ Clebsch-Gordan coefficients for the wavefunction
renormalisation of the decuplet and octet baryons.

$$
\halign{$#$\hfil\tabskip=5em&$#$\hfil\cr
\lambda_{\Delta}={25\over 27}\CH^2 +\CC^2&
\lambda_N=\CC^2+{17\over 3}D^2+15F^2-10DF\cr
\noalign{\bigskip\bigskip}
\lambda_{\Sigma^*}={40\over 27}\CH^2+{4\over 3}\CC^2&
\lambda_\Lambda=2\CC^2+{14\over 3}D^2+18F^2\cr
&\lambda_\Sigma={14\over 3}C^2+{26\over 3}D^2+6F^2\cr
\noalign{\bigskip\bigskip}
\lambda_{\Xi^*}={55\over 27}\CH^2+{5\over 3}\CC^2&\lambda_\Xi=
{13\over 3}C^2+{17\over 3}D^2+15F^2+10FD\cr
}
$$

\vfil\eject
\noindent{\bf Appendix B}
\medskip
\noindent B.1 \ \ \ \ Clebsch-Gordan coefficients for radiative decays of
the decuplet.

$$
\halign{$#$\hfil\tabskip=2em&$#$\hfil&$#$\hfil\cr
\Delta^+\to p\gamma&\cr
\beta_{\Delta^+\Delta^{++}p}={1\over\sqrt{3}}\CC\CH&
\beta_{\Delta^+\Delta^0p}={2\over 3\sqrt{3}}\CC\CH&
\beta_{\Delta^+np}=-{1\over\sqrt{3}}\CC(D+F)\cr
\beta_{\Delta^+\Sigma^{*0}p}={1\over3\sqrt{3}}\CC\CH&
\beta_{\Delta^+\Sigma^0p}=-{1\over\sqrt{3}}\CC(D-F)\cr
\noalign{\bigskip\bigskip}
\Sigma^{*+}\to\Sigma^+\gamma&\cr
\beta_{\Sigma^{*+}\Sigma^{*0}\Sigma^+}=-{1\over3\sqrt{3}}\CC\CH&
\beta_{\Sigma^{*+}\Sigma^{0}\Sigma^+}=-{1\over\sqrt{3}}\CC F&
\beta_{\Sigma^{*+}\Delta^{++}\Sigma^+}=-{1\over\sqrt{3}}\CC\CH\cr
\beta_{\Sigma^{*+}\Xi^{*0}\Sigma^+}=-{2\over3\sqrt{3}}\CC\CH&
\beta_{\Sigma^{*+}\Xi^{0}\Sigma^+}={1\over\sqrt{3}}\CC(D+F)&
\beta_{\Sigma^{*+}\Lambda\Sigma^+}={1\over\sqrt{3}}\CC D\cr
\noalign{\bigskip\bigskip}
\Sigma^{*0}\to\Sigma^0\gamma&\cr
\beta_{\Sigma^{*0}\Delta^+\Sigma^0}={2\over3\sqrt{3}}\CC\CH&
\beta_{\Sigma^{*0}\Sigma^{*+}\Sigma^0}=-{1\over3\sqrt{3}}\CC\CH&
\beta_{\Sigma^{*0}\Sigma^{*-}\Sigma^0}={1\over3\sqrt{3}}\CC\CH\cr
\beta_{\Sigma^{*0}\Xi^{*-}\Sigma^0}={1\over3\sqrt{3}}\CC\CH&
\beta_{\Sigma^{*0}p\Sigma^0}=-{1\over2\sqrt{3}}\CC(D-F)&
\beta_{\Sigma^{*0}\Sigma^{+}\Sigma^0}=-{1\over\sqrt{3}}\CC F\cr
\beta_{\Sigma^{*0}\Sigma^{-}\Sigma^0}={1\over\sqrt{3}}\CC F&
\beta_{\Sigma^{*0}\Xi^{-}\Sigma^0}=-{1\over2\sqrt{3}}\CC(D+F)\cr
\noalign{\bigskip\bigskip}
\Sigma^{*0}\to\Lambda\gamma&\cr
\beta_{\Sigma^{*0}\Sigma^{*+}\Lambda}=-{1\over3}\CC\CH&
\beta_{\Sigma^{*0}\Sigma^{*-}\Lambda}=-{1\over3}\CC\CH&
\beta_{\Sigma^{*0}\Xi^{*-}\Lambda}=-{1\over3}\CC\CH\cr
\beta_{\Sigma^{*0}p\Lambda}={1\over 6}\CC(D+3F)&
\beta_{\Sigma^{*0}\Sigma^{+}\Lambda}={1\over3}\CC D&
\beta_{\Sigma^{*0}\Sigma^{-}\Lambda}={1\over3}\CC D\cr
\beta_{\Sigma^{*0}\Xi^{-}\Lambda}={1\over6}\CC(D-3F)\cr
\noalign{\bigskip\bigskip}
\Sigma^{*-}\to\Sigma^-\gamma&\cr
\beta_{\Sigma^{*-}\Sigma^{*0}\Sigma^-}=-{1\over3\sqrt{3}}\CC\CH&
\beta_{\Sigma^{*-}\Sigma^{0}\Sigma^-}=-{1\over\sqrt{3}}\CC F&
\beta_{\Sigma^{*-}\Delta^{0}\Sigma^-}={1\over3\sqrt{3}}\CC\CH\cr
\beta_{\Sigma^{*-}n\Sigma^-}=-{1\over\sqrt{3}}\CC(D-F)&
\beta_{\Sigma^{*-}\Lambda\Sigma^-}={1\over\sqrt{3}}\CC D\cr
\noalign{\bigskip\bigskip}
\Xi^{*0}\to\Xi^0\gamma&\cr
\beta_{\Xi^{*0}\Sigma^{*+}\Xi^0}=-{2\over3\sqrt{3}}\CC\CH&
\beta_{\Xi^{*0}\Xi^{*-}\Xi^0}=-{1\over3\sqrt{3}}\CC\CH&
\beta_{\Xi^{*0}\Omega^-\Xi^0}=-{1\over\sqrt{3}}\CC\CH\cr
\beta_{\Xi^{*0}\Sigma^{+}\Xi^0}={1\over\sqrt{3}}\CC(D+F)&
\beta_{\Xi^{*0}\Xi^{-}\Xi^0}={1\over\sqrt{3}}\CC(D-F)\cr
\noalign{\bigskip\bigskip}
\Xi^{*-}\to\Xi^-\gamma&\cr
\beta_{\Xi^{*-}\Sigma^{*0}\Xi^-}={1\over3\sqrt{3}}\CC\CH&
\beta_{\Xi^{*-}\Xi^{*0}\Xi^-}=-{1\over3\sqrt{3}}\CC\CH&
\beta_{\Xi^{*-}\Sigma^0\Xi^-}=-{1\over2\sqrt{3}}\CC(D+F)\cr
\beta_{\Xi^{*-}\Xi^0\Xi^-}={1\over\sqrt{3}}\CC(D-F)&
\beta_{\Xi^{*-}\Lambda\Xi^-}=-{1\over2\sqrt{3}}\CC(D-3F)\cr
}
$$

\vfil\eject
\noindent B.2 \ \ \ Clebsch-Gordan coefficients for contribution to the
radiative decay from the counterterm.

$$
\halign{$#$\hfil\tabskip=2em&$#$\hfil&$#$\hfil\cr
Q_{\Delta^+ p} = {1\over\sqrt{3}} &
Q_{\Delta^0 n} = {1\over\sqrt{3}} &
Q_{\Sigma^{*0}\Lambda} = -{1\over 2} \cr
Q_{\Sigma^{*0}\Sigma^0} = {1\over \sqrt{12}} &
Q_{\Sigma^{*+}\Sigma^+} = -{1\over \sqrt{3}} &
Q_{\Sigma^{*-}\Sigma^-} =0 \cr
Q_{\Xi^{*0}\Xi^0} = -{1\over\sqrt{3}} &
Q_{\Xi^{*-}\Xi^-} = 0\cr}
$$

\vfil\eject

\listrefs
\listfigs
\end
The rate for one such process, $\Delta\to
N\gamma$, has been inferred from pion photoproduction experiments, with
an estimated branching ratio of only 0.56--0.66\%\ref\pdg{K.\ Hikasa
{\it et al.}, The Particle Data Group, Phys.\ Rev.\ {\bf D45}--II
(1992)}, and has been studied in the Skyrme model\ref\skyrme{B.A.\ Li, Phys.\
Rev.\ {\bf D37}, 1226 (1988).}, a light--cone constituent quark model
\ref\lccqm{H.\ Weber, Ann.\ Phys.\ {\bf 207}, 417 (1991).} and the
non--relativistic quark model\ref\nrqm{ }. Some hyperon decays of this
type have been studied in the MIT bag model and potential models
\ref\mitpot{E.\ Kaxiras, E.J.\ Moniz and M.\ Soyeur, Phys.\ Rev.\ {\bf D32},
695 (1985).}, specifically $\Sigma^{*0}\to \Sigma^0\gamma$, $\Sigma^{*0}\to
\Lambda\gamma$, $\Sigma^{*+}\to \Sigma^+\gamma$ and
$\Xi^{*0}\to\Xi^0\gamma$.
In this work, we wish to examine the whole family of
radiative decays, using the known $\Delta\to N\gamma$ rate as a constraint.
T

/$F2psDict 6400 dict def
$F2psDict begin
$F2psDict /mtrx matrix put
/l {lineto} bind def
/m {moveto} bind def
/s {stroke} bind def
/n {newpath} bind def
/gs {gsave} bind def
/gr {grestore} bind def
/clp {closepath} bind def
/graycol {dup dup currentrgbcolor 4 -2 roll mul 4 -2 roll mul
4 -2 roll mul setrgbcolor} bind def
/col-1 {} def
/col0 {0 0 0 setrgbcolor} bind def
/col1 {0 0 1 setrgbcolor} bind def
/col2 {0 1 0 setrgbcolor} bind def
/col3 {0 1 1 setrgbcolor} bind def
/col4 {1 0 0 setrgbcolor} bind def
/col5 {1 0 1 setrgbcolor} bind def
/col6 {1 1 0 setrgbcolor} bind def
/col7 {1 1 1 setrgbcolor} bind def
	end
	/$F2psBegin {$F2psDict begin /$F2psEnteredState save def} def
	/$F2psEnd {$F2psEnteredState restore end} def

$F2psBegin
0 setlinecap 0 setlinejoin
-1.0 763.0 translate 0.900 -0.900 scale
/Times-BoldItalic findfont 16.00 scalefont setfont
194 199 m
gs 1 -1 scale (K ) col-1 show gr
/Symbol findfont 16.00 scalefont setfont
214 199 m
gs 1 -1 scale (h) col-1 show gr
/Times-BoldItalic findfont 16.00 scalefont setfont
454 199 m
gs 1 -1 scale (K ) col-1 show gr
/Symbol findfont 16.00 scalefont setfont
474 199 m
gs 1 -1 scale (h) col-1 show gr
/Times-BoldItalic findfont 16.00 scalefont setfont
459 374 m
gs 1 -1 scale (K ) col-1 show gr
/Symbol findfont 16.00 scalefont setfont
479 374 m
gs 1 -1 scale (h) col-1 show gr
/Times-BoldItalic findfont 16.00 scalefont setfont
199 374 m
gs 1 -1 scale (K ) col-1 show gr
/Symbol findfont 16.00 scalefont setfont
219 374 m
gs 1 -1 scale (h) col-1 show gr
2.000 setlinewidth
	[8.000000] 0 setdash
n 212.500 408.500 57.502 -179.502 -0.498 arc
gs col-1 s gr
	[] 0 setdash
	[8.000000] 0 setdash
n 211.500 236.500 57.502 -179.502 -0.498 arc
gs col-1 s gr
	[] 0 setdash
	[8.000000] 0 setdash
n 472.500 406.500 57.502 -179.502 -0.498 arc
gs col-1 s gr
	[] 0 setdash
	[8.000000] 0 setdash
n 471.500 234.500 57.502 -179.502 -0.498 arc
gs col-1 s gr
	[] 0 setdash
7.000 setlinewidth
n 95 406 m 269 406 l gs 0.60 setgray fill gr
gs col-1 s gr
2.000 setlinewidth
n 268 407 m 326 407 l gs col-1 s gr
7.000 setlinewidth
n 95 234 m 155 234 l gs 0.60 setgray fill gr
gs col-1 s gr
2.000 setlinewidth
n 153 236 m 326 236 l gs col-1 s gr
7.000 setlinewidth
n 355 404 m 415 404 l gs 0.60 setgray fill gr
gs col-1 s gr
2.000 setlinewidth
n 520 405 m 586 405 l gs col-1 s gr
	[8.000000] 0 setdash
n 209 234 m 239 284 l gs col-1 s gr
	[] 0 setdash
	[8.000000] 0 setdash
n 214 409 m 214 409 l  244 454 l gs col-1 s gr
	[] 0 setdash
	[8.000000] 0 setdash
n 474 234 m 499 279 l gs col-1 s gr
	[] 0 setdash
	[8.000000] 0 setdash
n 469 404 m 509 454 l gs col-1 s gr
	[] 0 setdash
n 415 406 m 470 406 l gs col-1 s gr
7.000 setlinewidth
n 355 232 m 474 232 l gs col-1 s gr
n 240 574 m 339 574 l gs col-1 s gr
3.000 setlinewidth
	[10.000000] 0 setdash
n 340 575 m 383 631 l gs col-1 s gr
	[] 0 setdash
2.000 setlinewidth
n 413 234 m 586 234 l gs col-1 s gr
n 339 576 m 440 576 l gs col-1 s gr
7.000 setlinewidth
n 469 403 m 528 403 l gs col-1 s gr
3.000 setlinewidth
	[10.000000] 0 setdash
n 339 574 m
	323.882 563.493 317.632 558.493 314 554 curveto
	310.617 549.816 304.000 540.392 304 534 curveto
	304.000 527.608 309.661 517.993 314 514 curveto
	319.225 509.192 331.807 504.000 339 504 curveto
	346.193 504.000 358.775 509.192 364 514 curveto
	368.339 517.993 374.000 527.608 374 534 curveto
	374.000 540.392 367.382 549.816 364 554 curveto
	360.368 558.493 354.118 563.493 339 574 curveto
gs col-1 s gr
	[] 0 setdash
/Times-BoldItalic findfont 14.00 scalefont setfont
121 427 m
gs 1 -1 scale (T) col-1 show gr
/Times-BoldItalic findfont 14.00 scalefont setfont
295 428 m
gs 1 -1 scale (B) col-1 show gr
/Times-BoldItalic findfont 14.00 scalefont setfont
110 254 m
gs 1 -1 scale (T) col-1 show gr
/Times-BoldItalic findfont 14.00 scalefont setfont
297 254 m
gs 1 -1 scale (B) col-1 show gr
/Times-BoldItalic findfont 14.00 scalefont setfont
381 425 m
gs 1 -1 scale (T) col-1 show gr
/Times-BoldItalic findfont 14.00 scalefont setfont
555 426 m
gs 1 -1 scale (B) col-1 show gr
/Times-BoldItalic findfont 14.00 scalefont setfont
370 252 m
gs 1 -1 scale (T) col-1 show gr
/Times-BoldItalic findfont 14.00 scalefont setfont
557 252 m
gs 1 -1 scale (B) col-1 show gr
/Symbol findfont 20.00 scalefont setfont
249 289 m
gs 1 -1 scale (p) col-1 show gr
/Symbol findfont 20.00 scalefont setfont
514 289 m
gs 1 -1 scale (p) col-1 show gr
/Symbol findfont 20.00 scalefont setfont
519 464 m
gs 1 -1 scale (p) col-1 show gr
/Symbol findfont 20.00 scalefont setfont
259 464 m
gs 1 -1 scale (p) col-1 show gr
/Symbol findfont 20.00 scalefont setfont
399 639 m
gs 1 -1 scale (p) col-1 show gr
/Times-BoldItalic findfont 14.00 scalefont setfont
275 599 m
gs 1 -1 scale (T) col-1 show gr
/Times-BoldItalic findfont 14.00 scalefont setfont
400 598 m
gs 1 -1 scale (B) col-1 show gr
/Times-BoldItalic findfont 16.00 scalefont setfont
334 529 m
gs 1 -1 scale (K) col-1 show gr
showpage
$F2psEnd


/$F2psDict 6400 dict def
$F2psDict begin
$F2psDict /mtrx matrix put
/l {lineto} bind def
/m {moveto} bind def
/s {stroke} bind def
/n {newpath} bind def
/gs {gsave} bind def
/gr {grestore} bind def
/clp {closepath} bind def
/graycol {dup dup currentrgbcolor 4 -2 roll mul 4 -2 roll mul
4 -2 roll mul setrgbcolor} bind def
/col-1 {} def
/col0 {0 0 0 setrgbcolor} bind def
/col1 {0 0 1 setrgbcolor} bind def
/col2 {0 1 0 setrgbcolor} bind def
/col3 {0 1 1 setrgbcolor} bind def
/col4 {1 0 0 setrgbcolor} bind def
/col5 {1 0 1 setrgbcolor} bind def
/col6 {1 1 0 setrgbcolor} bind def
/col7 {1 1 1 setrgbcolor} bind def
	end
	/$F2psBegin {$F2psDict begin /$F2psEnteredState save def} def
	/$F2psEnd {$F2psEnteredState restore end} def

$F2psBegin
0 setlinecap 0 setlinejoin
12.5 664.0 translate 0.900 -0.900 scale
/Times-BoldItalic findfont 16.00 scalefont setfont
179 199 m
gs 1 -1 scale (K ) col-1 show gr
/Symbol findfont 16.00 scalefont setfont
199 199 m
gs 1 -1 scale (h) col-1 show gr
/Times-BoldItalic findfont 16.00 scalefont setfont
184 374 m
gs 1 -1 scale (K ) col-1 show gr
/Symbol findfont 16.00 scalefont setfont
204 374 m
gs 1 -1 scale (h) col-1 show gr
/Times-BoldItalic findfont 16.00 scalefont setfont
439 194 m
gs 1 -1 scale (K ) col-1 show gr
/Symbol findfont 16.00 scalefont setfont
459 194 m
gs 1 -1 scale (h) col-1 show gr
/Times-BoldItalic findfont 16.00 scalefont setfont
444 369 m
gs 1 -1 scale (K ) col-1 show gr
/Symbol findfont 16.00 scalefont setfont
464 369 m
gs 1 -1 scale (h) col-1 show gr
2.000 setlinewidth
	[8.000000] 0 setdash
n 197.500 403.500 57.502 -179.502 -0.498 arc
gs col-1 s gr
	[] 0 setdash
	[8.000000] 0 setdash
n 196.500 231.500 57.502 -179.502 -0.498 arc
gs col-1 s gr
	[] 0 setdash
	[8.000000] 0 setdash
n 457.500 401.500 57.502 -179.502 -0.498 arc
gs col-1 s gr
	[] 0 setdash
	[8.000000] 0 setdash
n 456.500 229.500 57.502 -179.502 -0.498 arc
gs col-1 s gr
	[] 0 setdash
7.000 setlinewidth
n 80 229 m 140 229 l gs 0.60 setgray fill gr
gs col-1 s gr
n 340 227 m 572 227 l gs col-1 s gr
2.000 setlinewidth
n 79 404 m 314 404 l gs col-1 s gr
n 347 402 m 570 401 l gs col-1 s gr
n 139 232 m 254 232 l gs col-1 s gr
7.000 setlinewidth
n 401 399 m 516 399 l gs col-1 s gr
n 254 228 m 300 229 l  302 227 l gs col-1 s gr
/Times-BoldItalic findfont 14.00 scalefont setfont
280 423 m
gs 1 -1 scale (B) col-1 show gr
/Times-BoldItalic findfont 14.00 scalefont setfont
95 249 m
gs 1 -1 scale (T) col-1 show gr
/Times-BoldItalic findfont 14.00 scalefont setfont
540 421 m
gs 1 -1 scale (B) col-1 show gr
/Times-BoldItalic findfont 14.00 scalefont setfont
355 247 m
gs 1 -1 scale (T) col-1 show gr
/Times-BoldItalic findfont 14.00 scalefont setfont
269 250 m
gs 1 -1 scale (T) col-1 show gr
/Times-BoldItalic findfont 14.00 scalefont setfont
452 248 m
gs 1 -1 scale (T) col-1 show gr
/Times-BoldItalic findfont 14.00 scalefont setfont
534 248 m
gs 1 -1 scale (T) col-1 show gr
/Times-BoldItalic findfont 14.00 scalefont setfont
458 421 m
gs 1 -1 scale (T) col-1 show gr
/Times-BoldItalic findfont 14.00 scalefont setfont
190 250 m
gs 1 -1 scale (B) col-1 show gr
/Times-BoldItalic findfont 14.00 scalefont setfont
195 425 m
gs 1 -1 scale (B) col-1 show gr
/Times-BoldItalic findfont 14.00 scalefont setfont
102 424 m
gs 1 -1 scale (B) col-1 show gr
/Times-BoldItalic findfont 14.00 scalefont setfont
365 421 m
gs 1 -1 scale (B) col-1 show gr
showpage
$F2psEnd




/__NXdef{1 index where{pop pop pop}{def}ifelse}bind def
/__NXbdef{1 index where{pop pop pop}{bind def}ifelse}bind def
/UserObjects 10 array __NXdef
/defineuserobject{
	exch dup 1 add dup UserObjects length gt{
		array dup 0 UserObjects putinterval
		/UserObjects exch def
	}{pop}ifelse UserObjects exch 3 -1 roll put
}__NXbdef
/undefineuserobject{UserObjects exch null put}__NXbdef
/execuserobject{UserObjects exch get exec}__NXbdef
/__NXRectPath{4 2 roll moveto 1 index 0 rlineto
0 exch rlineto neg 0 rlineto closepath}__NXbdef
/__NXProcessRectArgs{
	1 index type /arraytype eq{
		exch 0 4 2 index length 1 sub{
			dup 3 add 1 exch{1 index exch get exch}for
			5 1 roll 5 index exec
		}for pop pop
	}{exec}ifelse
}__NXbdef
/rectfill{gsave newpath {__NXRectPath fill} __NXProcessRectArgs
grestore}__NXbdef
/rectclip{newpath {__NXRectPath} __NXProcessRectArgs clip
newpath}__NXbdef
/rectstroke{
	gsave newpath dup type /arraytype eq{dup length 6 eq}{false}ifelse{
		{gsave __NXRectPath null concat stroke grestore}
		dup length array cvx copy dup 2 4 -1 roll put __NXProcessRectArgs
	}{{__NXRectPath stroke} __NXProcessRectArgs}ifelse grestore
}__NXbdef
/_NXLevel2 systemdict /languagelevel known {languagelevel 2 ge}{false}ifelse
__NXdef
/xyshow{
	0 1 3 index length 1 sub{
		currentpoint 4 index 3 index 1 getinterval show
		3 index 3 index 2 mul 1 add get add exch
		3 index	3 index 2 mul get add exch moveto pop
	}for pop pop
}__NXbdef
/xshow{
	0 1 3 index length 1 sub{
		currentpoint 4 index 3 index 1 getinterval show
		exch 3 index 3 index get add exch moveto pop
	}for pop pop
}__NXbdef
/yshow{
	0 1 3 index length 1 sub{
		currentpoint 4 index 3 index 1 getinterval show
		3 index 3 index get add moveto pop
	}for pop pop
}__NXbdef
/arct{arcto pop pop pop pop}__NXbdef
/setbbox{pop pop pop pop}__NXbdef
/ucache{}__NXbdef
/ucachestatus{mark 0 0 0 0 0}__NXbdef
/setucacheparams{cleartomark}__NXbdef
/uappend{systemdict begin cvx exec end}__NXbdef
/ueofill{gsave newpath uappend eofill grestore}__NXbdef
/ufill{gsave newpath uappend fill grestore}__NXbdef
/ustroke{
	gsave newpath dup length 6 eq
	{exch uappend concat}{uappend}ifelse
	stroke grestore
}__NXbdef
/__NXustrokepathMatrix dup where {pop pop}{matrix def}ifelse
/ustrokepath{
	newpath dup length 6 eq{
		exch uappend __NXustrokepathMatrix currentmatrix exch concat
		strokepath setmatrix
	}{uappend strokepath}ifelse
} __NXbdef
/upath{
	[exch {/ucache cvx}if pathbbox /setbbox cvx
	 {/moveto cvx}{/lineto cvx}{/curveto cvx}{/closepath cvx}pathforall]cvx
} __NXbdef
/setstrokeadjust{pop}__NXbdef
/currentstrokeadjust{false}__NXbdef
/selectfont{exch findfont exch
dup type /arraytype eq {makefont}{scalefont}ifelse setfont}__NXbdef
/_NXCombineArrays{
	counttomark dup 2 add index dup length 3 -1 roll {
		2 index length sub dup 4 1 roll 1 index exch 4 -1 roll putinterval exch
	}repeat pop pop pop
}__NXbdef
/flushgraphics{}def
/setwindowtype{pop pop}def
/currentwindowtype{pop 0}def
/setalpha{pop}def
/currentalpha{1.0}def
/hidecursor{}def
/obscurecursor{}def
/revealcursor{}def
/setcursor{4 {pop}repeat}bind def
/showcursor{}def
/NextStepEncoding where not{
/NextStepEncoding StandardEncoding 256 array copy def
0 [129/Agrave/Aacute/Acircumflex/Atilde/Adieresis/Aring/Ccedilla/Egrave
/Eacute/Ecircumflex/Edieresis/Igrave/Iacute/Icircumflex/Idieresis
/Eth/Ntilde/Ograve/Oacute/Ocircumflex/Otilde/Odieresis/Ugrave/Uacute
/Ucircumflex/Udieresis/Yacute/Thorn/mu/multiply/divide/copyright
176/registered 181/brokenbar 190/logicalnot 192/onesuperior 201/twosuperior
204/threesuperior 209/plusminus/onequarter/onehalf/threequarters/agrave
/aacute/acircumflex/atilde/adieresis/aring/ccedilla/egrave/eacute
/ecircumflex/edieresis/igrave 226/iacute 228/icircumflex/idieresis/eth
/ntilde 236/ograve/oacute/ocircumflex/otilde/odieresis 242/ugrave/uacute
/ucircumflex 246/udieresis/yacute 252/thorn/ydieresis]
{dup type /nametype eq
 {NextStepEncoding 2 index 2 index put pop 1 add}{exch pop}ifelse
}forall pop
/NextStepEncoding NextStepEncoding readonly def
/_NXfstr 128 string dup 0 (_NX) putinterval def
/_NXfindfont /findfont load def
/findfont{
 /currentshared where {pop currentshared} {false} ifelse
 {_NXfindfont}
 {dup _NXfstr 3 125 getinterval cvs length 3 add _NXfstr 0 3 -1 roll
  getinterval cvn exch FontDirectory 2 index known
  {pop FontDirectory exch get}
  {_NXfindfont dup /Encoding get StandardEncoding eq
   {	dup length dict exch
	{1 index /FID ne {2 index 3 1 roll put}{pop pop}ifelse}forall
	 dup /Encoding NextStepEncoding put definefont
	}{exch pop} ifelse
   }ifelse
 }ifelse
}bind def
}{pop}ifelse
/_NXImageString {/__NXImageString where{pop}{/__NXImageString 4000
string
__NXdef}ifelse __NXImageString}__NXbdef
/_NXDoImageOp{
	3 dict begin /parr 5 array def 1 index{dup}{1}ifelse /chans exch def
	chans 2 add 2 roll parr 0 chans getinterval astore pop
	5 index 4 index mul 2 index{1 sub 8 idiv 1 add mul}{mul 1 sub 8 idiv 1
add}ifelse
	4 index mul /totbytes exch def pop exch pop
	gsave matrix invertmatrix concat 0.5 setgray 0 0 4 2 roll rectfill grestore
	{0 1 chans 1 sub{parr exch get exec length totbytes exch sub /totbytes
exch
def}for totbytes 0 le{exit}if}loop end
}__NXbdef
/alphaimage{1 add _NXDoImageOp}def
_NXLevel2{
	/NXCalibratedRGBColorSpace where{pop}{
		/NXCalibratedRGBColorSpace
		{mark /NXCalibratedRGB /ColorSpace findresource exch
pop}stopped
		{cleartomark /NXCalibratedRGB[/CIEBasedABC 2 dict dup begin
		/MatrixLMN[.4124 .2126 .0193 .3576 .7152 .1192 .1805 .0722
.9505]def
		/WhitePoint[.9505 1 1.089] def end] /ColorSpace defineresource}if
def}ifelse
	/nxsetrgbcolor{NXCalibratedRGBColorSpace setcolorspace
setcolor}__NXbdef
	/nxsetgray{dup dup nxsetrgbcolor}__NXbdef
	/_NXCalibratedImage{exch{array astore dup length true}{false}ifelse
		8 -1 roll{NXCalibratedRGBColorSpace setcolorspace}if
		8 dict dup 9 1 roll begin /ImageType 1 def /MultipleDataSources
exch def
		currentcolorspace 0 get /Indexed eq{pop /Decode[0 2 6 index exp 1
sub]def}
		{2 mul dup array /Decode exch def 1 sub 0 1 3 -1 roll{Decode exch
dup 2 mod
put}for}ifelse
		/DataSource exch def /ImageMatrix exch def
		/BitsPerComponent exch def /Height exch def /Width exch def end
image}__NXbdef
} {
	/setcmykcolor{
		1.0 exch sub dup dup 6 -1 roll sub dup 0 lt{pop 0}if 5 1 roll
		4 -1 roll sub dup 0 lt{pop 0}if 3 1 roll exch sub dup 0 lt{pop 0}if
setrgbcolor}__NXbdef
	/currentcmykcolor{currentrgbcolor 3{1.0 exch sub 3 1 roll}repeat
0}__NXbdef
	/colorimage{_NXDoImageOp}__NXbdef
	/nxsetrgbcolor{setrgbcolor}__NXbdef /nxsetgray{setgray}__NXbdef
	/setpattern{pop .5 setgray}__NXbdef
	/_NXCalibratedImage{dup 1 eq {pop pop image}{colorimage}ifelse
pop}__NXbdef
} ifelse
/_NXSetCMYKOrRGB where{pop}{
	mark{systemdict /currentwindow get exec}stopped
	{{pop pop pop setcmykcolor}}{{nxsetrgbcolor pop pop pop pop}}ifelse
/_NXSetCMYKOrRGB exch def cleartomark
}ifelse

gsave
-20 -28 translate
 /__NXbasematrix matrix currentmatrix def
grestore
/oval {
    translate scale newpath 0.5 0.5 0.5 0 360 arc closepath
} def /line {
    moveto rlineto stroke
} def /setup {
    setlinewidth setlinecap setlinejoin gsave
} def /arrow {
    newpath moveto dup rotate -13 6 rlineto 4 -6 rlineto -4 -6 rlineto
closepath gsave 0 setlinejoin stroke grestore fill neg rotate
} def

/__NXsheetsavetoken save def
36 36 translate
gsave
-20 -28 translate
 /__NXbasematrix matrix currentmatrix def
grestore
gsave
0 0 translate
gsave
0 0 540 720 rectclip
0 0 540 720 rectclip
0 0 0 setup
gsave
/Courier findfont 16 scalefont [1 0 0 -1 0 0] makefont
12
exch
defineuserobject
12 execuserobject setfont
0 nxsetgray
[1 0 0 -1 0 1032] concat
/Symbol findfont 16 scalefont [1 0 0 -1 0 0] makefont
11
exch
defineuserobject
11 execuserobject setfont
0 nxsetgray
192 525 moveto (S) show
0 -6 rmoveto
(*+      ) show
0 6 rmoveto
(S) show
0 -6 rmoveto
(0) show
0 6 rmoveto
(p) show
0 -6 rmoveto
(+) show
0 6 rmoveto
grestore
grestore
0 0 0 setup
gsave
11 execuserobject setfont
0 nxsetgray
[1 0 0 -1 0 898] concat
11 execuserobject setfont
0 nxsetgray
201 458 moveto (D) show
0 -6 rmoveto
(++      ) show
0 6 rmoveto
12 execuserobject setfont
(P) show
11 execuserobject setfont
(p) show
0 -6 rmoveto
(+) show
0 6 rmoveto
grestore
grestore
0 0 0 setup
gsave
11 execuserobject setfont
0 nxsetgray
[1 0 0 -1 0 1026] concat
11 execuserobject setfont
0 nxsetgray
384 522 moveto (S) show
0 -6 rmoveto
(*+      ) show
0 6 rmoveto
(Lp) show
0 -6 rmoveto
(+) show
0 6 rmoveto
grestore
grestore
0 0 0 setup
gsave
11 execuserobject setfont
0 nxsetgray
[1 0 0 -1 0 900] concat
11 execuserobject setfont
0 nxsetgray
389 459 moveto (X) show
0 -6 rmoveto
(*0      ) show
0 6 rmoveto
(X) show
0 -6 rmoveto
(-) show
0 6 rmoveto
(p) show
0 -6 rmoveto
(+) show
0 6 rmoveto
grestore
grestore
0 0 1.627907 setup
0 0 1.627907 setup
gsave
163.333328 507.681305 translate
0.185185 0.054945 scale
0 nxsetgray
gsave
newpath
systemdict
begin
360 24 414 115 setbbox
363 115 moveto
0 -91 rlineto
-1 0 rlineto
-2 0 rlineto
54 44 rlineto
0 0 rlineto
0 0 rlineto
0 0 rlineto
-50 45 rlineto
-1 2 rlineto
end
fill
grestore
grestore
163.333328 507.681305 translate
0.185185 0.054945 scale
0 nxsetgray
gsave
newpath
systemdict
begin
360 24 414 115 setbbox
363 115 moveto
0 -91 rlineto
-1 0 rlineto
-2 0 rlineto
54 44 rlineto
0 0 rlineto
0 0 rlineto
0 0 rlineto
-50 45 rlineto
-1 2 rlineto
end
stroke
grestore
grestore
0 0 1.627907 setup
0 nxsetgray
23 0 214 512 line
grestore
grestore
0 0 1.627907 setup
0 0 1.627907 setup
gsave
171.333328 438.681305 translate
0.185185 0.054945 scale
0 nxsetgray
gsave
newpath
systemdict
begin
360 24 414 115 setbbox
363 115 moveto
0 -91 rlineto
-1 0 rlineto
-2 0 rlineto
54 44 rlineto
0 0 rlineto
0 0 rlineto
0 0 rlineto
-50 45 rlineto
-1 2 rlineto
end
fill
grestore
grestore
171.333328 438.681305 translate
0.185185 0.054945 scale
0 nxsetgray
gsave
newpath
systemdict
begin
360 24 414 115 setbbox
363 115 moveto
0 -91 rlineto
-1 0 rlineto
-2 0 rlineto
54 44 rlineto
0 0 rlineto
0 0 rlineto
0 0 rlineto
-50 45 rlineto
-1 2 rlineto
end
stroke
grestore
grestore
0 0 1.627907 setup
0 nxsetgray
23 0 222 443 line
grestore
grestore
0 0 1.627907 setup
0 0 1.627907 setup
gsave
354.333344 503.681305 translate
0.185185 0.054945 scale
0 nxsetgray
gsave
newpath
systemdict
begin
360 24 414 115 setbbox
363 115 moveto
0 -91 rlineto
-1 0 rlineto
-2 0 rlineto
54 44 rlineto
0 0 rlineto
0 0 rlineto
0 0 rlineto
-50 45 rlineto
-1 2 rlineto
end
fill
grestore
grestore
354.333344 503.681305 translate
0.185185 0.054945 scale
0 nxsetgray
gsave
newpath
systemdict
begin
360 24 414 115 setbbox
363 115 moveto
0 -91 rlineto
-1 0 rlineto
-2 0 rlineto
54 44 rlineto
0 0 rlineto
0 0 rlineto
0 0 rlineto
-50 45 rlineto
-1 2 rlineto
end
stroke
grestore
grestore
0 0 1.627907 setup
0 nxsetgray
23 0 405 508 line
grestore
grestore
0 0 1.627907 setup
0 0 1.627907 setup
gsave
317.487213 -7.846157 translate
0.206553 0.076923 scale
0 nxsetgray
gsave
newpath
systemdict
begin
360 24 414 115 setbbox
363 115 moveto
0 -91 rlineto
-1 0 rlineto
-2 0 rlineto
54 44 rlineto
0 0 rlineto
0 0 rlineto
0 0 rlineto
-50 45 rlineto
-1 2 rlineto
end
fill
grestore
grestore
317.487213 -7.846157 translate
0.206553 0.076923 scale
0 nxsetgray
gsave
newpath
systemdict
begin
360 24 414 115 setbbox
363 115 moveto
0 -91 rlineto
-1 0 rlineto
-2 0 rlineto
54 44 rlineto
0 0 rlineto
0 0 rlineto
0 0 rlineto
-50 45 rlineto
-1 2 rlineto
end
stroke
grestore
grestore
0 0 1.627907 setup
0 nxsetgray
25.653841 0 374 -1.8 line
grestore
grestore
0 0 1.627907 setup
0 0 1.627907 setup
gsave
359.333344 440.681305 translate
0.185185 0.054945 scale
0 nxsetgray
gsave
newpath
systemdict
begin
360 24 414 115 setbbox
363 115 moveto
0 -91 rlineto
-1 0 rlineto
-2 0 rlineto
54 44 rlineto
0 0 rlineto
0 0 rlineto
0 0 rlineto
-50 45 rlineto
-1 2 rlineto
end
fill
grestore
grestore
359.333344 440.681305 translate
0.185185 0.054945 scale
0 nxsetgray
gsave
newpath
systemdict
begin
360 24 414 115 setbbox
363 115 moveto
0 -91 rlineto
-1 0 rlineto
-2 0 rlineto
54 44 rlineto
0 0 rlineto
0 0 rlineto
0 0 rlineto
-50 45 rlineto
-1 2 rlineto
end
stroke
grestore
grestore
0 0 1.627907 setup
0 nxsetgray
23 0 410 444.999878 line
grestore
grestore
0 0 0 setup
gsave
11 execuserobject setfont
0 nxsetgray
[1 0 0 -1 0 286] concat
11 execuserobject setfont
0 nxsetgray
277 149 moveto (H) show
grestore
grestore
0 0 0 setup
gsave
11 execuserobject setfont
0 nxsetgray
[1 0 0 -1 0 776] concat
12 execuserobject setfont
0 nxsetgray
22 391 moveto (C) show
grestore
grestore
0 0 0 setup
30.000000 167.000000 transform
gsave __NXbasematrix setmatrix itransform translate
0 0 469 402 rectclip
gsave
1.493631 2.072165 scale

/__NXEPSSave save def /showpage {} def
_NXLevel2{/_NXsethsb where{pop}{/_NXsethsb /sethsbcolor load def}ifelse
/sethsbcolor{_NXsethsb currentrgbcolor nxsetrgbcolor}def
/setrgbcolor{nxsetrgbcolor}bind def /setgray{nxsetgray}bind def
/_NXcimage where{pop}{/_NXcimage /colorimage load def}ifelse
/colorimage{dup 3
eq{true 2 index{1 index}{1}ifelse 7 add 1 roll
_NXCalibratedImage}{_NXcimage}ifelse}def}if
0 setgray 0 setlinecap 1 setlinewidth
0 setlinejoin 10 setmiterlimit [] 0 setdash newpath count /__NXEPSOpCount
exch
def /__NXEPSDictCount countdictstack def
/Mnodistort true def
100 dict begin
/Mfixwid true def
/Mrot 0 def
/Mpstart {
    MathPictureStart
} bind def
/Mpend {
    MathPictureEnd
} bind def
/Mscale {
    0 1 0 1
    5 -1 roll
    MathScale
} bind def
/Plain	/Courier findfont def
/Bold	/Courier-Bold findfont def
/Italic /Courier-Oblique findfont def
/MathPictureStart {
	/Mimatrix
	 matrix currentmatrix
	def
	gsave
	newpath
	Mleft
	Mbottom
	translate
	1 -1 scale
	/Mtmatrix
	matrix currentmatrix
	def
	Plain
	Mfontsize scalefont
	setfont
	0 setgray
	0 setlinewidth
} bind def
/MathPictureEnd {
	grestore
} bind def
/MathSubStart {
        Mgmatrix Mtmatrix
        Mleft Mbottom
        Mwidth Mheight
        8 -2 roll
        moveto
        Mtmatrix setmatrix
        currentpoint
        Mgmatrix setmatrix
        10 -2 roll
        moveto
        Mtmatrix setmatrix
        currentpoint
        2 copy translate
        /Mtmatrix matrix currentmatrix def
        /Mleft 0 def
        /Mbottom 0 def
        3 -1 roll
        exch sub
        /Mheight exch def
        sub
        /Mwidth exch def
} bind def
/MathSubEnd {
        /Mheight exch def
        /Mwidth exch def
        /Mbottom exch def
        /Mleft exch def
        /Mtmatrix exch def
        dup setmatrix
        /Mgmatrix exch def
} bind def
/Mdot {
	moveto
	0 0 rlineto
	stroke
} bind def
/Mtetra {
	moveto
	lineto
	lineto
	lineto
	fill
} bind def
/Metetra {
	moveto
	lineto
	lineto
	lineto
	closepath
	gsave
	fill
	grestore
	0 setgray
	stroke
} bind def
/Mistroke {
	flattenpath
	0 0 0
	{
	4 2 roll
	pop pop
	}
	{
	4 -1 roll
	2 index
	sub dup mul
	4 -1 roll
	2 index
	sub dup mul
	add sqrt
	4 -1 roll
	add
	3 1 roll
	}
	{
	stop
	}
	{
	stop
	}
	pathforall
	pop pop
	currentpoint
	stroke
	moveto
	currentdash
	3 -1 roll
	add
	setdash
} bind def
/Mfstroke {
	stroke
	currentdash
	pop 0
	setdash
} bind def
/Mrotsboxa {
	gsave
	dup
	/Mrot
	exch def
	Mrotcheck
	Mtmatrix
	dup
	setmatrix
	7 1 roll
	4 index
	4 index
	translate
	rotate
	3 index
	-1 mul
	3 index
	-1 mul
	translate
	/Mtmatrix
	matrix
	currentmatrix
	def
	grestore
	Msboxa
	3  -1 roll
	/Mtmatrix
	exch def
	/Mrot
	0 def
} bind def
/Msboxa {
	newpath
	5 -1 roll
	Mvboxa
	pop
	Mboxout
	6 -1 roll
	5 -1 roll
	4 -1 roll
	Msboxa1
	5 -3 roll
	Msboxa1
	Mboxrot
	[
	7 -2 roll
	2 copy
	[
	3 1 roll
	10 -1 roll
	9 -1 roll
	]
	6 1 roll
	5 -2 roll
	]
} bind def
/Msboxa1 {
	sub
	2 div
	dup
	2 index
	1 add
	mul
	3 -1 roll
	-1 add
	3 -1 roll
	mul
} bind def
/Mvboxa	{
	Mfixwid
	{
	Mvboxa1
	}
	{
	dup
	Mwidthcal
	0 exch
	{
	add
	}
	forall
	exch
	Mvboxa1
	4 index
	7 -1 roll
	add
	4 -1 roll
	pop
	3 1 roll
	}
	ifelse
} bind def
/Mvboxa1 {
	gsave
	newpath
	[ true
	3 -1 roll
	{
	Mbbox
	5 -1 roll
	{
	0
	5 1 roll
	}
	{
	7 -1 roll
	exch sub
	(m) stringwidth pop
	.3 mul
	sub
	7 1 roll
	6 -1 roll
	4 -1 roll
	Mmin
	3 -1 roll
	5 index
	add
	5 -1 roll
	4 -1 roll
	Mmax
	4 -1 roll
	}
	ifelse
	false
	}
	forall
	{ stop } if
	counttomark
	1 add
	4 roll
	]
	grestore
} bind def
/Mbbox {
	0 0 moveto
	false charpath
	flattenpath
	pathbbox
	newpath
} bind def
/Mmin {
	2 copy
	gt
	{ exch } if
	pop
} bind def
/Mmax {
	2 copy
	lt
	{ exch } if
	pop
} bind def
/Mrotshowa {
	dup
	/Mrot
	exch def
	Mrotcheck
	Mtmatrix
	dup
	setmatrix
	7 1 roll
	4 index
	4 index
	translate
	rotate
	3 index
	-1 mul
	3 index
	-1 mul
	translate
	/Mtmatrix
	matrix
	currentmatrix
	def
	Mgmatrix setmatrix
	Mshowa
	/Mtmatrix
	exch def
	/Mrot 0 def
} bind def
/Mshowa {
	4 -2 roll
	moveto
	2 index
	Mtmatrix setmatrix
	Mvboxa
	7 1 roll
	Mboxout
	6 -1 roll
	5 -1 roll
	4 -1 roll
	Mshowa1
	4 1 roll
	Mshowa1
	rmoveto
	currentpoint
	Mfixwid
	{
	Mshowax
	}
	{
	Mshoway
	}
	ifelse
	pop pop pop pop
	Mgmatrix setmatrix
} bind def
/Mshowax {
	0 1
        4 index length
        -1 add
        {
        2 index
        4 index
        2 index
        get
        3 index
        add
        moveto
        4 index
        exch get
        show
        } for
} bind def
/Mshoway {
        3 index
        Mwidthcal
        5 1 roll
	0 1
	4 index length
	-1 add
	{
	2 index
	4 index
	2 index
	get
	3 index
	add
	moveto
	4 index
	exch get
	[
	6 index
	aload
	length
	2 add
	-1 roll
	{
	pop
	Strform
	stringwidth
	pop
	neg
	exch
	add
	0 rmoveto
	}
	exch
	kshow
	cleartomark
	} for
	pop
} bind def
/Mwidthcal {
	[
	exch
	{
	Mwidthcal1
	}
	forall
	]
	[
	exch
	dup
	Maxlen
	-1 add
	0 1
	3 -1 roll
	{
	[
	exch
	2 index
	{
	1 index
	Mget
	exch
	}
	forall
	pop
	Maxget
	exch
	}
	for
	pop
	]
	Mreva
} bind def
/Mreva	{
	[
	exch
	aload
	length
	-1 1
	{1 roll}
	for
	]
} bind def
/Mget	{
	1 index
	length
	-1 add
	1 index
	ge
	{
	get
	}
	{
	pop pop
	0
	}
	ifelse
} bind def
/Maxlen	{
	[
	exch
	{
	length
	}
	forall
	Maxget
} bind def
/Maxget	{
	counttomark
	-1 add
	1 1
	3 -1 roll
	{
	pop
	Mmax
	}
	for
	exch
	pop
} bind def
/Mwidthcal1 {
	[
	exch
	{
	Strform
	stringwidth
	pop
	}
	forall
	]
} bind def
/Strform {
	/tem (x) def
	tem 0
	3 -1 roll
	put
	tem
} bind def
/Mshowa1 {
	2 copy
	add
	4 1 roll
	sub
	mul
	sub
	-2 div
} bind def
/MathScale {
	Mwidth
	Mheight
	Mlp
	translate
	scale
	/Msaveaa exch def
	/Msavebb exch def
	/Msavecc exch def
	/Msavedd exch def
	/Mgmatrix
	matrix currentmatrix
	def
} bind def
/Mlp {
	3 copy
	Mlpfirst
	{
	Mnodistort
	{
	Mmin
	dup
	} if
	4 index
	2 index
	2 index
	Mlprun
	11 index
	11 -1 roll
	10 -4 roll
	Mlp1
	8 index
	9 -5 roll
	Mlp1
	4 -1 roll
	and
	{ exit } if
	3 -1 roll
	pop pop
	} loop
	exch
	3 1 roll
	7 -3 roll
	pop pop pop
} bind def
/Mlpfirst {
	3 -1 roll
	dup length
	2 copy
	-2 add
	get
	aload
	pop pop pop
	4 -2 roll
	-1 add
	get
	aload
	pop pop pop
	6 -1 roll
	3 -1 roll
	5 -1 roll
	sub
	dup /MsaveAx exch def
	div
	4 1 roll
	exch sub
	dup /MsaveAy exch def
	div
} bind def
/Mlprun {
	2 copy
	4 index
	0 get
	dup
	4 1 roll
	Mlprun1
	3 copy
	8 -2 roll
	9 -1 roll
	{
	3 copy
	Mlprun1
	3 copy
	11 -3 roll
	/gt Mlpminmax
	8 3 roll
	11 -3 roll
	/lt Mlpminmax
	8 3 roll
	} forall
	pop pop pop pop
	3 1 roll
	pop pop
	aload pop
	5 -1 roll
	aload pop
	exch
	6 -1 roll
	Mlprun2
	8 2 roll
	4 -1 roll
	Mlprun2
	6 2 roll
	3 -1 roll
	Mlprun2
	4 2 roll
	exch
	Mlprun2
	6 2 roll
} bind def
/Mlprun1 {
	aload pop
	exch
	6 -1 roll
	5 -1 roll
	mul add
	4 -2 roll
	mul
	3 -1 roll
	add
} bind def
/Mlprun2 {
	2 copy
	add 2 div
	3 1 roll
	exch sub
} bind def
/Mlpminmax {
	cvx
	2 index
	6 index
	2 index
	exec
	{
	7 -3 roll
	4 -1 roll
	} if
	1 index
	5 index
	3 -1 roll
	exec
	{
	4 1 roll
	pop
	5 -1 roll
	aload
	pop pop
	4 -1 roll
	aload pop
	[
	8 -2 roll
	pop
	5 -2 roll
	pop
	6 -2 roll
	pop
	5 -1 roll
	]
	4 1 roll
	pop
	}
	{
	pop pop pop
	} ifelse
} bind def
/Mlp1 {
	5 index
	3 index	sub
	5 index
	2 index mul
	1 index
	le
	1 index
	0 le
	or
	dup
	not
	{
	1 index
	3 index	div
	.99999 mul
	8 -1 roll
	pop
	7 1 roll
	}
	if
	8 -1 roll
	2 div
	7 -2 roll
	pop sub
	5 index
	6 -3 roll
	pop pop
	mul sub
	exch
} bind def
/intop 0 def
/inrht 0 def
/inflag 0 def
/outflag 0 def
/xadrht 0 def
/xadlft 0 def
/yadtop 0 def
/yadbot 0 def
/Minner {
	outflag
	1
	eq
	{
	/outflag 0 def
	/intop 0 def
	/inrht 0 def
	} if
	5 index
	gsave
	Mtmatrix setmatrix
	Mvboxa pop
	grestore
	3 -1 roll
	pop
	dup
	intop
	gt
	{
	/intop
	exch def
	}
	{ pop }
	ifelse
	dup
	inrht
	gt
	{
	/inrht
	exch def
	}
	{ pop }
	ifelse
	pop
	/inflag
	1 def
} bind def
/Mouter {
	/xadrht 0 def
	/xadlft 0 def
	/yadtop 0 def
	/yadbot 0 def
	inflag
	1 eq
	{
	dup
	0 lt
	{
	dup
	intop
	mul
	neg
	/yadtop
	exch def
	} if
	dup
	0 gt
	{
	dup
	intop
	mul
	/yadbot
	exch def
	}
	if
	pop
	dup
	0 lt
	{
	dup
	inrht
	mul
	neg
	/xadrht
	exch def
	} if
	dup
	0 gt
	{
	dup
	inrht
	mul
	/xadlft
	exch def
	} if
	pop
	/outflag 1 def
	}
	{ pop pop}
	ifelse
	/inflag 0 def
	/inrht 0 def
	/intop 0 def
} bind def
/Mboxout {
	outflag
	1
	eq
	{
	4 -1
	roll
	xadlft
	leadjust
	add
	sub
	4 1 roll
	3 -1
	roll
	yadbot
	leadjust
	add
	sub
	3 1
	roll
	exch
	xadrht
	leadjust
	add
	add
	exch
	yadtop
	leadjust
	add
	add
	/outflag 0 def
	/xadlft 0 def
	/yadbot 0 def
	/xadrht 0 def
	/yadtop 0 def
	} if
} bind def
/leadjust {
	(m) stringwidth pop
	.5 mul
} bind def
/Mrotcheck {
	dup
	90
	eq
	{
	yadbot
	/yadbot
	xadrht
	def
	/xadrht
	yadtop
	def
	/yadtop
	xadlft
	def
	/xadlft
	exch
	def
	}
	if
	dup
	cos
	1 index
	sin
	Checkaux
	dup
	cos
	1 index
	sin neg
	exch
	Checkaux
	3 1 roll
	pop pop
} bind def
/Checkaux {
	4 index
	exch
	4 index
	mul
	3 1 roll
	mul add
	4 1 roll
} bind def
/Mboxrot {
	Mrot
	90 eq
	{
	brotaux
	4 2
	roll
	}
	if
	Mrot
	180 eq
	{
	4 2
        roll
	brotaux
	4 2
	roll
	brotaux
	}
	if
	Mrot
	270 eq
	{
	4 2
	roll
	brotaux
	}
	if
} bind def
/brotaux {
	neg
	exch
	neg
} bind def
/Mabswid {
	Mimatrix
	0 get
	Mgmatrix
	0 get
	div
	mul
	setlinewidth
} bind def
/Mabsdash {
	exch
	Mimatrix
	0 get
	Mgmatrix
	0 get
	div
	[
	3 1 roll
	exch
	{
	exch
	dup
	3 -1 roll
	mul
	exch
	}
	forall
	pop ]
	exch
	setdash
} bind def
/colorimage where
{ pop }
{
/colorimage {
3 1 roll
 pop pop
 5 -1 roll
 mul
 4 1 roll
{
currentfile
1 index
readhexstring
pop }
image
} bind def
} ifelse
/sampledsound where
{ pop}
{ /sampledsound {
exch
pop
exch
5 1 roll
mul
4 idiv
mul
2 idiv
exch pop
exch
/Mtempproc exch def
{ Mtempproc pop }
repeat
} bind def
} ifelse

/Mleft		0.000000 def
/Mbottom	194.306152 def
/Mwidth		314.008667 def
/Mheight	194.306152 def
/Mfontsize 12 def
/Plain /Courier findfont def
0 Mbottom Mheight neg add 2 mul Mheight add translate
1 -1 scale
MathPictureStart
/Courier findfont 10  scalefont  setfont
0.5 0.121951 0.618034 0.412023 [
[(-4)] 0.0122 0 0 2 Msboxa
[(-2)] 0.2561 0 0 2 Msboxa
[(0)] 0.5 0 0 2 Msboxa
[(2)] 0.7439 0 0 2 Msboxa
[(4)] 0.9878 0 0 2 Msboxa
[(-1.4)] -0.0125 0.0412 1 0 Msboxa
[(-1.2)] -0.0125 0.12361 1 0 Msboxa
[(-1)] -0.0125 0.20601 1 0 Msboxa
[(-0.8)] -0.0125 0.28842 1 0 Msboxa
[(-0.6)] -0.0125 0.37082 1 0 Msboxa
[(-0.4)] -0.0125 0.45322 1 0 Msboxa
[(-0.2)] -0.0125 0.53563 1 0 Msboxa
[(0)] -0.0125 0.61803 1 0 Msboxa
[ -0.001 -0.001 0 0 ]
[ 1.001 0.61903 0 0 ]
] MathScale
1 setlinecap
1 setlinejoin
newpath
[ ] 0 setdash
0 setgray
gsave
gsave
0.002 setlinewidth
0.0122 0 moveto
0.0122 0.00625 lineto
stroke
grestore
[(-4)] 0.0122 0 0 2 Mshowa
gsave
0.002 setlinewidth
0.2561 0 moveto
0.2561 0.00625 lineto
stroke
grestore
[(-2)] 0.2561 0 0 2 Mshowa
gsave
0.002 setlinewidth
0.5 0 moveto
0.5 0.00625 lineto
stroke
grestore
[(0)] 0.5 0 0 2 Mshowa
gsave
0.002 setlinewidth
0.7439 0 moveto
0.7439 0.00625 lineto
stroke
grestore
[(2)] 0.7439 0 0 2 Mshowa
gsave
0.002 setlinewidth
0.9878 0 moveto
0.9878 0.00625 lineto
stroke
grestore
[(4)] 0.9878 0 0 2 Mshowa
gsave
0.001 setlinewidth
0.06098 0 moveto
0.06098 0.00375 lineto
stroke
grestore
gsave
0.001 setlinewidth
0.10976 0 moveto
0.10976 0.00375 lineto
stroke
grestore
gsave
0.001 setlinewidth
0.15854 0 moveto
0.15854 0.00375 lineto
stroke
grestore
gsave
0.001 setlinewidth
0.20732 0 moveto
0.20732 0.00375 lineto
stroke
grestore
gsave
0.001 setlinewidth
0.30488 0 moveto
0.30488 0.00375 lineto
stroke
grestore
gsave
0.001 setlinewidth
0.35366 0 moveto
0.35366 0.00375 lineto
stroke
grestore
gsave
0.001 setlinewidth
0.40244 0 moveto
0.40244 0.00375 lineto
stroke
grestore
gsave
0.001 setlinewidth
0.45122 0 moveto
0.45122 0.00375 lineto
stroke
grestore
gsave
0.001 setlinewidth
0.54878 0 moveto
0.54878 0.00375 lineto
stroke
grestore
gsave
0.001 setlinewidth
0.59756 0 moveto
0.59756 0.00375 lineto
stroke
grestore
gsave
0.001 setlinewidth
0.64634 0 moveto
0.64634 0.00375 lineto
stroke
grestore
gsave
0.001 setlinewidth
0.69512 0 moveto
0.69512 0.00375 lineto
stroke
grestore
gsave
0.001 setlinewidth
0.79268 0 moveto
0.79268 0.00375 lineto
stroke
grestore
gsave
0.001 setlinewidth
0.84146 0 moveto
0.84146 0.00375 lineto
stroke
grestore
gsave
0.001 setlinewidth
0.89024 0 moveto
0.89024 0.00375 lineto
stroke
grestore
gsave
0.001 setlinewidth
0.93902 0 moveto
0.93902 0.00375 lineto
stroke
grestore
gsave
0.002 setlinewidth
0 0 moveto
1 0 lineto
stroke
grestore
gsave
0.002 setlinewidth
0 0.0412 moveto
0.00625 0.0412 lineto
stroke
grestore
[(-1.4)] -0.0125 0.0412 1 0 Mshowa
gsave
0.002 setlinewidth
0 0.12361 moveto
0.00625 0.12361 lineto
stroke
grestore
[(-1.2)] -0.0125 0.12361 1 0 Mshowa
gsave
0.002 setlinewidth
0 0.20601 moveto
0.00625 0.20601 lineto
stroke
grestore
[(-1)] -0.0125 0.20601 1 0 Mshowa
gsave
0.002 setlinewidth
0 0.28842 moveto
0.00625 0.28842 lineto
stroke
grestore
[(-0.8)] -0.0125 0.28842 1 0 Mshowa
gsave
0.002 setlinewidth
0 0.37082 moveto
0.00625 0.37082 lineto
stroke
grestore
[(-0.6)] -0.0125 0.37082 1 0 Mshowa
gsave
0.002 setlinewidth
0 0.45322 moveto
0.00625 0.45322 lineto
stroke
grestore
[(-0.4)] -0.0125 0.45322 1 0 Mshowa
gsave
0.002 setlinewidth
0 0.53563 moveto
0.00625 0.53563 lineto
stroke
grestore
[(-0.2)] -0.0125 0.53563 1 0 Mshowa
gsave
0.002 setlinewidth
0 0.61803 moveto
0.00625 0.61803 lineto
stroke
grestore
[(0)] -0.0125 0.61803 1 0 Mshowa
gsave
0.001 setlinewidth
0 0.05768 moveto
0.00375 0.05768 lineto
stroke
grestore
gsave
0.001 setlinewidth
0 0.07416 moveto
0.00375 0.07416 lineto
stroke
grestore
gsave
0.001 setlinewidth
0 0.09064 moveto
0.00375 0.09064 lineto
stroke
grestore
gsave
0.001 setlinewidth
0 0.10713 moveto
0.00375 0.10713 lineto
stroke
grestore
gsave
0.001 setlinewidth
0 0.14009 moveto
0.00375 0.14009 lineto
stroke
grestore
gsave
0.001 setlinewidth
0 0.15657 moveto
0.00375 0.15657 lineto
stroke
grestore
gsave
0.001 setlinewidth
0 0.17305 moveto
0.00375 0.17305 lineto
stroke
grestore
gsave
0.001 setlinewidth
0 0.18953 moveto
0.00375 0.18953 lineto
stroke
grestore
gsave
0.001 setlinewidth
0 0.22249 moveto
0.00375 0.22249 lineto
stroke
grestore
gsave
0.001 setlinewidth
0 0.23897 moveto
0.00375 0.23897 lineto
stroke
grestore
gsave
0.001 setlinewidth
0 0.25545 moveto
0.00375 0.25545 lineto
stroke
grestore
gsave
0.001 setlinewidth
0 0.27193 moveto
0.00375 0.27193 lineto
stroke
grestore
gsave
0.001 setlinewidth
0 0.3049 moveto
0.00375 0.3049 lineto
stroke
grestore
gsave
0.001 setlinewidth
0 0.32138 moveto
0.00375 0.32138 lineto
stroke
grestore
gsave
0.001 setlinewidth
0 0.33786 moveto
0.00375 0.33786 lineto
stroke
grestore
gsave
0.001 setlinewidth
0 0.35434 moveto
0.00375 0.35434 lineto
stroke
grestore
gsave
0.001 setlinewidth
0 0.3873 moveto
0.00375 0.3873 lineto
stroke
grestore
gsave
0.001 setlinewidth
0 0.40378 moveto
0.00375 0.40378 lineto
stroke
grestore
gsave
0.001 setlinewidth
0 0.42026 moveto
0.00375 0.42026 lineto
stroke
grestore
gsave
0.001 setlinewidth
0 0.43674 moveto
0.00375 0.43674 lineto
stroke
grestore
gsave
0.001 setlinewidth
0 0.46971 moveto
0.00375 0.46971 lineto
stroke
grestore
gsave
0.001 setlinewidth
0 0.48619 moveto
0.00375 0.48619 lineto
stroke
grestore
gsave
0.001 setlinewidth
0 0.50267 moveto
0.00375 0.50267 lineto
stroke
grestore
gsave
0.001 setlinewidth
0 0.51915 moveto
0.00375 0.51915 lineto
stroke
grestore
gsave
0.001 setlinewidth
0 0.55211 moveto
0.00375 0.55211 lineto
stroke
grestore
gsave
0.001 setlinewidth
0 0.56859 moveto
0.00375 0.56859 lineto
stroke
grestore
gsave
0.001 setlinewidth
0 0.58507 moveto
0.00375 0.58507 lineto
stroke
grestore
gsave
0.001 setlinewidth
0 0.60155 moveto
0.00375 0.60155 lineto
stroke
grestore
gsave
0.001 setlinewidth
0 0.02472 moveto
0.00375 0.02472 lineto
stroke
grestore
gsave
0.001 setlinewidth
0 0.00824 moveto
0.00375 0.00824 lineto
stroke
grestore
gsave
0.002 setlinewidth
0 0 moveto
0 0.61803 lineto
stroke
grestore
grestore
gsave
gsave
0.002 setlinewidth
0.0122 0.61178 moveto
0.0122 0.61803 lineto
stroke
grestore
gsave
0.002 setlinewidth
0.2561 0.61178 moveto
0.2561 0.61803 lineto
stroke
grestore
gsave
0.002 setlinewidth
0.5 0.61178 moveto
0.5 0.61803 lineto
stroke
grestore
gsave
0.002 setlinewidth
0.7439 0.61178 moveto
0.7439 0.61803 lineto
stroke
grestore
gsave
0.002 setlinewidth
0.9878 0.61178 moveto
0.9878 0.61803 lineto
stroke
grestore
gsave
0.001 setlinewidth
0.06098 0.61428 moveto
0.06098 0.61803 lineto
stroke
grestore
gsave
0.001 setlinewidth
0.10976 0.61428 moveto
0.10976 0.61803 lineto
stroke
grestore
gsave
0.001 setlinewidth
0.15854 0.61428 moveto
0.15854 0.61803 lineto
stroke
grestore
gsave
0.001 setlinewidth
0.20732 0.61428 moveto
0.20732 0.61803 lineto
stroke
grestore
gsave
0.001 setlinewidth
0.30488 0.61428 moveto
0.30488 0.61803 lineto
stroke
grestore
gsave
0.001 setlinewidth
0.35366 0.61428 moveto
0.35366 0.61803 lineto
stroke
grestore
gsave
0.001 setlinewidth
0.40244 0.61428 moveto
0.40244 0.61803 lineto
stroke
grestore
gsave
0.001 setlinewidth
0.45122 0.61428 moveto
0.45122 0.61803 lineto
stroke
grestore
gsave
0.001 setlinewidth
0.54878 0.61428 moveto
0.54878 0.61803 lineto
stroke
grestore
gsave
0.001 setlinewidth
0.59756 0.61428 moveto
0.59756 0.61803 lineto
stroke
grestore
gsave
0.001 setlinewidth
0.64634 0.61428 moveto
0.64634 0.61803 lineto
stroke
grestore
gsave
0.001 setlinewidth
0.69512 0.61428 moveto
0.69512 0.61803 lineto
stroke
grestore
gsave
0.001 setlinewidth
0.79268 0.61428 moveto
0.79268 0.61803 lineto
stroke
grestore
gsave
0.001 setlinewidth
0.84146 0.61428 moveto
0.84146 0.61803 lineto
stroke
grestore
gsave
0.001 setlinewidth
0.89024 0.61428 moveto
0.89024 0.61803 lineto
stroke
grestore
gsave
0.001 setlinewidth
0.93902 0.61428 moveto
0.93902 0.61803 lineto
stroke
grestore
gsave
0.002 setlinewidth
0 0.61803 moveto
1 0.61803 lineto
stroke
grestore
gsave
0.002 setlinewidth
0.99375 0.0412 moveto
1 0.0412 lineto
stroke
grestore
gsave
0.002 setlinewidth
0.99375 0.12361 moveto
1 0.12361 lineto
stroke
grestore
gsave
0.002 setlinewidth
0.99375 0.20601 moveto
1 0.20601 lineto
stroke
grestore
gsave
0.002 setlinewidth
0.99375 0.28842 moveto
1 0.28842 lineto
stroke
grestore
gsave
0.002 setlinewidth
0.99375 0.37082 moveto
1 0.37082 lineto
stroke
grestore
gsave
0.002 setlinewidth
0.99375 0.45322 moveto
1 0.45322 lineto
stroke
grestore
gsave
0.002 setlinewidth
0.99375 0.53563 moveto
1 0.53563 lineto
stroke
grestore
gsave
0.001 setlinewidth
0.99625 0.05768 moveto
1 0.05768 lineto
stroke
grestore
gsave
0.001 setlinewidth
0.99625 0.07416 moveto
1 0.07416 lineto
stroke
grestore
gsave
0.001 setlinewidth
0.99625 0.09064 moveto
1 0.09064 lineto
stroke
grestore
gsave
0.001 setlinewidth
0.99625 0.10713 moveto
1 0.10713 lineto
stroke
grestore
gsave
0.001 setlinewidth
0.99625 0.14009 moveto
1 0.14009 lineto
stroke
grestore
gsave
0.001 setlinewidth
0.99625 0.15657 moveto
1 0.15657 lineto
stroke
grestore
gsave
0.001 setlinewidth
0.99625 0.17305 moveto
1 0.17305 lineto
stroke
grestore
gsave
0.001 setlinewidth
0.99625 0.18953 moveto
1 0.18953 lineto
stroke
grestore
gsave
0.001 setlinewidth
0.99625 0.22249 moveto
1 0.22249 lineto
stroke
grestore
gsave
0.001 setlinewidth
0.99625 0.23897 moveto
1 0.23897 lineto
stroke
grestore
gsave
0.001 setlinewidth
0.99625 0.25545 moveto
1 0.25545 lineto
stroke
grestore
gsave
0.001 setlinewidth
0.99625 0.27193 moveto
1 0.27193 lineto
stroke
grestore
gsave
0.001 setlinewidth
0.99625 0.3049 moveto
1 0.3049 lineto
stroke
grestore
gsave
0.001 setlinewidth
0.99625 0.32138 moveto
1 0.32138 lineto
stroke
grestore
gsave
0.001 setlinewidth
0.99625 0.33786 moveto
1 0.33786 lineto
stroke
grestore
gsave
0.001 setlinewidth
0.99625 0.35434 moveto
1 0.35434 lineto
stroke
grestore
gsave
0.001 setlinewidth
0.99625 0.3873 moveto
1 0.3873 lineto
stroke
grestore
gsave
0.001 setlinewidth
0.99625 0.40378 moveto
1 0.40378 lineto
stroke
grestore
gsave
0.001 setlinewidth
0.99625 0.42026 moveto
1 0.42026 lineto
stroke
grestore
gsave
0.001 setlinewidth
0.99625 0.43674 moveto
1 0.43674 lineto
stroke
grestore
gsave
0.001 setlinewidth
0.99625 0.46971 moveto
1 0.46971 lineto
stroke
grestore
gsave
0.001 setlinewidth
0.99625 0.48619 moveto
1 0.48619 lineto
stroke
grestore
gsave
0.001 setlinewidth
0.99625 0.50267 moveto
1 0.50267 lineto
stroke
grestore
gsave
0.001 setlinewidth
0.99625 0.51915 moveto
1 0.51915 lineto
stroke
grestore
gsave
0.001 setlinewidth
0.99625 0.55211 moveto
1 0.55211 lineto
stroke
grestore
gsave
0.001 setlinewidth
0.99625 0.56859 moveto
1 0.56859 lineto
stroke
grestore
gsave
0.001 setlinewidth
0.99625 0.58507 moveto
1 0.58507 lineto
stroke
grestore
gsave
0.001 setlinewidth
0.99625 0.60155 moveto
1 0.60155 lineto
stroke
grestore
gsave
0.001 setlinewidth
0.99625 0.02472 moveto
1 0.02472 lineto
stroke
grestore
gsave
0.001 setlinewidth
0.99625 0.00824 moveto
1 0.00824 lineto
stroke
grestore
gsave
0.002 setlinewidth
1 0 moveto
1 0.61803 lineto
stroke
grestore
grestore
gsave
grestore
0 0 moveto
1 0 lineto
1 0.61803 lineto
0 0.61803 lineto
closepath
clip
newpath
gsave
gsave
0.5 setgray
0.0122 0.02217 moveto
0.02439 0.02986 lineto
0.03659 0.03736 lineto
0.04878 0.04467 lineto
0.06098 0.05181 lineto
0.07317 0.05876 lineto
0.08537 0.06553 lineto
0.09756 0.07213 lineto
0.10976 0.07854 lineto
0.12195 0.08477 lineto
0.13415 0.09083 lineto
0.14634 0.09671 lineto
0.15854 0.10242 lineto
0.17073 0.10795 lineto
0.18293 0.11332 lineto
0.19512 0.11853 lineto
0.20732 0.12356 lineto
0.21951 0.12844 lineto
0.23171 0.13315 lineto
0.2439 0.13771 lineto
0.2561 0.14212 lineto
0.26829 0.14637 lineto
0.28049 0.15047 lineto
0.29268 0.15443 lineto
0.30488 0.15824 lineto
0.31707 0.16191 lineto
0.32927 0.16544 lineto
0.34146 0.16884 lineto
0.35366 0.1721 lineto
0.36585 0.17523 lineto
0.37805 0.17823 lineto
0.39024 0.18111 lineto
0.40244 0.18385 lineto
0.41463 0.18648 lineto
0.42683 0.18899 lineto
0.43902 0.19138 lineto
0.45122 0.19365 lineto
0.46341 0.19581 lineto
0.47561 0.19785 lineto
0.4878 0.19979 lineto
0.5 0.20161 lineto
0.5122 0.20333 lineto
0.52439 0.20494 lineto
0.53659 0.20645 lineto
0.54878 0.20785 lineto
0.56098 0.20915 lineto
0.57317 0.21035 lineto
0.58537 0.21145 lineto
0.59756 0.21245 lineto
0.60976 0.21336 lineto
0.62195 0.21416 lineto
0.63415 0.21487 lineto
0.64634 0.21549 lineto
0.65854 0.216 lineto
0.67073 0.21643 lineto
0.68293 0.21676 lineto
0.69512 0.21699 lineto
0.70732 0.21713 lineto
0.71951 0.21718 lineto
0.73171 0.21713 lineto
0.7439 0.21699 lineto
0.7561 0.21676 lineto
0.76829 0.21643 lineto
0.78049 0.216 lineto
0.79268 0.21549 lineto
0.80488 0.21487 lineto
0.81707 0.21416 lineto
0.82927 0.21336 lineto
0.84146 0.21245 lineto
0.85366 0.21145 lineto
0.86585 0.21035 lineto
0.87805 0.20915 lineto
0.89024 0.20785 lineto
0.90244 0.20645 lineto
0.91463 0.20494 lineto
0.92683 0.20333 lineto
0.93902 0.20161 lineto
0.95122 0.19979 lineto
0.96341 0.19785 lineto
0.97561 0.19581 lineto
0.9878 0.19365 lineto
0.9878 0.14725 lineto
0.97561 0.14944 lineto
0.96341 0.15152 lineto
0.95122 0.15348 lineto
0.93902 0.15534 lineto
0.92683 0.15709 lineto
0.91463 0.15873 lineto
0.90244 0.16027 lineto
0.89024 0.1617 lineto
0.87805 0.16303 lineto
0.86585 0.16425 lineto
0.85366 0.16538 lineto
0.84146 0.1664 lineto
0.82927 0.16733 lineto
0.81707 0.16815 lineto
0.80488 0.16888 lineto
0.79268 0.16951 lineto
0.78049 0.17004 lineto
0.76829 0.17047 lineto
0.7561 0.17081 lineto
0.7439 0.17105 lineto
0.73171 0.1712 lineto
0.71951 0.17124 lineto
0.70732 0.1712 lineto
0.69512 0.17105 lineto
0.68293 0.17081 lineto
0.67073 0.17047 lineto
0.65854 0.17004 lineto
0.64634 0.16951 lineto
0.63415 0.16888 lineto
0.62195 0.16815 lineto
0.60976 0.16733 lineto
0.59756 0.1664 lineto
0.58537 0.16538 lineto
0.57317 0.16425 lineto
0.56098 0.16303 lineto
0.54878 0.1617 lineto
0.53659 0.16027 lineto
0.52439 0.15873 lineto
0.5122 0.15709 lineto
0.5 0.15534 lineto
0.4878 0.15348 lineto
0.47561 0.15152 lineto
0.46341 0.14944 lineto
0.45122 0.14725 lineto
0.43902 0.14495 lineto
0.42683 0.14253 lineto
0.41463 0.14 lineto
0.40244 0.13734 lineto
0.39024 0.13457 lineto
0.37805 0.13167 lineto
0.36585 0.12865 lineto
0.35366 0.12551 lineto
0.34146 0.12223 lineto
0.32927 0.11883 lineto
0.31707 0.1153 lineto
0.30488 0.11163 lineto
0.29268 0.10782 lineto
0.28049 0.10388 lineto
0.26829 0.0998 lineto
0.2561 0.09557 lineto
0.2439 0.09121 lineto
0.23171 0.08669 lineto
0.21951 0.08203 lineto
0.20732 0.07722 lineto
0.19512 0.07226 lineto
0.18293 0.06714 lineto
0.17073 0.06187 lineto
0.15854 0.05644 lineto
0.14634 0.05086 lineto
0.13415 0.04511 lineto
0.12195 0.0392 lineto
0.10976 0.03313 lineto
0.09756 0.0269 lineto
0.08537 0.0205 lineto
0.07317 0.01394 lineto
0.06098 0.00721 lineto
0.04878 0.00031 lineto
0.03659 -0.00675 lineto
0.02439 -0.01398 lineto
0.0122 -0.02138 lineto
fill
grestore
gsave
0.5 setgray
0.18293 0.51915 moveto
0.18293 0.55211 lineto
0.2561 0.55211 lineto
0.2561 0.51915 lineto
fill
grestore
gsave
0.1 setgray
0.0122 0.16061 moveto
0.02439 0.15575 lineto
0.03659 0.15098 lineto
0.04878 0.1463 lineto
0.06098 0.14171 lineto
0.07317 0.13723 lineto
0.08537 0.13286 lineto
0.09756 0.12861 lineto
0.10976 0.12448 lineto
0.12195 0.12048 lineto
0.13415 0.11662 lineto
0.14634 0.1129 lineto
0.15854 0.10934 lineto
0.17073 0.10592 lineto
0.18293 0.10267 lineto
0.19512 0.09958 lineto
0.20732 0.09667 lineto
0.21951 0.09393 lineto
0.23171 0.09137 lineto
0.2439 0.08899 lineto
0.2561 0.0868 lineto
0.26829 0.08481 lineto
0.28049 0.08301 lineto
0.29268 0.08141 lineto
0.30488 0.08001 lineto
0.31707 0.07882 lineto
0.32927 0.07783 lineto
0.34146 0.07704 lineto
0.35366 0.07647 lineto
0.36585 0.0761 lineto
0.37805 0.07595 lineto
0.39024 0.076 lineto
0.40244 0.07627 lineto
0.41463 0.07674 lineto
0.42683 0.07743 lineto
0.43902 0.07832 lineto
0.45122 0.07942 lineto
0.46341 0.08072 lineto
0.47561 0.08223 lineto
0.4878 0.08394 lineto
0.5 0.08584 lineto
0.5122 0.08793 lineto
0.52439 0.09022 lineto
0.53659 0.09269 lineto
0.54878 0.09535 lineto
0.56098 0.09818 lineto
0.57317 0.10119 lineto
0.58537 0.10436 lineto
0.59756 0.1077 lineto
0.60976 0.1112 lineto
0.62195 0.11485 lineto
0.63415 0.11864 lineto
0.64634 0.12257 lineto
0.65854 0.12664 lineto
0.67073 0.13084 lineto
0.68293 0.13515 lineto
0.69512 0.13958 lineto
0.70732 0.14412 lineto
0.71951 0.14876 lineto
0.73171 0.15349 lineto
0.7439 0.1583 lineto
0.7561 0.1632 lineto
0.76829 0.16817 lineto
0.78049 0.1732 lineto
0.79268 0.1783 lineto
0.80488 0.18344 lineto
0.81707 0.18863 lineto
0.82927 0.19386 lineto
0.84146 0.19912 lineto
0.85366 0.20441 lineto
0.86585 0.20972 lineto
0.87805 0.21504 lineto
0.89024 0.22037 lineto
0.90244 0.2257 lineto
0.91463 0.23103 lineto
0.92683 0.23635 lineto
0.93902 0.24166 lineto
0.95122 0.24695 lineto
0.96341 0.25222 lineto
0.97561 0.25746 lineto
0.9878 0.26267 lineto
0.9878 0.26567 lineto
0.97561 0.26049 lineto
0.96341 0.25529 lineto
0.95122 0.25005 lineto
0.93902 0.2448 lineto
0.92683 0.23952 lineto
0.91463 0.23424 lineto
0.90244 0.22894 lineto
0.89024 0.22364 lineto
0.87805 0.21835 lineto
0.86585 0.21306 lineto
0.85366 0.20778 lineto
0.84146 0.20252 lineto
0.82927 0.19729 lineto
0.81707 0.19209 lineto
0.80488 0.18693 lineto
0.79268 0.18181 lineto
0.78049 0.17675 lineto
0.76829 0.17174 lineto
0.7561 0.1668 lineto
0.7439 0.16193 lineto
0.73171 0.15713 lineto
0.71951 0.15242 lineto
0.70732 0.14781 lineto
0.69512 0.14329 lineto
0.68293 0.13888 lineto
0.67073 0.13459 lineto
0.65854 0.13041 lineto
0.64634 0.12636 lineto
0.63415 0.12244 lineto
0.62195 0.11867 lineto
0.60976 0.11503 lineto
0.59756 0.11155 lineto
0.58537 0.10823 lineto
0.57317 0.10506 lineto
0.56098 0.10207 lineto
0.54878 0.09924 lineto
0.53659 0.0966 lineto
0.52439 0.09413 lineto
0.5122 0.09185 lineto
0.5 0.08977 lineto
0.4878 0.08787 lineto
0.47561 0.08617 lineto
0.46341 0.08467 lineto
0.45122 0.08337 lineto
0.43902 0.08228 lineto
0.42683 0.08139 lineto
0.41463 0.0807 lineto
0.40244 0.08023 lineto
0.39024 0.07996 lineto
0.37805 0.07991 lineto
0.36585 0.08006 lineto
0.35366 0.08043 lineto
0.34146 0.081 lineto
0.32927 0.08178 lineto
0.31707 0.08277 lineto
0.30488 0.08396 lineto
0.29268 0.08535 lineto
0.28049 0.08695 lineto
0.26829 0.08874 lineto
0.2561 0.09073 lineto
0.2439 0.09291 lineto
0.23171 0.09528 lineto
0.21951 0.09783 lineto
0.20732 0.10056 lineto
0.19512 0.10346 lineto
0.18293 0.10654 lineto
0.17073 0.10978 lineto
0.15854 0.11318 lineto
0.14634 0.11673 lineto
0.13415 0.12043 lineto
0.12195 0.12428 lineto
0.10976 0.12826 lineto
0.09756 0.13237 lineto
0.08537 0.1366 lineto
0.07317 0.14096 lineto
0.06098 0.14542 lineto
0.04878 0.14998 lineto
0.03659 0.15464 lineto
0.02439 0.15939 lineto
0.0122 0.16422 lineto
fill
grestore
gsave
0.1 setgray
0.18293 0.41202 moveto
0.18293 0.44498 lineto
0.2561 0.44498 lineto
0.2561 0.41202 lineto
fill
grestore
gsave
0.3 setgray
0.0122 0.18769 moveto
0.02439 0.18257 lineto
0.03659 0.17759 lineto
0.04878 0.17275 lineto
0.06098 0.16806 lineto
0.07317 0.16353 lineto
0.08537 0.15918 lineto
0.09756 0.15499 lineto
0.10976 0.151 lineto
0.12195 0.1472 lineto
0.13415 0.14361 lineto
0.14634 0.14023 lineto
0.15854 0.13706 lineto
0.17073 0.13412 lineto
0.18293 0.13141 lineto
0.19512 0.12894 lineto
0.20732 0.1267 lineto
0.21951 0.12471 lineto
0.23171 0.12297 lineto
0.2439 0.12149 lineto
0.2561 0.12026 lineto
0.26829 0.11928 lineto
0.28049 0.11857 lineto
0.29268 0.11812 lineto
0.30488 0.11793 lineto
0.31707 0.11801 lineto
0.32927 0.11835 lineto
0.34146 0.11895 lineto
0.35366 0.11981 lineto
0.36585 0.12093 lineto
0.37805 0.12231 lineto
0.39024 0.12394 lineto
0.40244 0.12583 lineto
0.41463 0.12796 lineto
0.42683 0.13033 lineto
0.43902 0.13294 lineto
0.45122 0.13578 lineto
0.46341 0.13886 lineto
0.47561 0.14215 lineto
0.4878 0.14565 lineto
0.5 0.14936 lineto
0.5122 0.15327 lineto
0.52439 0.15738 lineto
0.53659 0.16166 lineto
0.54878 0.16612 lineto
0.56098 0.17074 lineto
0.57317 0.17552 lineto
0.58537 0.18044 lineto
0.59756 0.1855 lineto
0.60976 0.19068 lineto
0.62195 0.19598 lineto
0.63415 0.20139 lineto
0.64634 0.20689 lineto
0.65854 0.21247 lineto
0.67073 0.21812 lineto
0.68293 0.22384 lineto
0.69512 0.22962 lineto
0.70732 0.23544 lineto
0.71951 0.24129 lineto
0.73171 0.24716 lineto
0.7439 0.25305 lineto
0.7561 0.25895 lineto
0.76829 0.26484 lineto
0.78049 0.27072 lineto
0.79268 0.27659 lineto
0.80488 0.28243 lineto
0.81707 0.28824 lineto
0.82927 0.29401 lineto
0.84146 0.29973 lineto
0.85366 0.3054 lineto
0.86585 0.31102 lineto
0.87805 0.31658 lineto
0.89024 0.32207 lineto
0.90244 0.3275 lineto
0.91463 0.33285 lineto
0.92683 0.33813 lineto
0.93902 0.34333 lineto
0.95122 0.34845 lineto
0.96341 0.35349 lineto
0.97561 0.35845 lineto
0.9878 0.36333 lineto
0.9878 0.37412 lineto
0.97561 0.36943 lineto
0.96341 0.36466 lineto
0.95122 0.3598 lineto
0.93902 0.35486 lineto
0.92683 0.34985 lineto
0.91463 0.34475 lineto
0.90244 0.33959 lineto
0.89024 0.33435 lineto
0.87805 0.32904 lineto
0.86585 0.32367 lineto
0.85366 0.31823 lineto
0.84146 0.31274 lineto
0.82927 0.3072 lineto
0.81707 0.30161 lineto
0.80488 0.29598 lineto
0.79268 0.29031 lineto
0.78049 0.28462 lineto
0.76829 0.2789 lineto
0.7561 0.27317 lineto
0.7439 0.26744 lineto
0.73171 0.2617 lineto
0.71951 0.25598 lineto
0.70732 0.25028 lineto
0.69512 0.2446 lineto
0.68293 0.23896 lineto
0.67073 0.23338 lineto
0.65854 0.22785 lineto
0.64634 0.22238 lineto
0.63415 0.217 lineto
0.62195 0.2117 lineto
0.60976 0.20651 lineto
0.59756 0.20142 lineto
0.58537 0.19645 lineto
0.57317 0.19162 lineto
0.56098 0.18692 lineto
0.54878 0.18237 lineto
0.53659 0.17798 lineto
0.52439 0.17376 lineto
0.5122 0.16972 lineto
0.5 0.16587 lineto
0.4878 0.1622 lineto
0.47561 0.15874 lineto
0.46341 0.15549 lineto
0.45122 0.15246 lineto
0.43902 0.14965 lineto
0.42683 0.14707 lineto
0.41463 0.14472 lineto
0.40244 0.14261 lineto
0.39024 0.14075 lineto
0.37805 0.13914 lineto
0.36585 0.13777 lineto
0.35366 0.13666 lineto
0.34146 0.13581 lineto
0.32927 0.13521 lineto
0.31707 0.13488 lineto
0.30488 0.1348 lineto
0.29268 0.13499 lineto
0.28049 0.13543 lineto
0.26829 0.13614 lineto
0.2561 0.1371 lineto
0.2439 0.13832 lineto
0.23171 0.13979 lineto
0.21951 0.14151 lineto
0.20732 0.14348 lineto
0.19512 0.14569 lineto
0.18293 0.14814 lineto
0.17073 0.15082 lineto
0.15854 0.15372 lineto
0.14634 0.15685 lineto
0.13415 0.16019 lineto
0.12195 0.16374 lineto
0.10976 0.16748 lineto
0.09756 0.17142 lineto
0.08537 0.17554 lineto
0.07317 0.17983 lineto
0.06098 0.18429 lineto
0.04878 0.1889 lineto
0.03659 0.19366 lineto
0.02439 0.19855 lineto
0.0122 0.20357 lineto
fill
grestore
gsave
0.3 setgray
0.62195 0.51915 moveto
0.62195 0.55211 lineto
0.69512 0.55211 lineto
0.69512 0.51915 lineto
fill
grestore
gsave
0.7 setgray
0.0122 0.11311 moveto
0.02439 0.11293 lineto
0.03659 0.11289 lineto
0.04878 0.11298 lineto
0.06098 0.11319 lineto
0.07317 0.11354 lineto
0.08537 0.11402 lineto
0.09756 0.11464 lineto
0.10976 0.11538 lineto
0.12195 0.11625 lineto
0.13415 0.11725 lineto
0.14634 0.11838 lineto
0.15854 0.11963 lineto
0.17073 0.12102 lineto
0.18293 0.12252 lineto
0.19512 0.12416 lineto
0.20732 0.12591 lineto
0.21951 0.12779 lineto
0.23171 0.12979 lineto
0.2439 0.13191 lineto
0.2561 0.13415 lineto
0.26829 0.1365 lineto
0.28049 0.13897 lineto
0.29268 0.14155 lineto
0.30488 0.14424 lineto
0.31707 0.14704 lineto
0.32927 0.14994 lineto
0.34146 0.15295 lineto
0.35366 0.15606 lineto
0.36585 0.15926 lineto
0.37805 0.16256 lineto
0.39024 0.16596 lineto
0.40244 0.16944 lineto
0.41463 0.17302 lineto
0.42683 0.17667 lineto
0.43902 0.18041 lineto
0.45122 0.18422 lineto
0.46341 0.1881 lineto
0.47561 0.19206 lineto
0.4878 0.19608 lineto
0.5 0.20016 lineto
0.5122 0.2043 lineto
0.52439 0.2085 lineto
0.53659 0.21274 lineto
0.54878 0.21704 lineto
0.56098 0.22137 lineto
0.57317 0.22575 lineto
0.58537 0.23015 lineto
0.59756 0.23459 lineto
0.60976 0.23906 lineto
0.62195 0.24354 lineto
0.63415 0.24805 lineto
0.64634 0.25257 lineto
0.65854 0.25709 lineto
0.67073 0.26163 lineto
0.68293 0.26617 lineto
0.69512 0.2707 lineto
0.70732 0.27523 lineto
0.71951 0.27976 lineto
0.73171 0.28427 lineto
0.7439 0.28876 lineto
0.7561 0.29324 lineto
0.76829 0.2977 lineto
0.78049 0.30213 lineto
0.79268 0.30654 lineto
0.80488 0.31091 lineto
0.81707 0.31526 lineto
0.82927 0.31957 lineto
0.84146 0.32384 lineto
0.85366 0.32807 lineto
0.86585 0.33227 lineto
0.87805 0.33642 lineto
0.89024 0.34053 lineto
0.90244 0.34459 lineto
0.91463 0.34861 lineto
0.92683 0.35257 lineto
0.93902 0.35649 lineto
0.95122 0.36036 lineto
0.96341 0.36418 lineto
0.97561 0.36795 lineto
0.9878 0.37167 lineto
0.9878 0.38403 lineto
0.97561 0.38046 lineto
0.96341 0.37684 lineto
0.95122 0.37318 lineto
0.93902 0.36946 lineto
0.92683 0.36569 lineto
0.91463 0.36187 lineto
0.90244 0.358 lineto
0.89024 0.35408 lineto
0.87805 0.35012 lineto
0.86585 0.34612 lineto
0.85366 0.34207 lineto
0.84146 0.33797 lineto
0.82927 0.33384 lineto
0.81707 0.32967 lineto
0.80488 0.32546 lineto
0.79268 0.32122 lineto
0.78049 0.31694 lineto
0.76829 0.31264 lineto
0.7561 0.30831 lineto
0.7439 0.30395 lineto
0.73171 0.29958 lineto
0.71951 0.29518 lineto
0.70732 0.29077 lineto
0.69512 0.28635 lineto
0.68293 0.28191 lineto
0.67073 0.27748 lineto
0.65854 0.27304 lineto
0.64634 0.2686 lineto
0.63415 0.26417 lineto
0.62195 0.25975 lineto
0.60976 0.25534 lineto
0.59756 0.25095 lineto
0.58537 0.24658 lineto
0.57317 0.24224 lineto
0.56098 0.23793 lineto
0.54878 0.23365 lineto
0.53659 0.22942 lineto
0.52439 0.22522 lineto
0.5122 0.22107 lineto
0.5 0.21697 lineto
0.4878 0.21293 lineto
0.47561 0.20894 lineto
0.46341 0.20502 lineto
0.45122 0.20116 lineto
0.43902 0.19737 lineto
0.42683 0.19366 lineto
0.41463 0.19002 lineto
0.40244 0.18647 lineto
0.39024 0.183 lineto
0.37805 0.17961 lineto
0.36585 0.17632 lineto
0.35366 0.17312 lineto
0.34146 0.17002 lineto
0.32927 0.16701 lineto
0.31707 0.16411 lineto
0.30488 0.16132 lineto
0.29268 0.15863 lineto
0.28049 0.15604 lineto
0.26829 0.15358 lineto
0.2561 0.15122 lineto
0.2439 0.14898 lineto
0.23171 0.14686 lineto
0.21951 0.14485 lineto
0.20732 0.14297 lineto
0.19512 0.14121 lineto
0.18293 0.13957 lineto
0.17073 0.13805 lineto
0.15854 0.13667 lineto
0.14634 0.13541 lineto
0.13415 0.13427 lineto
0.12195 0.13327 lineto
0.10976 0.1324 lineto
0.09756 0.13165 lineto
0.08537 0.13104 lineto
0.07317 0.13055 lineto
0.06098 0.1302 lineto
0.04878 0.12998 lineto
0.03659 0.12989 lineto
0.02439 0.12994 lineto
0.0122 0.13011 lineto
fill
grestore
gsave
0.7 setgray
0.62195 0.41202 moveto
0.62195 0.44498 lineto
0.69512 0.44498 lineto
0.69512 0.41202 lineto
fill
grestore
grestore
MathPictureEnd
end


count __NXEPSOpCount sub {pop} repeat countdictstack __NXEPSDictCount
sub {end}
repeat __NXEPSSave restore
grestore
grestore
grestore
0 setgray
0.333333 setgray
gsave
0 0 540 720 rectclip
[1 0 0 -1 0 720] concat
grestore
grestore
grestore
showpage
__NXsheetsavetoken restore


/$F2psDict 6400 dict def
$F2psDict begin
$F2psDict /mtrx matrix put
/l {lineto} bind def
/m {moveto} bind def
/s {stroke} bind def
/n {newpath} bind def
/gs {gsave} bind def
/gr {grestore} bind def
/clp {closepath} bind def
/graycol {dup dup currentrgbcolor 4 -2 roll mul 4 -2 roll mul
4 -2 roll mul setrgbcolor} bind def
/col-1 {} def
/col0 {0 0 0 setrgbcolor} bind def
/col1 {0 0 1 setrgbcolor} bind def
/col2 {0 1 0 setrgbcolor} bind def
/col3 {0 1 1 setrgbcolor} bind def
/col4 {1 0 0 setrgbcolor} bind def
/col5 {1 0 1 setrgbcolor} bind def
/col6 {1 1 0 setrgbcolor} bind def
/col7 {1 1 1 setrgbcolor} bind def
	end
	/$F2psBegin {$F2psDict begin /$F2psEnteredState save def} def
	/$F2psEnd {$F2psEnteredState restore end} def

$F2psBegin
0 setlinecap 0 setlinejoin
19.0 706.5 translate 0.900 -0.900 scale
/Times-BoldItalic findfont 20.00 scalefont setfont
304 479 m
gs 1 -1 scale (K) col-1 show gr
/Symbol findfont 20.00 scalefont setfont
279 479 m
gs 1 -1 scale (p,) col-1 show gr
/Times-BoldItalic findfont 20.00 scalefont setfont
304 239 m
gs 1 -1 scale (K) col-1 show gr
/Symbol findfont 20.00 scalefont setfont
279 239 m
gs 1 -1 scale (p,) col-1 show gr
2.000 setlinewidth
	[8.000000] 0 setdash
n 321.000 522.000 80.000 180.000 0.000 arc
gs col-1 s gr
	[] 0 setdash
	[8.000000] 0 setdash
n 320.000 282.000 80.000 180.000 0.000 arc
gs col-1 s gr
	[] 0 setdash
7.000 setlinewidth
n 158 519 m 400 519 l gs 0.60 setgray fill gr
gs col-1 s gr
2.000 setlinewidth
n 399 521 m 480 521 l gs col-1 s gr
7.000 setlinewidth
n 157 279 m 241 279 l gs 0.60 setgray fill gr
gs col-1 s gr
2.000 setlinewidth
n 239 281 m 480 281 l gs col-1 s gr
n 321 442 m
	318.938 436.562 318.938 434.062 321 432 curveto
	323.750 429.250 328.250 434.750 331 432 curveto
	333.750 429.250 328.250 424.750 331 422 curveto
	333.750 419.250 338.250 424.750 341 422 curveto
	343.750 419.250 338.250 414.750 341 412 curveto
	343.750 409.250 348.250 414.750 351 412 curveto
	353.750 409.250 348.250 404.750 351 402 curveto
	353.750 399.250 358.250 404.750 361 402 curveto
	363.750 399.250 358.250 394.750 361 392 curveto
	363.750 389.250 368.250 394.750 371 392 curveto
	373.750 389.250 368.250 384.750 371 382 curveto
	373.062 379.938 375.562 379.938 381 382 curveto
gs col-1 s gr
n 320 202 m
	317.938 196.562 317.938 194.062 320 192 curveto
	322.750 189.250 327.250 194.750 330 192 curveto
	332.750 189.250 327.250 184.750 330 182 curveto
	332.750 179.250 337.250 184.750 340 182 curveto
	342.750 179.250 337.250 174.750 340 172 curveto
	342.750 169.250 347.250 174.750 350 172 curveto
	352.750 169.250 347.250 164.750 350 162 curveto
	352.750 159.250 357.250 164.750 360 162 curveto
	362.750 159.250 357.250 154.750 360 152 curveto
	362.750 149.250 367.250 154.750 370 152 curveto
	372.750 149.250 367.250 144.750 370 142 curveto
	372.062 139.938 374.562 139.938 380 142 curveto
gs col-1 s gr
/Times-BoldItalic findfont 20.00 scalefont setfont
316 550 m
gs 1 -1 scale (T) col-1 show gr
/Times-BoldItalic findfont 20.00 scalefont setfont
194 549 m
gs 1 -1 scale (T) col-1 show gr
/Times-BoldItalic findfont 20.00 scalefont setfont
437 550 m
gs 1 -1 scale (B) col-1 show gr
/Times-BoldItalic findfont 20.00 scalefont setfont
179 307 m
gs 1 -1 scale (T) col-1 show gr
/Times-BoldItalic findfont 20.00 scalefont setfont
313 308 m
gs 1 -1 scale (B) col-1 show gr
/Times-BoldItalic findfont 20.00 scalefont setfont
440 307 m
gs 1 -1 scale (B) col-1 show gr
/Symbol findfont 20.00 scalefont setfont
387 161 m
gs 1 -1 scale (g) col-1 show gr
/Symbol findfont 20.00 scalefont setfont
382 403 m
gs 1 -1 scale (g) col-1 show gr
showpage
$F2psEnd




/__NXdef{1 index where{pop pop pop}{def}ifelse}bind def
/__NXbdef{1 index where{pop pop pop}{bind def}ifelse}bind def
/UserObjects 10 array __NXdef
/defineuserobject{
	exch dup 1 add dup UserObjects length gt{
		array dup 0 UserObjects putinterval
		/UserObjects exch def
	}{pop}ifelse UserObjects exch 3 -1 roll put
}__NXbdef
/undefineuserobject{UserObjects exch null put}__NXbdef
/execuserobject{UserObjects exch get exec}__NXbdef
/__NXRectPath{4 2 roll moveto 1 index 0 rlineto
0 exch rlineto neg 0 rlineto closepath}__NXbdef
/__NXProcessRectArgs{
	1 index type /arraytype eq{
		exch 0 4 2 index length 1 sub{
			dup 3 add 1 exch{1 index exch get exch}for
			5 1 roll 5 index exec
		}for pop pop
	}{exec}ifelse
}__NXbdef
/rectfill{gsave newpath {__NXRectPath fill} __NXProcessRectArgs
grestore}__NXbdef
/rectclip{newpath {__NXRectPath} __NXProcessRectArgs clip
newpath}__NXbdef
/rectstroke{
	gsave newpath dup type /arraytype eq{dup length 6 eq}{false}ifelse{
		{gsave __NXRectPath null concat stroke grestore}
		dup length array cvx copy dup 2 4 -1 roll put __NXProcessRectArgs
	}{{__NXRectPath stroke} __NXProcessRectArgs}ifelse grestore
}__NXbdef
/_NXLevel2 systemdict /languagelevel known {languagelevel 2 ge}{false}ifelse
__NXdef
/xyshow{
	0 1 3 index length 1 sub{
		currentpoint 4 index 3 index 1 getinterval show
		3 index 3 index 2 mul 1 add get add exch
		3 index	3 index 2 mul get add exch moveto pop
	}for pop pop
}__NXbdef
/xshow{
	0 1 3 index length 1 sub{
		currentpoint 4 index 3 index 1 getinterval show
		exch 3 index 3 index get add exch moveto pop
	}for pop pop
}__NXbdef
/yshow{
	0 1 3 index length 1 sub{
		currentpoint 4 index 3 index 1 getinterval show
		3 index 3 index get add moveto pop
	}for pop pop
}__NXbdef
/arct{arcto pop pop pop pop}__NXbdef
/setbbox{pop pop pop pop}__NXbdef
/ucache{}__NXbdef
/ucachestatus{mark 0 0 0 0 0}__NXbdef
/setucacheparams{cleartomark}__NXbdef
/uappend{systemdict begin cvx exec end}__NXbdef
/ueofill{gsave newpath uappend eofill grestore}__NXbdef
/ufill{gsave newpath uappend fill grestore}__NXbdef
/ustroke{
	gsave newpath dup length 6 eq
	{exch uappend concat}{uappend}ifelse
	stroke grestore
}__NXbdef
/__NXustrokepathMatrix dup where {pop pop}{matrix def}ifelse
/ustrokepath{
	newpath dup length 6 eq{
		exch uappend __NXustrokepathMatrix currentmatrix exch concat
		strokepath setmatrix
	}{uappend strokepath}ifelse
} __NXbdef
/upath{
	[exch {/ucache cvx}if pathbbox /setbbox cvx
	 {/moveto cvx}{/lineto cvx}{/curveto cvx}{/closepath cvx}pathforall]cvx
} __NXbdef
/setstrokeadjust{pop}__NXbdef
/currentstrokeadjust{false}__NXbdef
/selectfont{exch findfont exch
dup type /arraytype eq {makefont}{scalefont}ifelse setfont}__NXbdef
/_NXCombineArrays{
	counttomark dup 2 add index dup length 3 -1 roll {
		2 index length sub dup 4 1 roll 1 index exch 4 -1 roll putinterval exch
	}repeat pop pop pop
}__NXbdef
/flushgraphics{}def
/setwindowtype{pop pop}def
/currentwindowtype{pop 0}def
/setalpha{pop}def
/currentalpha{1.0}def
/hidecursor{}def
/obscurecursor{}def
/revealcursor{}def
/setcursor{4 {pop}repeat}bind def
/showcursor{}def
/NextStepEncoding where not{
/NextStepEncoding StandardEncoding 256 array copy def
0 [129/Agrave/Aacute/Acircumflex/Atilde/Adieresis/Aring/Ccedilla/Egrave
/Eacute/Ecircumflex/Edieresis/Igrave/Iacute/Icircumflex/Idieresis
/Eth/Ntilde/Ograve/Oacute/Ocircumflex/Otilde/Odieresis/Ugrave/Uacute
/Ucircumflex/Udieresis/Yacute/Thorn/mu/multiply/divide/copyright
176/registered 181/brokenbar 190/logicalnot 192/onesuperior 201/twosuperior
204/threesuperior 209/plusminus/onequarter/onehalf/threequarters/agrave
/aacute/acircumflex/atilde/adieresis/aring/ccedilla/egrave/eacute
/ecircumflex/edieresis/igrave 226/iacute 228/icircumflex/idieresis/eth
/ntilde 236/ograve/oacute/ocircumflex/otilde/odieresis 242/ugrave/uacute
/ucircumflex 246/udieresis/yacute 252/thorn/ydieresis]
{dup type /nametype eq
 {NextStepEncoding 2 index 2 index put pop 1 add}{exch pop}ifelse
}forall pop
/NextStepEncoding NextStepEncoding readonly def
/_NXfstr 128 string dup 0 (_NX) putinterval def
/_NXfindfont /findfont load def
/findfont{
 /currentshared where {pop currentshared} {false} ifelse
 {_NXfindfont}
 {dup _NXfstr 3 125 getinterval cvs length 3 add _NXfstr 0 3 -1 roll
  getinterval cvn exch FontDirectory 2 index known
  {pop FontDirectory exch get}
  {_NXfindfont dup /Encoding get StandardEncoding eq
   {	dup length dict exch
	{1 index /FID ne {2 index 3 1 roll put}{pop pop}ifelse}forall
	 dup /Encoding NextStepEncoding put definefont
	}{exch pop} ifelse
   }ifelse
 }ifelse
}bind def
}{pop}ifelse
/_NXImageString {/__NXImageString where{pop}{/__NXImageString 4000
string
__NXdef}ifelse __NXImageString}__NXbdef
/_NXDoImageOp{
	3 dict begin /parr 5 array def 1 index{dup}{1}ifelse /chans exch def
	chans 2 add 2 roll parr 0 chans getinterval astore pop
	5 index 4 index mul 2 index{1 sub 8 idiv 1 add mul}{mul 1 sub 8 idiv 1
add}ifelse
	4 index mul /totbytes exch def pop exch pop
	gsave matrix invertmatrix concat 0.5 setgray 0 0 4 2 roll rectfill grestore
	{0 1 chans 1 sub{parr exch get exec length totbytes exch sub /totbytes
exch
def}for totbytes 0 le{exit}if}loop end
}__NXbdef
/alphaimage{1 add _NXDoImageOp}def
_NXLevel2{
	/NXCalibratedRGBColorSpace where{pop}{
		/NXCalibratedRGBColorSpace
		{mark /NXCalibratedRGB /ColorSpace findresource exch
pop}stopped
		{cleartomark /NXCalibratedRGB[/CIEBasedABC 2 dict dup begin
		/MatrixLMN[.4124 .2126 .0193 .3576 .7152 .1192 .1805 .0722
.9505]def
		/WhitePoint[.9505 1 1.089] def end] /ColorSpace defineresource}if
def}ifelse
	/nxsetrgbcolor{NXCalibratedRGBColorSpace setcolorspace
setcolor}__NXbdef
	/nxsetgray{dup dup nxsetrgbcolor}__NXbdef
	/_NXCalibratedImage{exch{array astore dup length true}{false}ifelse
		8 -1 roll{NXCalibratedRGBColorSpace setcolorspace}if
		8 dict dup 9 1 roll begin /ImageType 1 def /MultipleDataSources
exch def
		currentcolorspace 0 get /Indexed eq{pop /Decode[0 2 6 index exp 1
sub]def}
		{2 mul dup array /Decode exch def 1 sub 0 1 3 -1 roll{Decode exch
dup 2 mod
put}for}ifelse
		/DataSource exch def /ImageMatrix exch def
		/BitsPerComponent exch def /Height exch def /Width exch def end
image}__NXbdef
} {
	/setcmykcolor{
		1.0 exch sub dup dup 6 -1 roll sub dup 0 lt{pop 0}if 5 1 roll
		4 -1 roll sub dup 0 lt{pop 0}if 3 1 roll exch sub dup 0 lt{pop 0}if
setrgbcolor}__NXbdef
	/currentcmykcolor{currentrgbcolor 3{1.0 exch sub 3 1 roll}repeat
0}__NXbdef
	/colorimage{_NXDoImageOp}__NXbdef
	/nxsetrgbcolor{setrgbcolor}__NXbdef /nxsetgray{setgray}__NXbdef
	/setpattern{pop .5 setgray}__NXbdef
	/_NXCalibratedImage{dup 1 eq {pop pop image}{colorimage}ifelse
pop}__NXbdef
} ifelse
/_NXSetCMYKOrRGB where{pop}{
	mark{systemdict /currentwindow get exec}stopped
	{{pop pop pop setcmykcolor}}{{nxsetrgbcolor pop pop pop pop}}ifelse
/_NXSetCMYKOrRGB exch def cleartomark
}ifelse

gsave
-20 -28 translate
 /__NXbasematrix matrix currentmatrix def
grestore
/oval {
    translate scale newpath 0.5 0.5 0.5 0 360 arc closepath
} def /line {
    moveto rlineto stroke
} def /setup {
    setlinewidth setlinecap setlinejoin gsave
} def /arrow {
    newpath moveto dup rotate -13 6 rlineto 4 -6 rlineto -4 -6 rlineto
closepath gsave 0 setlinejoin stroke grestore fill neg rotate
} def

/__NXsheetsavetoken save def
36 36 translate
gsave
-20 -28 translate
 /__NXbasematrix matrix currentmatrix def
grestore
gsave
0 0 translate
gsave
0 0 540 720 rectclip
0 0 540 720 rectclip
0 0 10 setup
0 0 10 setup
0 0 0 setup
50.586678 307.676331 transform
gsave __NXbasematrix setmatrix itransform translate
0 0 144.124512 140.225677 rectclip
gsave
0.52219 0.488591 scale

/__NXEPSSave save def /showpage {} def
_NXLevel2{/_NXsethsb where{pop}{/_NXsethsb /sethsbcolor load def}ifelse
/sethsbcolor{_NXsethsb currentrgbcolor nxsetrgbcolor}def
/setrgbcolor{nxsetrgbcolor}bind def /setgray{nxsetgray}bind def
/_NXcimage where{pop}{/_NXcimage /colorimage load def}ifelse
/colorimage{dup 3
eq{true 2 index{1 index}{1}ifelse 7 add 1 roll
_NXCalibratedImage}{_NXcimage}ifelse}def}if
0 setgray 0 setlinecap 1 setlinewidth
0 setlinejoin 10 setmiterlimit [] 0 setdash newpath count /__NXEPSOpCount
exch
def /__NXEPSDictCount countdictstack def
/Mnodistort true def
100 dict begin
/Mfixwid true def
/Mrot 0 def
/Mpstart {
    MathPictureStart
} bind def
/Mpend {
    MathPictureEnd
} bind def
/Mscale {
    0 1 0 1
    5 -1 roll
    MathScale
} bind def
/Plain	/Courier findfont def
/Bold	/Courier-Bold findfont def
/Italic /Courier-Oblique findfont def
/MathPictureStart {
	/Mimatrix
	 matrix currentmatrix
	def
	gsave
	newpath
	Mleft
	Mbottom
	translate
	1 -1 scale
	/Mtmatrix
	matrix currentmatrix
	def
	Plain
	Mfontsize scalefont
	setfont
	0 setgray
	0 setlinewidth
} bind def
/MathPictureEnd {
	grestore
} bind def
/MathSubStart {
        Mgmatrix Mtmatrix
        Mleft Mbottom
        Mwidth Mheight
        8 -2 roll
        moveto
        Mtmatrix setmatrix
        currentpoint
        Mgmatrix setmatrix
        10 -2 roll
        moveto
        Mtmatrix setmatrix
        currentpoint
        2 copy translate
        /Mtmatrix matrix currentmatrix def
        /Mleft 0 def
        /Mbottom 0 def
        3 -1 roll
        exch sub
        /Mheight exch def
        sub
        /Mwidth exch def
} bind def
/MathSubEnd {
        /Mheight exch def
        /Mwidth exch def
        /Mbottom exch def
        /Mleft exch def
        /Mtmatrix exch def
        dup setmatrix
        /Mgmatrix exch def
} bind def
/Mdot {
	moveto
	0 0 rlineto
	stroke
} bind def
/Mtetra {
	moveto
	lineto
	lineto
	lineto
	fill
} bind def
/Metetra {
	moveto
	lineto
	lineto
	lineto
	closepath
	gsave
	fill
	grestore
	0 setgray
	stroke
} bind def
/Mistroke {
	flattenpath
	0 0 0
	{
	4 2 roll
	pop pop
	}
	{
	4 -1 roll
	2 index
	sub dup mul
	4 -1 roll
	2 index
	sub dup mul
	add sqrt
	4 -1 roll
	add
	3 1 roll
	}
	{
	stop
	}
	{
	stop
	}
	pathforall
	pop pop
	currentpoint
	stroke
	moveto
	currentdash
	3 -1 roll
	add
	setdash
} bind def
/Mfstroke {
	stroke
	currentdash
	pop 0
	setdash
} bind def
/Mrotsboxa {
	gsave
	dup
	/Mrot
	exch def
	Mrotcheck
	Mtmatrix
	dup
	setmatrix
	7 1 roll
	4 index
	4 index
	translate
	rotate
	3 index
	-1 mul
	3 index
	-1 mul
	translate
	/Mtmatrix
	matrix
	currentmatrix
	def
	grestore
	Msboxa
	3  -1 roll
	/Mtmatrix
	exch def
	/Mrot
	0 def
} bind def
/Msboxa {
	newpath
	5 -1 roll
	Mvboxa
	pop
	Mboxout
	6 -1 roll
	5 -1 roll
	4 -1 roll
	Msboxa1
	5 -3 roll
	Msboxa1
	Mboxrot
	[
	7 -2 roll
	2 copy
	[
	3 1 roll
	10 -1 roll
	9 -1 roll
	]
	6 1 roll
	5 -2 roll
	]
} bind def
/Msboxa1 {
	sub
	2 div
	dup
	2 index
	1 add
	mul
	3 -1 roll
	-1 add
	3 -1 roll
	mul
} bind def
/Mvboxa	{
	Mfixwid
	{
	Mvboxa1
	}
	{
	dup
	Mwidthcal
	0 exch
	{
	add
	}
	forall
	exch
	Mvboxa1
	4 index
	7 -1 roll
	add
	4 -1 roll
	pop
	3 1 roll
	}
	ifelse
} bind def
/Mvboxa1 {
	gsave
	newpath
	[ true
	3 -1 roll
	{
	Mbbox
	5 -1 roll
	{
	0
	5 1 roll
	}
	{
	7 -1 roll
	exch sub
	(m) stringwidth pop
	.3 mul
	sub
	7 1 roll
	6 -1 roll
	4 -1 roll
	Mmin
	3 -1 roll
	5 index
	add
	5 -1 roll
	4 -1 roll
	Mmax
	4 -1 roll
	}
	ifelse
	false
	}
	forall
	{ stop } if
	counttomark
	1 add
	4 roll
	]
	grestore
} bind def
/Mbbox {
	0 0 moveto
	false charpath
	flattenpath
	pathbbox
	newpath
} bind def
/Mmin {
	2 copy
	gt
	{ exch } if
	pop
} bind def
/Mmax {
	2 copy
	lt
	{ exch } if
	pop
} bind def
/Mrotshowa {
	dup
	/Mrot
	exch def
	Mrotcheck
	Mtmatrix
	dup
	setmatrix
	7 1 roll
	4 index
	4 index
	translate
	rotate
	3 index
	-1 mul
	3 index
	-1 mul
	translate
	/Mtmatrix
	matrix
	currentmatrix
	def
	Mgmatrix setmatrix
	Mshowa
	/Mtmatrix
	exch def
	/Mrot 0 def
} bind def
/Mshowa {
	4 -2 roll
	moveto
	2 index
	Mtmatrix setmatrix
	Mvboxa
	7 1 roll
	Mboxout
	6 -1 roll
	5 -1 roll
	4 -1 roll
	Mshowa1
	4 1 roll
	Mshowa1
	rmoveto
	currentpoint
	Mfixwid
	{
	Mshowax
	}
	{
	Mshoway
	}
	ifelse
	pop pop pop pop
	Mgmatrix setmatrix
} bind def
/Mshowax {
	0 1
        4 index length
        -1 add
        {
        2 index
        4 index
        2 index
        get
        3 index
        add
        moveto
        4 index
        exch get
        show
        } for
} bind def
/Mshoway {
        3 index
        Mwidthcal
        5 1 roll
	0 1
	4 index length
	-1 add
	{
	2 index
	4 index
	2 index
	get
	3 index
	add
	moveto
	4 index
	exch get
	[
	6 index
	aload
	length
	2 add
	-1 roll
	{
	pop
	Strform
	stringwidth
	pop
	neg
	exch
	add
	0 rmoveto
	}
	exch
	kshow
	cleartomark
	} for
	pop
} bind def
/Mwidthcal {
	[
	exch
	{
	Mwidthcal1
	}
	forall
	]
	[
	exch
	dup
	Maxlen
	-1 add
	0 1
	3 -1 roll
	{
	[
	exch
	2 index
	{
	1 index
	Mget
	exch
	}
	forall
	pop
	Maxget
	exch
	}
	for
	pop
	]
	Mreva
} bind def
/Mreva	{
	[
	exch
	aload
	length
	-1 1
	{1 roll}
	for
	]
} bind def
/Mget	{
	1 index
	length
	-1 add
	1 index
	ge
	{
	get
	}
	{
	pop pop
	0
	}
	ifelse
} bind def
/Maxlen	{
	[
	exch
	{
	length
	}
	forall
	Maxget
} bind def
/Maxget	{
	counttomark
	-1 add
	1 1
	3 -1 roll
	{
	pop
	Mmax
	}
	for
	exch
	pop
} bind def
/Mwidthcal1 {
	[
	exch
	{
	Strform
	stringwidth
	pop
	}
	forall
	]
} bind def
/Strform {
	/tem (x) def
	tem 0
	3 -1 roll
	put
	tem
} bind def
/Mshowa1 {
	2 copy
	add
	4 1 roll
	sub
	mul
	sub
	-2 div
} bind def
/MathScale {
	Mwidth
	Mheight
	Mlp
	translate
	scale
	/Msaveaa exch def
	/Msavebb exch def
	/Msavecc exch def
	/Msavedd exch def
	/Mgmatrix
	matrix currentmatrix
	def
} bind def
/Mlp {
	3 copy
	Mlpfirst
	{
	Mnodistort
	{
	Mmin
	dup
	} if
	4 index
	2 index
	2 index
	Mlprun
	11 index
	11 -1 roll
	10 -4 roll
	Mlp1
	8 index
	9 -5 roll
	Mlp1
	4 -1 roll
	and
	{ exit } if
	3 -1 roll
	pop pop
	} loop
	exch
	3 1 roll
	7 -3 roll
	pop pop pop
} bind def
/Mlpfirst {
	3 -1 roll
	dup length
	2 copy
	-2 add
	get
	aload
	pop pop pop
	4 -2 roll
	-1 add
	get
	aload
	pop pop pop
	6 -1 roll
	3 -1 roll
	5 -1 roll
	sub
	dup /MsaveAx exch def
	div
	4 1 roll
	exch sub
	dup /MsaveAy exch def
	div
} bind def
/Mlprun {
	2 copy
	4 index
	0 get
	dup
	4 1 roll
	Mlprun1
	3 copy
	8 -2 roll
	9 -1 roll
	{
	3 copy
	Mlprun1
	3 copy
	11 -3 roll
	/gt Mlpminmax
	8 3 roll
	11 -3 roll
	/lt Mlpminmax
	8 3 roll
	} forall
	pop pop pop pop
	3 1 roll
	pop pop
	aload pop
	5 -1 roll
	aload pop
	exch
	6 -1 roll
	Mlprun2
	8 2 roll
	4 -1 roll
	Mlprun2
	6 2 roll
	3 -1 roll
	Mlprun2
	4 2 roll
	exch
	Mlprun2
	6 2 roll
} bind def
/Mlprun1 {
	aload pop
	exch
	6 -1 roll
	5 -1 roll
	mul add
	4 -2 roll
	mul
	3 -1 roll
	add
} bind def
/Mlprun2 {
	2 copy
	add 2 div
	3 1 roll
	exch sub
} bind def
/Mlpminmax {
	cvx
	2 index
	6 index
	2 index
	exec
	{
	7 -3 roll
	4 -1 roll
	} if
	1 index
	5 index
	3 -1 roll
	exec
	{
	4 1 roll
	pop
	5 -1 roll
	aload
	pop pop
	4 -1 roll
	aload pop
	[
	8 -2 roll
	pop
	5 -2 roll
	pop
	6 -2 roll
	pop
	5 -1 roll
	]
	4 1 roll
	pop
	}
	{
	pop pop pop
	} ifelse
} bind def
/Mlp1 {
	5 index
	3 index	sub
	5 index
	2 index mul
	1 index
	le
	1 index
	0 le
	or
	dup
	not
	{
	1 index
	3 index	div
	.99999 mul
	8 -1 roll
	pop
	7 1 roll
	}
	if
	8 -1 roll
	2 div
	7 -2 roll
	pop sub
	5 index
	6 -3 roll
	pop pop
	mul sub
	exch
} bind def
/intop 0 def
/inrht 0 def
/inflag 0 def
/outflag 0 def
/xadrht 0 def
/xadlft 0 def
/yadtop 0 def
/yadbot 0 def
/Minner {
	outflag
	1
	eq
	{
	/outflag 0 def
	/intop 0 def
	/inrht 0 def
	} if
	5 index
	gsave
	Mtmatrix setmatrix
	Mvboxa pop
	grestore
	3 -1 roll
	pop
	dup
	intop
	gt
	{
	/intop
	exch def
	}
	{ pop }
	ifelse
	dup
	inrht
	gt
	{
	/inrht
	exch def
	}
	{ pop }
	ifelse
	pop
	/inflag
	1 def
} bind def
/Mouter {
	/xadrht 0 def
	/xadlft 0 def
	/yadtop 0 def
	/yadbot 0 def
	inflag
	1 eq
	{
	dup
	0 lt
	{
	dup
	intop
	mul
	neg
	/yadtop
	exch def
	} if
	dup
	0 gt
	{
	dup
	intop
	mul
	/yadbot
	exch def
	}
	if
	pop
	dup
	0 lt
	{
	dup
	inrht
	mul
	neg
	/xadrht
	exch def
	} if
	dup
	0 gt
	{
	dup
	inrht
	mul
	/xadlft
	exch def
	} if
	pop
	/outflag 1 def
	}
	{ pop pop}
	ifelse
	/inflag 0 def
	/inrht 0 def
	/intop 0 def
} bind def
/Mboxout {
	outflag
	1
	eq
	{
	4 -1
	roll
	xadlft
	leadjust
	add
	sub
	4 1 roll
	3 -1
	roll
	yadbot
	leadjust
	add
	sub
	3 1
	roll
	exch
	xadrht
	leadjust
	add
	add
	exch
	yadtop
	leadjust
	add
	add
	/outflag 0 def
	/xadlft 0 def
	/yadbot 0 def
	/xadrht 0 def
	/yadtop 0 def
	} if
} bind def
/leadjust {
	(m) stringwidth pop
	.5 mul
} bind def
/Mrotcheck {
	dup
	90
	eq
	{
	yadbot
	/yadbot
	xadrht
	def
	/xadrht
	yadtop
	def
	/yadtop
	xadlft
	def
	/xadlft
	exch
	def
	}
	if
	dup
	cos
	1 index
	sin
	Checkaux
	dup
	cos
	1 index
	sin neg
	exch
	Checkaux
	3 1 roll
	pop pop
} bind def
/Checkaux {
	4 index
	exch
	4 index
	mul
	3 1 roll
	mul add
	4 1 roll
} bind def
/Mboxrot {
	Mrot
	90 eq
	{
	brotaux
	4 2
	roll
	}
	if
	Mrot
	180 eq
	{
	4 2
        roll
	brotaux
	4 2
	roll
	brotaux
	}
	if
	Mrot
	270 eq
	{
	4 2
	roll
	brotaux
	}
	if
} bind def
/brotaux {
	neg
	exch
	neg
} bind def
/Mabswid {
	Mimatrix
	0 get
	Mgmatrix
	0 get
	div
	mul
	setlinewidth
} bind def
/Mabsdash {
	exch
	Mimatrix
	0 get
	Mgmatrix
	0 get
	div
	[
	3 1 roll
	exch
	{
	exch
	dup
	3 -1 roll
	mul
	exch
	}
	forall
	pop ]
	exch
	setdash
} bind def
/colorimage where
{ pop }
{
/colorimage {
3 1 roll
 pop pop
 5 -1 roll
 mul
 4 1 roll
{
currentfile
1 index
readhexstring
pop }
image
} bind def
} ifelse
/sampledsound where
{ pop}
{ /sampledsound {
exch
pop
exch
5 1 roll
mul
4 idiv
mul
2 idiv
exch pop
exch
/Mtempproc exch def
{ Mtempproc pop }
repeat
} bind def
} ifelse

/Mleft		0.000000 def
/Mbottom	287.000000 def
/Mwidth		276.941498 def
/Mheight	287.000000 def
/Mfontsize 12 def
/Plain /Courier findfont def
0 Mbottom Mheight neg add 2 mul Mheight add translate
1 -1 scale
MathPictureStart
/Courier findfont 10  scalefont  setfont
0.5 0.035727 0.51816 0.035727 [
[ 0 0 0 0 ]
[ 1 1.03632 0 0 ]
] MathScale
1 setlinecap
1 setlinejoin
newpath
[ ] 0 setdash
0 setgray
gsave
grestore
0 0 moveto
1 0 lineto
1 1.03632 lineto
0 1.03632 lineto
closepath
clip
newpath
gsave
gsave
6 Mabswid
0.02381 0.99435 moveto
0.05307 1.00377 lineto
0.06676 1.00754 lineto
0.07322 1.00904 lineto
0.07938 1.01024 lineto
0.08519 1.01109 lineto
0.08795 1.01137 lineto
0.09062 1.01156 lineto
0.09319 1.01165 lineto
0.09565 1.01163 lineto
0.09801 1.0115 lineto
0.10025 1.01127 lineto
0.10238 1.01092 lineto
0.1044 1.01046 lineto
0.1063 1.00988 lineto
0.10809 1.00918 lineto
0.10975 1.00837 lineto
0.11129 1.00743 lineto
0.11272 1.00637 lineto
0.11402 1.00519 lineto
0.1152 1.00389 lineto
0.11625 1.00247 lineto
0.11719 1.00093 lineto
0.11801 0.99926 lineto
0.1187 0.99748 lineto
0.11928 0.99558 lineto
0.11974 0.99356 lineto
0.12009 0.99143 lineto
0.12033 0.98918 lineto
0.12045 0.98683 lineto
0.12047 0.98436 lineto
0.12038 0.9818 lineto
0.1202 0.97913 lineto
0.11991 0.97636 lineto
0.11906 0.97055 lineto
0.11787 0.9644 lineto
0.11636 0.95794 lineto
0.11259 0.94424 lineto
0.10317 0.91499 lineto
0.09376 0.88573 lineto
0.08999 0.87203 lineto
0.08848 0.86557 lineto
0.08729 0.85942 lineto
0.08682 0.85647 lineto
0.08644 0.85361 lineto
0.08615 0.85084 lineto
0.08597 0.84817 lineto
0.08588 0.84561 lineto
0.0859 0.84314 lineto
0.08602 0.84079 lineto
Mistroke
0.08626 0.83854 lineto
0.0866 0.83641 lineto
0.08707 0.83439 lineto
0.08765 0.83249 lineto
0.08834 0.83071 lineto
0.09009 0.8275 lineto
0.09115 0.82608 lineto
0.09233 0.82478 lineto
0.18254 0.83562 lineto
0.2118 0.84504 lineto
0.22549 0.84881 lineto
0.23195 0.85031 lineto
0.23811 0.85151 lineto
0.24106 0.85198 lineto
0.24392 0.85236 lineto
0.24668 0.85264 lineto
0.24935 0.85283 lineto
0.25192 0.85292 lineto
0.25438 0.8529 lineto
0.25674 0.85277 lineto
0.25898 0.85254 lineto
0.26111 0.85219 lineto
0.26313 0.85173 lineto
0.26503 0.85115 lineto
0.26682 0.85045 lineto
0.27002 0.8487 lineto
0.27145 0.84764 lineto
0.27275 0.84646 lineto
0.2619 0.75626 lineto
0.25249 0.727 lineto
0.24872 0.7133 lineto
0.24721 0.70684 lineto
0.24602 0.70069 lineto
0.24555 0.69774 lineto
0.24517 0.69488 lineto
0.24488 0.69211 lineto
0.2447 0.68944 lineto
0.24461 0.68688 lineto
0.24463 0.68441 lineto
0.24475 0.68206 lineto
0.24499 0.67981 lineto
0.24533 0.67768 lineto
0.2458 0.67566 lineto
0.24707 0.67198 lineto
0.24789 0.67031 lineto
0.24883 0.66877 lineto
0.24988 0.66735 lineto
0.25106 0.66605 lineto
0.25379 0.66381 lineto
0.25533 0.66287 lineto
Mistroke
0.25699 0.66206 lineto
0.25878 0.66136 lineto
0.26068 0.66078 lineto
0.2627 0.66032 lineto
0.26483 0.65997 lineto
0.26707 0.65974 lineto
0.26943 0.65961 lineto
0.27189 0.65959 lineto
0.27446 0.65968 lineto
0.27713 0.65987 lineto
0.27989 0.66016 lineto
0.28275 0.66053 lineto
0.2857 0.661 lineto
0.34127 0.67689 lineto
0.37053 0.6863 lineto
0.38422 0.69008 lineto
0.39068 0.69158 lineto
0.39684 0.69278 lineto
0.39979 0.69325 lineto
0.40265 0.69363 lineto
0.40541 0.69391 lineto
0.40808 0.6941 lineto
0.41065 0.69419 lineto
0.41311 0.69417 lineto
0.41547 0.69404 lineto
0.41771 0.69381 lineto
0.41984 0.69346 lineto
0.42186 0.693 lineto
0.42376 0.69242 lineto
0.42555 0.69172 lineto
0.42721 0.69091 lineto
0.42875 0.68997 lineto
0.43018 0.68891 lineto
0.43148 0.68773 lineto
0.43371 0.68501 lineto
0.43465 0.68347 lineto
0.43547 0.6818 lineto
0.43616 0.68002 lineto
0.43674 0.67812 lineto
0.4372 0.6761 lineto
0.43755 0.67397 lineto
0.43779 0.67172 lineto
0.43791 0.66937 lineto
0.43793 0.6669 lineto
0.43784 0.66434 lineto
0.43766 0.66167 lineto
0.43737 0.6589 lineto
0.43699 0.65604 lineto
0.43652 0.65309 lineto
0.42063 0.59753 lineto
Mistroke
0.41122 0.56827 lineto
0.40745 0.55457 lineto
0.40594 0.54811 lineto
0.40475 0.54196 lineto
0.40428 0.53901 lineto
0.4039 0.53615 lineto
0.40361 0.53338 lineto
0.40343 0.53071 lineto
0.40334 0.52815 lineto
0.40336 0.52568 lineto
0.40348 0.52333 lineto
0.40372 0.52108 lineto
0.40407 0.51895 lineto
0.40453 0.51693 lineto
0.40511 0.51503 lineto
0.4058 0.51325 lineto
0.40756 0.51004 lineto
0.40861 0.50862 lineto
0.40979 0.50732 lineto
0.41109 0.50614 lineto
0.41252 0.50508 lineto
0.41406 0.50414 lineto
0.41572 0.50333 lineto
0.41751 0.50263 lineto
0.41941 0.50205 lineto
0.42143 0.50159 lineto
0.42356 0.50124 lineto
0.4258 0.50101 lineto
0.42816 0.50088 lineto
0.43062 0.50086 lineto
0.43319 0.50095 lineto
0.43586 0.50114 lineto
0.43862 0.50142 lineto
0.44443 0.50227 lineto
0.45059 0.50347 lineto
0.45705 0.50497 lineto
0.47074 0.50875 lineto
0.5 0.51816 lineto
0.52926 0.52757 lineto
0.54295 0.53135 lineto
0.54941 0.53285 lineto
0.55557 0.53405 lineto
0.55852 0.53452 lineto
0.56138 0.5349 lineto
0.56414 0.53518 lineto
0.56681 0.53537 lineto
0.56938 0.53546 lineto
0.57184 0.53544 lineto
0.5742 0.53531 lineto
0.57644 0.53508 lineto
Mistroke
0.57857 0.53473 lineto
0.58059 0.53427 lineto
0.58249 0.53369 lineto
0.58428 0.53299 lineto
0.58748 0.53124 lineto
0.58891 0.53018 lineto
0.59021 0.529 lineto
0.59139 0.5277 lineto
0.59244 0.52628 lineto
0.59338 0.52474 lineto
0.5942 0.52307 lineto
0.59489 0.52129 lineto
0.59547 0.51939 lineto
0.59593 0.51737 lineto
0.59628 0.51524 lineto
0.59652 0.51299 lineto
0.59664 0.51064 lineto
0.59666 0.50817 lineto
0.59657 0.50561 lineto
0.59639 0.50294 lineto
0.5961 0.50017 lineto
0.59525 0.49436 lineto
0.59406 0.48821 lineto
0.59255 0.48175 lineto
0.58878 0.46805 lineto
0.57937 0.4388 lineto
0.56995 0.40954 lineto
0.56618 0.39584 lineto
0.56467 0.38938 lineto
0.56348 0.38323 lineto
0.56301 0.38028 lineto
0.56263 0.37742 lineto
0.56234 0.37465 lineto
0.56216 0.37198 lineto
0.56207 0.36942 lineto
0.56209 0.36695 lineto
0.56221 0.3646 lineto
0.56245 0.36235 lineto
0.5628 0.36022 lineto
0.56326 0.3582 lineto
0.56384 0.3563 lineto
0.56453 0.35452 lineto
0.56535 0.35285 lineto
0.56629 0.35131 lineto
0.56734 0.34989 lineto
0.56852 0.34859 lineto
0.57125 0.34635 lineto
0.57279 0.34541 lineto
0.57445 0.3446 lineto
0.57624 0.3439 lineto
Mistroke
0.57814 0.34332 lineto
0.58016 0.34286 lineto
0.58229 0.34251 lineto
0.58453 0.34228 lineto
0.58689 0.34215 lineto
0.58935 0.34213 lineto
0.59192 0.34222 lineto
0.59459 0.34241 lineto
0.59735 0.34269 lineto
0.60021 0.34307 lineto
0.60316 0.34354 lineto
0.65873 0.35943 lineto
0.68799 0.36884 lineto
0.70168 0.37262 lineto
0.70814 0.37412 lineto
0.7143 0.37532 lineto
0.71725 0.37579 lineto
0.72011 0.37617 lineto
0.72287 0.37645 lineto
0.72554 0.37664 lineto
0.72811 0.37673 lineto
0.73057 0.37671 lineto
0.73293 0.37658 lineto
0.73517 0.37635 lineto
0.7373 0.376 lineto
0.73932 0.37554 lineto
0.74122 0.37496 lineto
0.74301 0.37426 lineto
0.74467 0.37345 lineto
0.74621 0.37251 lineto
0.74894 0.37027 lineto
0.75012 0.36897 lineto
0.75117 0.36755 lineto
0.75211 0.36601 lineto
0.75293 0.36434 lineto
0.75362 0.36256 lineto
0.7542 0.36066 lineto
0.75467 0.35864 lineto
0.75501 0.35651 lineto
0.75525 0.35426 lineto
0.75537 0.35191 lineto
0.75539 0.34944 lineto
0.7553 0.34688 lineto
0.75512 0.34421 lineto
0.75483 0.34144 lineto
0.75398 0.33563 lineto
0.75279 0.32948 lineto
0.75128 0.32302 lineto
0.74751 0.30932 lineto
0.7381 0.28006 lineto
Mistroke
0.72868 0.25081 lineto
0.72491 0.23711 lineto
0.7234 0.23065 lineto
0.72221 0.2245 lineto
0.72174 0.22155 lineto
0.72136 0.21869 lineto
0.72107 0.21592 lineto
0.72089 0.21325 lineto
0.7208 0.21069 lineto
0.72082 0.20822 lineto
0.72094 0.20587 lineto
0.72118 0.20362 lineto
0.72153 0.20149 lineto
0.72199 0.19947 lineto
0.72257 0.19757 lineto
0.72326 0.19579 lineto
0.72502 0.19258 lineto
0.72607 0.19116 lineto
0.72725 0.18986 lineto
0.81746 0.2007 lineto
0.84672 0.21011 lineto
0.86041 0.21389 lineto
0.86687 0.21539 lineto
0.87303 0.21659 lineto
0.87598 0.21706 lineto
0.87884 0.21744 lineto
0.8816 0.21772 lineto
0.88427 0.21791 lineto
0.88684 0.218 lineto
0.8893 0.21798 lineto
0.89166 0.21785 lineto
0.8939 0.21762 lineto
0.89603 0.21727 lineto
0.89805 0.21681 lineto
0.89995 0.21623 lineto
0.90174 0.21553 lineto
0.90494 0.21378 lineto
0.90637 0.21272 lineto
0.90767 0.21154 lineto
0.89683 0.12133 lineto
0.88741 0.09208 lineto
0.88364 0.07838 lineto
0.88213 0.07192 lineto
0.88094 0.06577 lineto
0.88047 0.06282 lineto
0.88009 0.05996 lineto
0.8798 0.05719 lineto
0.87962 0.05452 lineto
0.87953 0.05196 lineto
0.87955 0.04949 lineto
Mistroke
0.87967 0.04714 lineto
0.87991 0.04489 lineto
0.88026 0.04276 lineto
0.88072 0.04074 lineto
0.8813 0.03884 lineto
0.88199 0.03706 lineto
0.88281 0.03539 lineto
0.88375 0.03385 lineto
0.8848 0.03243 lineto
0.88598 0.03113 lineto
0.88728 0.02995 lineto
0.88871 0.02889 lineto
0.89025 0.02795 lineto
0.89191 0.02714 lineto
0.8937 0.02644 lineto
0.8956 0.02586 lineto
0.89762 0.0254 lineto
0.89975 0.02505 lineto
0.90199 0.02482 lineto
0.90435 0.02469 lineto
0.90681 0.02467 lineto
0.90938 0.02476 lineto
0.91205 0.02495 lineto
0.91481 0.02523 lineto
0.92062 0.02608 lineto
0.92678 0.02728 lineto
0.93324 0.02878 lineto
0.94693 0.03256 lineto
0.97619 0.04197 lineto
Mfstroke
grestore
grestore
MathPictureEnd
end


count __NXEPSOpCount sub {pop} repeat countdictstack __NXEPSDictCount
sub {end}
repeat __NXEPSSave restore
grestore
grestore
grestore
0 0 2.906977 setup
0 nxsetgray
161.333328 0 28 311.24115 line
grestore
0 0 2.906977 setup
0 nxsetgray
161.333328 0 350.666718 311.24115 line
grestore
0 0 10 setup
0 nxsetgray
161.333328 0 189.333344 311.24115 line
grestore
0 0 0 setup
343.137878 307.676331 transform
gsave __NXbasematrix setmatrix itransform translate
0 0 144.124512 142.602448 rectclip
gsave
0.52219 0.496873 scale

/__NXEPSSave save def /showpage {} def
_NXLevel2{/_NXsethsb where{pop}{/_NXsethsb /sethsbcolor load def}ifelse
/sethsbcolor{_NXsethsb currentrgbcolor nxsetrgbcolor}def
/setrgbcolor{nxsetrgbcolor}bind def /setgray{nxsetgray}bind def
/_NXcimage where{pop}{/_NXcimage /colorimage load def}ifelse
/colorimage{dup 3
eq{true 2 index{1 index}{1}ifelse 7 add 1 roll
_NXCalibratedImage}{_NXcimage}ifelse}def}if
0 setgray 0 setlinecap 1 setlinewidth
0 setlinejoin 10 setmiterlimit [] 0 setdash newpath count /__NXEPSOpCount
exch
def /__NXEPSDictCount countdictstack def
/Mnodistort true def
100 dict begin
/Mfixwid true def
/Mrot 0 def
/Mpstart {
    MathPictureStart
} bind def
/Mpend {
    MathPictureEnd
} bind def
/Mscale {
    0 1 0 1
    5 -1 roll
    MathScale
} bind def
/Plain	/Courier findfont def
/Bold	/Courier-Bold findfont def
/Italic /Courier-Oblique findfont def
/MathPictureStart {
	/Mimatrix
	 matrix currentmatrix
	def
	gsave
	newpath
	Mleft
	Mbottom
	translate
	1 -1 scale
	/Mtmatrix
	matrix currentmatrix
	def
	Plain
	Mfontsize scalefont
	setfont
	0 setgray
	0 setlinewidth
} bind def
/MathPictureEnd {
	grestore
} bind def
/MathSubStart {
        Mgmatrix Mtmatrix
        Mleft Mbottom
        Mwidth Mheight
        8 -2 roll
        moveto
        Mtmatrix setmatrix
        currentpoint
        Mgmatrix setmatrix
        10 -2 roll
        moveto
        Mtmatrix setmatrix
        currentpoint
        2 copy translate
        /Mtmatrix matrix currentmatrix def
        /Mleft 0 def
        /Mbottom 0 def
        3 -1 roll
        exch sub
        /Mheight exch def
        sub
        /Mwidth exch def
} bind def
/MathSubEnd {
        /Mheight exch def
        /Mwidth exch def
        /Mbottom exch def
        /Mleft exch def
        /Mtmatrix exch def
        dup setmatrix
        /Mgmatrix exch def
} bind def
/Mdot {
	moveto
	0 0 rlineto
	stroke
} bind def
/Mtetra {
	moveto
	lineto
	lineto
	lineto
	fill
} bind def
/Metetra {
	moveto
	lineto
	lineto
	lineto
	closepath
	gsave
	fill
	grestore
	0 setgray
	stroke
} bind def
/Mistroke {
	flattenpath
	0 0 0
	{
	4 2 roll
	pop pop
	}
	{
	4 -1 roll
	2 index
	sub dup mul
	4 -1 roll
	2 index
	sub dup mul
	add sqrt
	4 -1 roll
	add
	3 1 roll
	}
	{
	stop
	}
	{
	stop
	}
	pathforall
	pop pop
	currentpoint
	stroke
	moveto
	currentdash
	3 -1 roll
	add
	setdash
} bind def
/Mfstroke {
	stroke
	currentdash
	pop 0
	setdash
} bind def
/Mrotsboxa {
	gsave
	dup
	/Mrot
	exch def
	Mrotcheck
	Mtmatrix
	dup
	setmatrix
	7 1 roll
	4 index
	4 index
	translate
	rotate
	3 index
	-1 mul
	3 index
	-1 mul
	translate
	/Mtmatrix
	matrix
	currentmatrix
	def
	grestore
	Msboxa
	3  -1 roll
	/Mtmatrix
	exch def
	/Mrot
	0 def
} bind def
/Msboxa {
	newpath
	5 -1 roll
	Mvboxa
	pop
	Mboxout
	6 -1 roll
	5 -1 roll
	4 -1 roll
	Msboxa1
	5 -3 roll
	Msboxa1
	Mboxrot
	[
	7 -2 roll
	2 copy
	[
	3 1 roll
	10 -1 roll
	9 -1 roll
	]
	6 1 roll
	5 -2 roll
	]
} bind def
/Msboxa1 {
	sub
	2 div
	dup
	2 index
	1 add
	mul
	3 -1 roll
	-1 add
	3 -1 roll
	mul
} bind def
/Mvboxa	{
	Mfixwid
	{
	Mvboxa1
	}
	{
	dup
	Mwidthcal
	0 exch
	{
	add
	}
	forall
	exch
	Mvboxa1
	4 index
	7 -1 roll
	add
	4 -1 roll
	pop
	3 1 roll
	}
	ifelse
} bind def
/Mvboxa1 {
	gsave
	newpath
	[ true
	3 -1 roll
	{
	Mbbox
	5 -1 roll
	{
	0
	5 1 roll
	}
	{
	7 -1 roll
	exch sub
	(m) stringwidth pop
	.3 mul
	sub
	7 1 roll
	6 -1 roll
	4 -1 roll
	Mmin
	3 -1 roll
	5 index
	add
	5 -1 roll
	4 -1 roll
	Mmax
	4 -1 roll
	}
	ifelse
	false
	}
	forall
	{ stop } if
	counttomark
	1 add
	4 roll
	]
	grestore
} bind def
/Mbbox {
	0 0 moveto
	false charpath
	flattenpath
	pathbbox
	newpath
} bind def
/Mmin {
	2 copy
	gt
	{ exch } if
	pop
} bind def
/Mmax {
	2 copy
	lt
	{ exch } if
	pop
} bind def
/Mrotshowa {
	dup
	/Mrot
	exch def
	Mrotcheck
	Mtmatrix
	dup
	setmatrix
	7 1 roll
	4 index
	4 index
	translate
	rotate
	3 index
	-1 mul
	3 index
	-1 mul
	translate
	/Mtmatrix
	matrix
	currentmatrix
	def
	Mgmatrix setmatrix
	Mshowa
	/Mtmatrix
	exch def
	/Mrot 0 def
} bind def
/Mshowa {
	4 -2 roll
	moveto
	2 index
	Mtmatrix setmatrix
	Mvboxa
	7 1 roll
	Mboxout
	6 -1 roll
	5 -1 roll
	4 -1 roll
	Mshowa1
	4 1 roll
	Mshowa1
	rmoveto
	currentpoint
	Mfixwid
	{
	Mshowax
	}
	{
	Mshoway
	}
	ifelse
	pop pop pop pop
	Mgmatrix setmatrix
} bind def
/Mshowax {
	0 1
        4 index length
        -1 add
        {
        2 index
        4 index
        2 index
        get
        3 index
        add
        moveto
        4 index
        exch get
        show
        } for
} bind def
/Mshoway {
        3 index
        Mwidthcal
        5 1 roll
	0 1
	4 index length
	-1 add
	{
	2 index
	4 index
	2 index
	get
	3 index
	add
	moveto
	4 index
	exch get
	[
	6 index
	aload
	length
	2 add
	-1 roll
	{
	pop
	Strform
	stringwidth
	pop
	neg
	exch
	add
	0 rmoveto
	}
	exch
	kshow
	cleartomark
	} for
	pop
} bind def
/Mwidthcal {
	[
	exch
	{
	Mwidthcal1
	}
	forall
	]
	[
	exch
	dup
	Maxlen
	-1 add
	0 1
	3 -1 roll
	{
	[
	exch
	2 index
	{
	1 index
	Mget
	exch
	}
	forall
	pop
	Maxget
	exch
	}
	for
	pop
	]
	Mreva
} bind def
/Mreva	{
	[
	exch
	aload
	length
	-1 1
	{1 roll}
	for
	]
} bind def
/Mget	{
	1 index
	length
	-1 add
	1 index
	ge
	{
	get
	}
	{
	pop pop
	0
	}
	ifelse
} bind def
/Maxlen	{
	[
	exch
	{
	length
	}
	forall
	Maxget
} bind def
/Maxget	{
	counttomark
	-1 add
	1 1
	3 -1 roll
	{
	pop
	Mmax
	}
	for
	exch
	pop
} bind def
/Mwidthcal1 {
	[
	exch
	{
	Strform
	stringwidth
	pop
	}
	forall
	]
} bind def
/Strform {
	/tem (x) def
	tem 0
	3 -1 roll
	put
	tem
} bind def
/Mshowa1 {
	2 copy
	add
	4 1 roll
	sub
	mul
	sub
	-2 div
} bind def
/MathScale {
	Mwidth
	Mheight
	Mlp
	translate
	scale
	/Msaveaa exch def
	/Msavebb exch def
	/Msavecc exch def
	/Msavedd exch def
	/Mgmatrix
	matrix currentmatrix
	def
} bind def
/Mlp {
	3 copy
	Mlpfirst
	{
	Mnodistort
	{
	Mmin
	dup
	} if
	4 index
	2 index
	2 index
	Mlprun
	11 index
	11 -1 roll
	10 -4 roll
	Mlp1
	8 index
	9 -5 roll
	Mlp1
	4 -1 roll
	and
	{ exit } if
	3 -1 roll
	pop pop
	} loop
	exch
	3 1 roll
	7 -3 roll
	pop pop pop
} bind def
/Mlpfirst {
	3 -1 roll
	dup length
	2 copy
	-2 add
	get
	aload
	pop pop pop
	4 -2 roll
	-1 add
	get
	aload
	pop pop pop
	6 -1 roll
	3 -1 roll
	5 -1 roll
	sub
	dup /MsaveAx exch def
	div
	4 1 roll
	exch sub
	dup /MsaveAy exch def
	div
} bind def
/Mlprun {
	2 copy
	4 index
	0 get
	dup
	4 1 roll
	Mlprun1
	3 copy
	8 -2 roll
	9 -1 roll
	{
	3 copy
	Mlprun1
	3 copy
	11 -3 roll
	/gt Mlpminmax
	8 3 roll
	11 -3 roll
	/lt Mlpminmax
	8 3 roll
	} forall
	pop pop pop pop
	3 1 roll
	pop pop
	aload pop
	5 -1 roll
	aload pop
	exch
	6 -1 roll
	Mlprun2
	8 2 roll
	4 -1 roll
	Mlprun2
	6 2 roll
	3 -1 roll
	Mlprun2
	4 2 roll
	exch
	Mlprun2
	6 2 roll
} bind def
/Mlprun1 {
	aload pop
	exch
	6 -1 roll
	5 -1 roll
	mul add
	4 -2 roll
	mul
	3 -1 roll
	add
} bind def
/Mlprun2 {
	2 copy
	add 2 div
	3 1 roll
	exch sub
} bind def
/Mlpminmax {
	cvx
	2 index
	6 index
	2 index
	exec
	{
	7 -3 roll
	4 -1 roll
	} if
	1 index
	5 index
	3 -1 roll
	exec
	{
	4 1 roll
	pop
	5 -1 roll
	aload
	pop pop
	4 -1 roll
	aload pop
	[
	8 -2 roll
	pop
	5 -2 roll
	pop
	6 -2 roll
	pop
	5 -1 roll
	]
	4 1 roll
	pop
	}
	{
	pop pop pop
	} ifelse
} bind def
/Mlp1 {
	5 index
	3 index	sub
	5 index
	2 index mul
	1 index
	le
	1 index
	0 le
	or
	dup
	not
	{
	1 index
	3 index	div
	.99999 mul
	8 -1 roll
	pop
	7 1 roll
	}
	if
	8 -1 roll
	2 div
	7 -2 roll
	pop sub
	5 index
	6 -3 roll
	pop pop
	mul sub
	exch
} bind def
/intop 0 def
/inrht 0 def
/inflag 0 def
/outflag 0 def
/xadrht 0 def
/xadlft 0 def
/yadtop 0 def
/yadbot 0 def
/Minner {
	outflag
	1
	eq
	{
	/outflag 0 def
	/intop 0 def
	/inrht 0 def
	} if
	5 index
	gsave
	Mtmatrix setmatrix
	Mvboxa pop
	grestore
	3 -1 roll
	pop
	dup
	intop
	gt
	{
	/intop
	exch def
	}
	{ pop }
	ifelse
	dup
	inrht
	gt
	{
	/inrht
	exch def
	}
	{ pop }
	ifelse
	pop
	/inflag
	1 def
} bind def
/Mouter {
	/xadrht 0 def
	/xadlft 0 def
	/yadtop 0 def
	/yadbot 0 def
	inflag
	1 eq
	{
	dup
	0 lt
	{
	dup
	intop
	mul
	neg
	/yadtop
	exch def
	} if
	dup
	0 gt
	{
	dup
	intop
	mul
	/yadbot
	exch def
	}
	if
	pop
	dup
	0 lt
	{
	dup
	inrht
	mul
	neg
	/xadrht
	exch def
	} if
	dup
	0 gt
	{
	dup
	inrht
	mul
	/xadlft
	exch def
	} if
	pop
	/outflag 1 def
	}
	{ pop pop}
	ifelse
	/inflag 0 def
	/inrht 0 def
	/intop 0 def
} bind def
/Mboxout {
	outflag
	1
	eq
	{
	4 -1
	roll
	xadlft
	leadjust
	add
	sub
	4 1 roll
	3 -1
	roll
	yadbot
	leadjust
	add
	sub
	3 1
	roll
	exch
	xadrht
	leadjust
	add
	add
	exch
	yadtop
	leadjust
	add
	add
	/outflag 0 def
	/xadlft 0 def
	/yadbot 0 def
	/xadrht 0 def
	/yadtop 0 def
	} if
} bind def
/leadjust {
	(m) stringwidth pop
	.5 mul
} bind def
/Mrotcheck {
	dup
	90
	eq
	{
	yadbot
	/yadbot
	xadrht
	def
	/xadrht
	yadtop
	def
	/yadtop
	xadlft
	def
	/xadlft
	exch
	def
	}
	if
	dup
	cos
	1 index
	sin
	Checkaux
	dup
	cos
	1 index
	sin neg
	exch
	Checkaux
	3 1 roll
	pop pop
} bind def
/Checkaux {
	4 index
	exch
	4 index
	mul
	3 1 roll
	mul add
	4 1 roll
} bind def
/Mboxrot {
	Mrot
	90 eq
	{
	brotaux
	4 2
	roll
	}
	if
	Mrot
	180 eq
	{
	4 2
        roll
	brotaux
	4 2
	roll
	brotaux
	}
	if
	Mrot
	270 eq
	{
	4 2
	roll
	brotaux
	}
	if
} bind def
/brotaux {
	neg
	exch
	neg
} bind def
/Mabswid {
	Mimatrix
	0 get
	Mgmatrix
	0 get
	div
	mul
	setlinewidth
} bind def
/Mabsdash {
	exch
	Mimatrix
	0 get
	Mgmatrix
	0 get
	div
	[
	3 1 roll
	exch
	{
	exch
	dup
	3 -1 roll
	mul
	exch
	}
	forall
	pop ]
	exch
	setdash
} bind def
/colorimage where
{ pop }
{
/colorimage {
3 1 roll
 pop pop
 5 -1 roll
 mul
 4 1 roll
{
currentfile
1 index
readhexstring
pop }
image
} bind def
} ifelse
/sampledsound where
{ pop}
{ /sampledsound {
exch
pop
exch
5 1 roll
mul
4 idiv
mul
2 idiv
exch pop
exch
/Mtempproc exch def
{ Mtempproc pop }
repeat
} bind def
} ifelse

/Mleft		0.000000 def
/Mbottom	287.000000 def
/Mwidth		276.941498 def
/Mheight	287.000000 def
/Mfontsize 12 def
/Plain /Courier findfont def
0 Mbottom Mheight neg add 2 mul Mheight add translate
1 -1 scale
MathPictureStart
/Courier findfont 10  scalefont  setfont
0.5 0.035727 0.51816 0.035727 [
[ 0 0 0 0 ]
[ 1 1.03632 0 0 ]
] MathScale
1 setlinecap
1 setlinejoin
newpath
[ ] 0 setdash
0 setgray
gsave
grestore
0 0 moveto
1 0 lineto
1 1.03632 lineto
0 1.03632 lineto
closepath
clip
newpath
gsave
gsave
6 Mabswid
0.02381 0.04197 moveto
0.05307 0.03256 lineto
0.06676 0.02878 lineto
0.07322 0.02728 lineto
0.07938 0.02608 lineto
0.08519 0.02523 lineto
0.08795 0.02495 lineto
0.09062 0.02476 lineto
0.09319 0.02467 lineto
0.09565 0.02469 lineto
0.09801 0.02482 lineto
0.10025 0.02505 lineto
0.10238 0.0254 lineto
0.1044 0.02586 lineto
0.1063 0.02644 lineto
0.10809 0.02714 lineto
0.10975 0.02795 lineto
0.11129 0.02889 lineto
0.11272 0.02995 lineto
0.11402 0.03113 lineto
0.1152 0.03243 lineto
0.11625 0.03385 lineto
0.11719 0.03539 lineto
0.11801 0.03706 lineto
0.1187 0.03884 lineto
0.11928 0.04074 lineto
0.11974 0.04276 lineto
0.12009 0.04489 lineto
0.12033 0.04714 lineto
0.12045 0.04949 lineto
0.12047 0.05196 lineto
0.12038 0.05452 lineto
0.1202 0.05719 lineto
0.11991 0.05996 lineto
0.11906 0.06577 lineto
0.11787 0.07192 lineto
0.11636 0.07838 lineto
0.11259 0.09208 lineto
0.10317 0.12133 lineto
0.09376 0.15059 lineto
0.08999 0.16429 lineto
0.08848 0.17075 lineto
0.08729 0.1769 lineto
0.08682 0.17985 lineto
0.08644 0.18271 lineto
0.08615 0.18548 lineto
0.08597 0.18815 lineto
0.08588 0.19071 lineto
0.0859 0.19318 lineto
0.08602 0.19553 lineto
Mistroke
0.08626 0.19778 lineto
0.0866 0.19991 lineto
0.08707 0.20193 lineto
0.08765 0.20383 lineto
0.08834 0.20561 lineto
0.09009 0.20882 lineto
0.09115 0.21024 lineto
0.09233 0.21154 lineto
0.18254 0.2007 lineto
0.2118 0.19129 lineto
0.22549 0.18751 lineto
0.23195 0.18601 lineto
0.23811 0.18481 lineto
0.24106 0.18434 lineto
0.24392 0.18396 lineto
0.24668 0.18368 lineto
0.24935 0.18349 lineto
0.25192 0.1834 lineto
0.25438 0.18342 lineto
0.25674 0.18355 lineto
0.25898 0.18378 lineto
0.26111 0.18413 lineto
0.26313 0.18459 lineto
0.26503 0.18517 lineto
0.26682 0.18587 lineto
0.27002 0.18762 lineto
0.27145 0.18868 lineto
0.27275 0.18986 lineto
0.2619 0.28006 lineto
0.25249 0.30932 lineto
0.24872 0.32302 lineto
0.24721 0.32948 lineto
0.24602 0.33563 lineto
0.24555 0.33858 lineto
0.24517 0.34144 lineto
0.24488 0.34421 lineto
0.2447 0.34688 lineto
0.24461 0.34944 lineto
0.24463 0.35191 lineto
0.24475 0.35426 lineto
0.24499 0.35651 lineto
0.24533 0.35864 lineto
0.2458 0.36066 lineto
0.24707 0.36434 lineto
0.24789 0.36601 lineto
0.24883 0.36755 lineto
0.24988 0.36897 lineto
0.25106 0.37027 lineto
0.25379 0.37251 lineto
0.25533 0.37345 lineto
Mistroke
0.25699 0.37426 lineto
0.25878 0.37496 lineto
0.26068 0.37554 lineto
0.2627 0.376 lineto
0.26483 0.37635 lineto
0.26707 0.37658 lineto
0.26943 0.37671 lineto
0.27189 0.37673 lineto
0.27446 0.37664 lineto
0.27713 0.37645 lineto
0.27989 0.37617 lineto
0.28275 0.37579 lineto
0.2857 0.37532 lineto
0.34127 0.35943 lineto
0.37053 0.35002 lineto
0.38422 0.34624 lineto
0.39068 0.34474 lineto
0.39684 0.34354 lineto
0.39979 0.34307 lineto
0.40265 0.34269 lineto
0.40541 0.34241 lineto
0.40808 0.34222 lineto
0.41065 0.34213 lineto
0.41311 0.34215 lineto
0.41547 0.34228 lineto
0.41771 0.34251 lineto
0.41984 0.34286 lineto
0.42186 0.34332 lineto
0.42376 0.3439 lineto
0.42555 0.3446 lineto
0.42721 0.34541 lineto
0.42875 0.34635 lineto
0.43018 0.34741 lineto
0.43148 0.34859 lineto
0.43371 0.35131 lineto
0.43465 0.35285 lineto
0.43547 0.35452 lineto
0.43616 0.3563 lineto
0.43674 0.3582 lineto
0.4372 0.36022 lineto
0.43755 0.36235 lineto
0.43779 0.3646 lineto
0.43791 0.36695 lineto
0.43793 0.36942 lineto
0.43784 0.37198 lineto
0.43766 0.37465 lineto
0.43737 0.37742 lineto
0.43699 0.38028 lineto
0.43652 0.38323 lineto
0.42063 0.4388 lineto
Mistroke
0.41122 0.46805 lineto
0.40745 0.48175 lineto
0.40594 0.48821 lineto
0.40475 0.49436 lineto
0.40428 0.49731 lineto
0.4039 0.50017 lineto
0.40361 0.50294 lineto
0.40343 0.50561 lineto
0.40334 0.50817 lineto
0.40336 0.51064 lineto
0.40348 0.51299 lineto
0.40372 0.51524 lineto
0.40407 0.51737 lineto
0.40453 0.51939 lineto
0.40511 0.52129 lineto
0.4058 0.52307 lineto
0.40756 0.52628 lineto
0.40861 0.5277 lineto
0.40979 0.529 lineto
0.41109 0.53018 lineto
0.41252 0.53124 lineto
0.41406 0.53218 lineto
0.41572 0.53299 lineto
0.41751 0.53369 lineto
0.41941 0.53427 lineto
0.42143 0.53473 lineto
0.42356 0.53508 lineto
0.4258 0.53531 lineto
0.42816 0.53544 lineto
0.43062 0.53546 lineto
0.43319 0.53537 lineto
0.43586 0.53518 lineto
0.43862 0.5349 lineto
0.44443 0.53405 lineto
0.45059 0.53285 lineto
0.45705 0.53135 lineto
0.47074 0.52757 lineto
0.5 0.51816 lineto
0.52926 0.50875 lineto
0.54295 0.50497 lineto
0.54941 0.50347 lineto
0.55557 0.50227 lineto
0.55852 0.5018 lineto
0.56138 0.50142 lineto
0.56414 0.50114 lineto
0.56681 0.50095 lineto
0.56938 0.50086 lineto
0.57184 0.50088 lineto
0.5742 0.50101 lineto
0.57644 0.50124 lineto
Mistroke
0.57857 0.50159 lineto
0.58059 0.50205 lineto
0.58249 0.50263 lineto
0.58428 0.50333 lineto
0.58748 0.50508 lineto
0.58891 0.50614 lineto
0.59021 0.50732 lineto
0.59139 0.50862 lineto
0.59244 0.51004 lineto
0.59338 0.51158 lineto
0.5942 0.51325 lineto
0.59489 0.51503 lineto
0.59547 0.51693 lineto
0.59593 0.51895 lineto
0.59628 0.52108 lineto
0.59652 0.52333 lineto
0.59664 0.52568 lineto
0.59666 0.52815 lineto
0.59657 0.53071 lineto
0.59639 0.53338 lineto
0.5961 0.53615 lineto
0.59525 0.54196 lineto
0.59406 0.54811 lineto
0.59255 0.55457 lineto
0.58878 0.56827 lineto
0.57937 0.59753 lineto
0.56995 0.62678 lineto
0.56618 0.64048 lineto
0.56467 0.64694 lineto
0.56348 0.65309 lineto
0.56301 0.65604 lineto
0.56263 0.6589 lineto
0.56234 0.66167 lineto
0.56216 0.66434 lineto
0.56207 0.6669 lineto
0.56209 0.66937 lineto
0.56221 0.67172 lineto
0.56245 0.67397 lineto
0.5628 0.6761 lineto
0.56326 0.67812 lineto
0.56384 0.68002 lineto
0.56453 0.6818 lineto
0.56535 0.68347 lineto
0.56629 0.68501 lineto
0.56734 0.68643 lineto
0.56852 0.68773 lineto
0.57125 0.68997 lineto
0.57279 0.69091 lineto
0.57445 0.69172 lineto
0.57624 0.69242 lineto
Mistroke
0.57814 0.693 lineto
0.58016 0.69346 lineto
0.58229 0.69381 lineto
0.58453 0.69404 lineto
0.58689 0.69417 lineto
0.58935 0.69419 lineto
0.59192 0.6941 lineto
0.59459 0.69391 lineto
0.59735 0.69363 lineto
0.60021 0.69325 lineto
0.60316 0.69278 lineto
0.65873 0.67689 lineto
0.68799 0.66748 lineto
0.70168 0.6637 lineto
0.70814 0.6622 lineto
0.7143 0.661 lineto
0.71725 0.66053 lineto
0.72011 0.66016 lineto
0.72287 0.65987 lineto
0.72554 0.65968 lineto
0.72811 0.65959 lineto
0.73057 0.65961 lineto
0.73293 0.65974 lineto
0.73517 0.65997 lineto
0.7373 0.66032 lineto
0.73932 0.66078 lineto
0.74122 0.66136 lineto
0.74301 0.66206 lineto
0.74467 0.66287 lineto
0.74621 0.66381 lineto
0.74894 0.66605 lineto
0.75012 0.66735 lineto
0.75117 0.66877 lineto
0.75211 0.67031 lineto
0.75293 0.67198 lineto
0.75362 0.67376 lineto
0.7542 0.67566 lineto
0.75467 0.67768 lineto
0.75501 0.67981 lineto
0.75525 0.68206 lineto
0.75537 0.68441 lineto
0.75539 0.68688 lineto
0.7553 0.68944 lineto
0.75512 0.69211 lineto
0.75483 0.69488 lineto
0.75398 0.70069 lineto
0.75279 0.70684 lineto
0.75128 0.7133 lineto
0.74751 0.727 lineto
0.7381 0.75626 lineto
Mistroke
0.72868 0.78551 lineto
0.72491 0.79921 lineto
0.7234 0.80567 lineto
0.72221 0.81182 lineto
0.72174 0.81477 lineto
0.72136 0.81763 lineto
0.72107 0.8204 lineto
0.72089 0.82307 lineto
0.7208 0.82563 lineto
0.72082 0.8281 lineto
0.72094 0.83045 lineto
0.72118 0.8327 lineto
0.72153 0.83483 lineto
0.72199 0.83685 lineto
0.72257 0.83875 lineto
0.72326 0.84053 lineto
0.72502 0.84374 lineto
0.72607 0.84516 lineto
0.72725 0.84646 lineto
0.81746 0.83562 lineto
0.84672 0.82621 lineto
0.86041 0.82243 lineto
0.86687 0.82093 lineto
0.87303 0.81973 lineto
0.87598 0.81926 lineto
0.87884 0.81889 lineto
0.8816 0.8186 lineto
0.88427 0.81841 lineto
0.88684 0.81833 lineto
0.8893 0.81834 lineto
0.89166 0.81847 lineto
0.8939 0.8187 lineto
0.89603 0.81905 lineto
0.89805 0.81951 lineto
0.89995 0.82009 lineto
0.90174 0.82079 lineto
0.90494 0.82254 lineto
0.90637 0.8236 lineto
0.90767 0.82478 lineto
0.89683 0.91499 lineto
0.88741 0.94424 lineto
0.88364 0.95794 lineto
0.88213 0.9644 lineto
0.88094 0.97055 lineto
0.88047 0.9735 lineto
0.88009 0.97636 lineto
0.8798 0.97913 lineto
0.87962 0.9818 lineto
0.87953 0.98436 lineto
0.87955 0.98683 lineto
Mistroke
0.87967 0.98918 lineto
0.87991 0.99143 lineto
0.88026 0.99356 lineto
0.88072 0.99558 lineto
0.8813 0.99748 lineto
0.88199 0.99926 lineto
0.88281 1.00093 lineto
0.88375 1.00247 lineto
0.8848 1.00389 lineto
0.88598 1.00519 lineto
0.88728 1.00637 lineto
0.88871 1.00743 lineto
0.89025 1.00837 lineto
0.89191 1.00918 lineto
0.8937 1.00988 lineto
0.8956 1.01046 lineto
0.89762 1.01092 lineto
0.89975 1.01127 lineto
0.90199 1.0115 lineto
0.90435 1.01163 lineto
0.90681 1.01165 lineto
0.90938 1.01156 lineto
0.91205 1.01137 lineto
0.91481 1.01109 lineto
0.92062 1.01024 lineto
0.92678 1.00904 lineto
0.93324 1.00754 lineto
0.94693 1.00377 lineto
0.97619 0.99435 lineto
Mfstroke
grestore
grestore
MathPictureEnd
end


count __NXEPSOpCount sub {pop} repeat countdictstack __NXEPSDictCount
sub {end}
repeat __NXEPSSave restore
grestore
grestore
grestore
0 0 0 setup
gsave
0.345866 nxsetgray
57.004436 59.417675 162.444366 280.344238 oval fill
grestore
1 nxsetgray
57.004436 59.417675 162.444366 280.344238 oval matrix defaultmatrix
setmatrix
stroke
grestore
0 0 0 setup
gsave
0.345866 nxsetgray
57.004436 59.417675 320.550751 281.532623 oval fill
grestore
0.345866 nxsetgray
57.004436 59.417675 320.550751 281.532623 oval matrix defaultmatrix
setmatrix
stroke
grestore
grestore
0 0 0 setup
gsave
/Symbol findfont 24 scalefont [1 0 0 -1 0 0] makefont
113
exch
defineuserobject
113 execuserobject setfont
0 nxsetgray
[1 0 0 -1 0 559.491821] concat
113 execuserobject setfont
0 nxsetgray
262.274353 285 moveto (D) show
grestore
grestore
0 0 0 setup
gsave
113 execuserobject setfont
0 nxsetgray
[1 0 0 -1 0 561.885315] concat
113 execuserobject setfont
0 nxsetgray
32.288429 286.196747 moveto (N) show
grestore
grestore
0 0 0 setup
gsave
113 execuserobject setfont
0 nxsetgray
[1 0 0 -1 0 564.278687] concat
113 execuserobject setfont
0 nxsetgray
484.25061 287.393433 moveto (N) show
grestore
grestore
0 0 0 setup
gsave
113 execuserobject setfont
0 nxsetgray
[1 0 0 -1 0 918.508423] concat
113 execuserobject setfont
0 nxsetgray
33.432655 464.508301 moveto (g) show
grestore
grestore
0 0 0 setup
gsave
113 execuserobject setfont
0 nxsetgray
[1 0 0 -1 0 918.508423] concat
113 execuserobject setfont
0 nxsetgray
492.259735 464.508301 moveto (g) show
grestore
grestore
grestore
0 setgray
0.333333 setgray
gsave
0 0 540 720 rectclip
[1 0 0 -1 0 720] concat
grestore
grestore
grestore
showpage
__NXsheetsavetoken restore
